\documentclass[9pt,twocolumn,twoside]{osajnl}
\journal{pr} 
\setboolean{shortarticle}{false}
\usepackage{graphicx} 
\usepackage{amssymb}
\usepackage{lineno}
\usepackage{comment}
\renewcommand{\thesubsection}{\thesection.\Alph{subsection}}
\usepackage{graphicx} % for figures
\usepackage{amsmath,amsbsy,amsthm,mathrsfs,amsfonts,dsfont}
\usepackage{xcolor}
\usepackage{braket}
\usepackage{multirow}% multiculumn multirow
\usepackage{multicol}
\usepackage{array}     % column centering
\usepackage{comment}
\usepackage[normalem]{ulem}
\usepackage[colorlinks = true]{hyperref}
\usepackage{float}
\usepackage{aligned-overset}
\usepackage{comment}
\usepackage{cuted}
\newcommand{\tr}{\mathrm{Tr}}

\newtheorem{theorem}{Theorem}

\newtheorem{proposition}{Proposition}
\newtheorem*{thminf*}{Theorem (informal)}

\graphicspath{{./Figures/}}

\newcommand{\mc}{\mathcal}

\newcommand{\mbb}{\mathbb}
\newcommand{\mr}{\mathrm}

\title{Pilot-reference-free continuous-variable quantum key distribution with efficient decoy-state analysis}

\author[1,2]{Anran Jin}
\author[3,4,5]{Xingjian Zhang}
\author[6]{Liang Jiang}
\author[2]{Richard V.~Penty}
\author[6,*]{Pei Zeng}

\affil[1]{Yau Mathematical Sciences Center, Tsinghua University, Beijing 100084, China}
\affil[2]{Electrical Engineering Division, Department of Engineering, University of Cambridge, CAPE Building 9 JJ Thomson Avenue, CB3 0FA Cambridge, UK}
\affil[3]{QICI Quantum Information and Computation Initiative, Department of Computer Science, School of Computing and Data Science, The University of Hong Kong, Pokfulam Road, Hong Kong}
\affil[4]{Shenzhen Futian SUSTech Institute for Quantum Technology and Engineering, Shenzhen 518055, China}
\affil[5]{International Quantum Academy, Shenzhen 518048, China}
\affil[6]{Pritzker School of Molecular Engineering, The University of Chicago, Illinois 60637, USA}
\affil[*]{peizeng@uchicago.edu}

\begin{abstract} 
 Continuous-variable quantum key distribution (CV QKD) using optical coherent detectors is practically favorable due to its low implementation cost, flexibility of wavelength division multiplexing, and compatibility with standard coherent communication technologies. However, the security analysis and parameter estimation of CV QKD are complicated due to the infinite-dimensional latent Hilbert space. Also, the transmission of strong reference pulses undermines the security and complicates the experiments. In this work, we tackle these two problems by presenting a time-bin-encoding CV protocol with a simple phase-error-based security analysis valid under general coherent attacks. With the key encoded into the relative intensity between two optical modes, the need for global references is removed. Furthermore, phase randomization can be introduced to decouple the security analysis of different photon-number components. We can hence tag the photon number for each round, effectively estimate the associated privacy using a carefully designed coherent-detection method, and independently extract encryption keys from each component. Simulations manifest that the protocol using multi-photon components increases the key rate by two orders of magnitude compared to the one using only the single-photon component. Meanwhile, the protocol with four-intensity decoy analysis is sufficient to yield tight parameter estimation with a short-distance key-rate performance comparable to the best Bennett-Brassard-1984 implementation.

\end{abstract}

\setboolean{displaycopyright}{false}

\begin{document}	
\maketitle
\fancyhead{} %Clear header

% \printinunitsof{pt}\prntlen{\linewidth}

%%%%%%%%%%%%%%%%%%%%%%%%%%  body  %%%%%%%%%%%%%%%%%%%%%%%%%%
\section{Introduction}
Quantum key distribution (QKD) allows the generation of random secure keys between distant communication parties, of which the security is guaranteed by quantum physical laws. Apart from its theoretical advances, QKD is also one of the few quantum information processing technologies that can be robustly deployed in the fields, where photonic systems are considered the most suitable carriers of QKD operation. In general, two types of QKD protocols exist based on the detection methods: discrete-variable (DV) QKD~\cite{bennett1984quantum,Bennett1992} uses the single-photon detector or photon-number-resolving detector to generate discrete detection information, while continuous-variable (CV) QKD~\cite{Grosshans2002,Lin2019asymptotic,Matsuura2021Finite} applies optical homodyne or heterodyne detection to generate continuous measurement information.  

CV QKD has its advantages over DV QKD in short distances, mainly attributing to the distinct features of the coherent detectors used. The homodyne and heterodyne detectors are compatible with the standard classical communication and can be operated at much milder conditions than single-photon detectors. The spatial-temporal filtering of the local oscillators (LO) allows dense wavelength-division multiplexing with intense classical channels~\cite{Qi2010feasibility,Kumar2015coexistence,Eriksson2019WDM}, and the high quantum efficiency and operation rate give CV QKD high key rates in metropolitan distances~\cite{Wang2018high,Wang2020highspeed,Wang2022Gbps}. Moreover, the feasibility of on-chip implementations of the coherent detectors~\cite{Zhang2019integrated} promises large-scale integrated quantum networks. CV QKD is therefore considered highly practical and promising.  

However, there exist two major limitations to the reliability of CV QKD. First, the transmission of the strong local oscillators is usually necessary to set up the phase reference between the communication parties, yet this complicates the implementation in the multiplexing separation and the relative phase shift calibration with the signals~\cite{Soh2015selfreferenced}. The LO transmission also opens up security loopholes where the eavesdropper Eve can affect the estimation of the signal variance by manipulating the LO intensity~\cite{Ma2013LO,Fan2023quantum}, input time~\cite{Jouguet2013preventing} and wavelength~\cite{Huang2013quantum}. The ``local" local oscillator scheme~\cite{Soh2015selfreferenced,Qi2015generating} is a valid solution, yet it still requires the transmission of pilot pulses and compensation in the classical post-processing layer, which increase the experimental complexity. Second, the security of CV QKD in the finite-data regime under coherent attacks is still incomplete. In fact, for the traditional entanglement-distillation approach~\cite{Devetak2005}, the finite-size coherent attack security is only tackled for Gaussian-modulated CV QKD~\cite{Leverrier2017security}, which is, however, impractical since continuous modulation is never possible in reality. 

Recently, several remarkable works on CV QKD have been proposed, aiming at closing its security loopholes. In Ref.~\cite{Matsuura2021Finite,MatsuuraRefined2023}, Matsuura et al.~proposed a DV-like security analysis for the binary phase shift keying CV QKD. Their analysis covers finite size and coherent attack intrinsically since it follows the phase-error complementarity approach~\cite{koashi2009simple}, yet their protocol still assumes the transmission of local oscillators. Qi~\cite{Qi2021BB84} and Primaatmaja et al.~\cite{Primaatmaja2022Discrete} respectively proposed CV QKD protocols with two-mode encoding, generating dual-rail qubits that do not require global references. Their security analyses, however, do not cover the finite-data regime and coherent attacks. In fact, Qi's analysis requires repeated measurements and is only valid for individual attacks, and Primaatmaja et al.'s analysis is based on Devetak-Winter formula~\cite{Devetak2005} and is only valid for collective attacks. Hence, the gap in CV QKD between theory and practice is still a challenging problem to be tackled. 

In this work, we close this gap by proposing a new time-bin-encoding CV QKD protocol that enjoys both simple security proof and practical implementation. We remove the necessity of LO transmission by the two-mode encoding, hence closing the security loophole while simplifying the experimental setups. We follow the phase-error complementarity approach~\cite{lo1999Unconditional,shor2000Simple,koashi2009simple} so that the security naturally covers the general coherent-attack case. What is more, the intensity-based encoding allows phase randomization to be applied, where we can group the received signals based on the transmitted photon numbers and restore the tagging-based security analysis~\cite{GLLP2004,Ma2008PhD} and the decoy-state method~\cite{Lo2005Decoy,wang2005decoy}. Instead of generating secure key bits from all raw key bits, we can thus take advantage of the photon-number tags and distill key bits from the rounds with low phase-error rate. Our tagging-based security analysis builds a direct connection between CV QKD and the normal Bennett-Brassard-1984 (BB84) protocol: we clearly show how the multi-photon components in the CV QKD protocol contribute to a higher key rate in short distances. Compared with a similar protocol in Ref.~\cite{Primaatmaja2022Discrete} with numerical optimization under collective attacks, our protocol generates higher key rates under coherent attacks with simpler parameter estimation using four decoy levels, while also demonstrating good finite-size performance with a data size of $N=10^{10}$.

We will start with the protocol description of the time-bin-encoding CV QKD in Sec.~\ref{Sec:protocol}. We present its security analysis based on phase error correction~\cite{lo1999Unconditional,shor2000Simple,koashi2009simple} in Sec.~\ref{Sec::security}, identifying an equivalent protocol squashing the optical modes into qubits with identical key mapping statistics~\cite{GLLP2004,Beaudry2008squashing,Matsuura2021Finite} in Sec.~\ref{Subsec::EBsquashing}. We exploit the block-diagonal structures of both the source and the receiver in Sec.~\ref{Sec::Photon number tagging}, thus invoking the photon-number tagging technique~\cite{GLLP2004,Ma2008PhD} standard in DV QKD in this CV protocol. In Sec.~\ref{Sec::PER calculation}, we calculate the parameters in the key-rate formula with quantities on optical modes. The estimation of these quantities will be explained in Sec.~\ref{Sec::Parameter estimation} with homodyne tomography~\cite{DAriano2007HomoTomo} and decoy method~\cite{Lo2005Decoy,wang2005decoy}. We finally simulate the asymptotic and finite-size performances of the time-bin CV QKD under realistic fiber-channel setups in Sec.~\ref{Sec::results}. 

\section{Protocol Description}
\label{Sec:protocol}
\begin{figure*}[htbp!] 
    \centering
    \includegraphics[width = 0.8\linewidth]{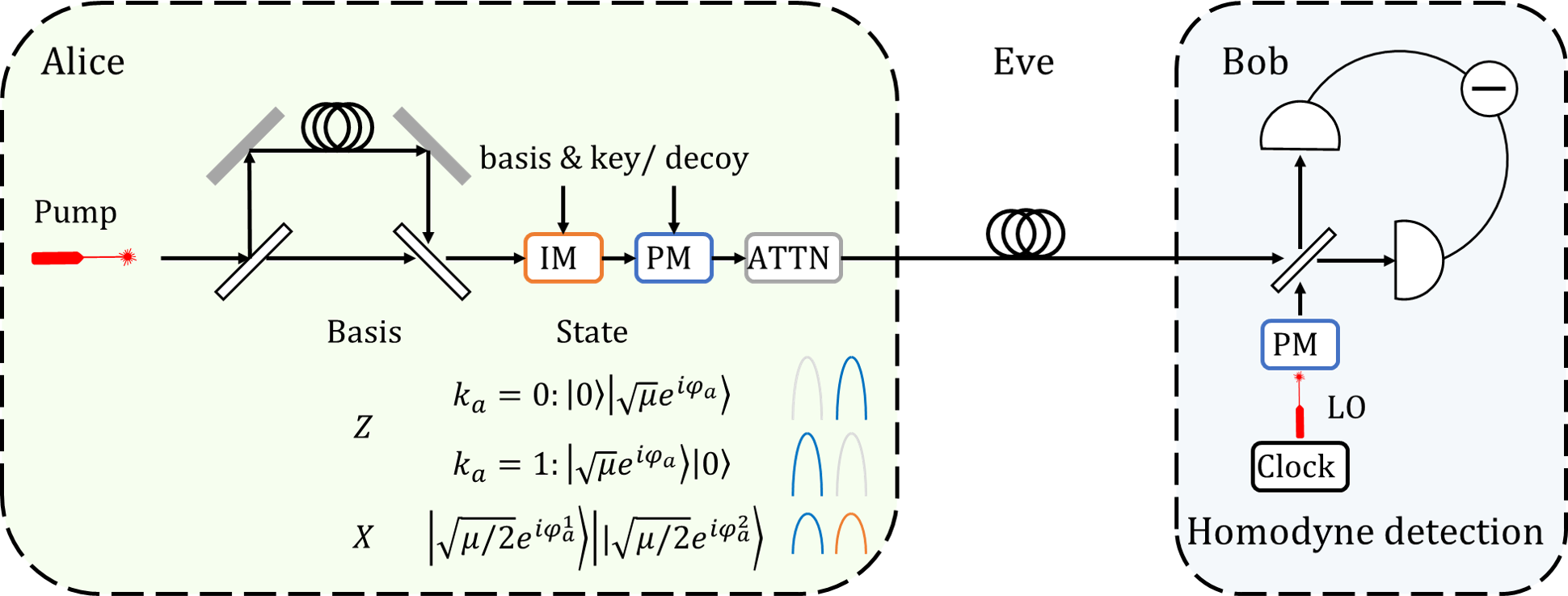}
    \captionsetup{justification = raggedright}
    \caption{Schematic diagram of the experimental setup. The setups of Alice and Bob are shaded in green and blue, respectively. Alice prepares two-mode phase-randomized states according to the basis choice and raw key value in key generation rounds, as shown in the table. In this work, we consider a time-bin encoding, where one obtains two modes via time delay. The state modulation consists of intensity modulation (IM), phase modulation (PM), and necessary attenuation (ATTN). Upon receiving the state, Bob measures each mode with homodyne detectors. He uses a synchronized clock to distinguish adjacent modes and applies phase modulation (PM) to the local oscillator (LO).}
    \label{Fig::ExpImplement}
\end{figure*}

We present the proposed time-bin-encoding CV QKD protocol in Table~\ref{table:protocol} and depict its schematic diagram in Fig.~\ref{Fig::ExpImplement}. The two communication parties, Alice and Bob, employ the time-bin degree of freedom to encode keys. They use $Z$-basis for key generation and $X$-basis for parameter estimation. At the moment, we do not present the details of the $X$-basis parameter settings. We will specify the choices of the random phase factors, $\varphi_a^1,\varphi_a^2,\varphi_b^1$ and $\varphi_b^2$, and the light intensity, $\mu_a\in\{\mu,\nu_1,\nu_2,0\}$, in Sec.~\ref{Sec::Parameter estimation}.

\begin{table*}[htbp!] 
  \centering
  \caption{Phase-randomized time-bin-encoding CV QKD}\label{table:protocol}
  \begin{tabular}{p{0.95\linewidth}}%{p{20cm}}
  \hline
  \hline
  \begin{enumerate}
  \item On Alice's side (source):
  \begin{itemize}
  \item \emph{$Z$-basis}:
  \begin{enumerate}
    \item Randomly select a key bit $k_a \in \{0,1\}$, a phase factor $\varphi_a\in [0,2\pi)$, and a light intensity $\mu_a\in\{\mu,\nu_1,\nu_2,0\}$.
    \item Prepare a coherent state of $\ket{0}_{A1} \ket{\sqrt{\mu_a}e^{i\varphi_a}}_{A2}$ for $k_a=0$ or $\ket{\sqrt{\mu_a}e^{i\varphi_a}}_{A1} \ket{0}_{A2}$ for $k_a=1$.
  \end{enumerate}
  \item \emph{$X$-basis}:
  \begin{enumerate}
    \item Randomly select two phase factors $\varphi_a^1$ and $\varphi_a^2$ and a light intensity $\mu_a\in\{\mu,\nu_1,\nu_2,0\}$.
    \item Prepare a coherent state of $\ket{\sqrt{\mu_a/2}e^{i\varphi_a^1}}\ket{\sqrt{\mu_a/2}e^{i\varphi_a^2}}$.
  \end{enumerate}
\end{itemize}
  \item Alice sends the state through an authenticated channel to Bob.
  \item On Bob's side (detection):
  \begin{itemize}
  \item \emph{$Z$-basis}:
  \begin{enumerate}
    \item Randomly select a phase factor $\varphi_b\in [0,2\pi)$.
    \item Use homodyne detectors both with LO phases $\varphi_b$ to measure the modes and obtain quadratures $q_1$ and $q_2$.
    \item Decode the key bit as $0$ if $|q_1|<\tau\wedge|q_2|>\tau$, $1$ if $|q_1|>\tau\wedge|q_2|<\tau$, and $\emptyset$ otherwise.
  \end{enumerate}
  \item \emph{$X$-basis}:
  \begin{enumerate}
    \item Randomly select two phases $\varphi_b^1$ and $\varphi_b^2$ independently.
    \item Use homodyne detectors with LO phases $\varphi_b^1$ and $\varphi_b^2$ to measure the modes and obtain quadratures $q_1$ and $q_2$.
    \item Use $\varphi_1,\varphi_2$ and $q_1,q_2$ for phase-error estimation (see Sec.~\ref{Sec::Parameter estimation}).
  \end{enumerate}
\end{itemize}
\item Alice and Bob perform basis sifting, where they obtain raw keys in the rounds they both choose $Z$-basis with light intensity $\mu_a=\mu$ and $k_b\neq\emptyset$.
\item Based on parameter estimation, Alice and Bob perform information reconciliation and privacy amplification to obtain final keys.
    \end{enumerate} \\
\hline
\hline
\end{tabular}
\end{table*}

Here, we briefly explain the idea behind the protocol design. The source states of our scheme resemble the ones in the time-bin-encoding BB84 protocol with coherent states~\cite{bennett1984quantum}, where the light intensities of the consecutive pulses naturally encode the key-bit information. As the key information is encoded in the relative intensity between the two modes, Alice does not need to send a pilot phase reference as in common CV QKD. The source states also resemble those in the coherent-one-way (COW) protocol~\cite{stucki2005fast,lavie2022improved}. The difference is that we randomize the phase of the coherent state pulses and employ homodyne detection for the measurement devices.

In our scheme, Bob decodes the key bit information, namely the $Z$-basis information, by measuring the light intensity of the pulses using the homodyne detectors. For instance, when Bob observes $q_1$ to be close to $0$ and $q_2$ to be far away from $0$, he may naturally guess that the original state sent by Alice corresponds to $k_a=0$. However, unlike the key decoding with photon-number detectors, the result of measuring a coherent state's quadrature is subjected to a Gaussian distribution rather than a fixed value. The inherent shot noise of the homodyne detection introduces an intrinsic error in distinguishing a vacuum state from a pulse with a non-zero intensity~\cite{weedbrook2012gaussian}. To suppress the bit error, we introduce a threshold value, $\tau$, in key decoding. The pulse intensity will be considered non-zero only when the quadrature magnitude is larger than $\tau$. So a bit 0 will be decoded if $|q_1|<\tau$ and $|q_2|>\tau$ and bit 1 if $|q_1|>\tau$ and $|q_2|<\tau$. The choice of $\tau$ should be optimized with respect to the channel transmittance and pulse intensity. 
As we will show in Appendix~\ref{Sec:refinedkeymapping}, although this key mapping scheme can be further optimized, the performance of the simple key mapping scheme is already near-optimal.

The $X$-basis is designed to estimate the information leakage of different photon-number components of the $Z$-basis. Thanks to the phase randomization for both the sources and the detectors, the $Z$-basis states are block-diagonal on the total photon-number basis on the two optical modes after emitted from the source and before being measured by the homodyne detectors. As we will clarify in Sec.~\ref{Sec::security}, we can equivalently introduce total photon-number measurements at these two locations. As a result, Eve's eavesdropping strategy is effectively ``twirled'' to a photonic channel that only acts on the states incoherently with respect to the total photon numbers. One can thus virtually tag the emitted and received pulses according to the photon-number space, allowing the Gottesman-L\"utkenhaus-Lo-Preskill (GLLP) framework~\cite{GLLP2004} for analyzing the key privacy contained in each photon-number subspace. In particular, dealing with photon-number spaces effectively brings our security analysis to the DV regime. In Sec.~\ref{Sec::security}, we shall construct observables to estimate the $m$-photon component phase-error rates $e^X_{m,m}$ for privacy estimation. Intuitively, the phase-error rates $e^X_{m,m}$ provide upper bounds on the key information leakage to the eavesdropper, Eve.

To estimate the $m$-photon component phase-error rates, $e^X_{m,m}$, ideally, we need a source emitting the photon-number cat states, $(\ket{0}\ket{m}\pm\ket{m}\ket{0})/\sqrt{2}$, and photon-number-resolving measurements that distinguish the cat states. While this is not directly implementable, we can use only coherent states and homodyne measurements to establish unbiased estimators of $e^X_{m,m}$, as shown in Table~\ref{tab:Realization}. On the source side, we employ a generalized decoy-state method to estimate the behaviors of photon-number-cat states using coherent states with various intensities~\cite{Lo2005Decoy,wang2005decoy}, which shall be discussed in Sec.~\ref{Subsec::decoy}. On the detection side, ideally, we also want to measure the photon-number-cat states to obtain unbiased estimation of the phase-error rates, $e_{m,m}^X$. While this is not directly measurable in practice, we employ the homodyne tomography technique and estimate the photon-number-cat state measurement via quadrature measurement results~\cite{vogel1989determination,smithey1993measurement,d1994detection,d1995homodyne,leonhardt1995measuring,d1995tomographic}, which shall be discussed in Sec.~\ref{Sec:Homotomo}.

\begin{table*}[htbp!]
\centering
\captionsetup{justification = raggedright}
\caption{State preparation and detection settings in the ideal implementation and the realistic implementation. For brevity, we omit the subscripts of modes and express the detection with the measurement operators. In key generation rounds, Bob applies phase-randomized homodyne detection for key-decoding. The expression of measurement operator $\hat{\Pi}(q_1,q_2)$ is given in \eqref{Eq::homo2fock}, where $q_1$ and $q_2$ represent the quadratures of the two modes. The operator is block-diagonal on the total photon-number basis. For parameter estimation, ideally, Alice sends photon-number-cat states, and Bob performs a corresponding projective measurement. 
%For simplicity, we denote the projector defined by the cat state as $\hat{\Pi}\left(\ket{n_1}\ket{n_2}\pm\ket{n_2}\ket{n_1}\right)$. 
In the realistic setting, Alice can only prepare phase-randomized weak coherent states, and Bob can only perform phase-randomized homodyne measurements. The homodyne measurement operator, $\hat{Q}_{\varphi_1}\otimes\hat{Q}_{\varphi_2}$, is given in \eqref{eq:QuadratureObs}. Afterward, Bob estimates the photon-number-cat state measurement expectations via homodyne tomography methods, as shown in \eqref{eq:HomoTomoEst}.}
\begin{tabular}{ccccc}
\hline
\hline
\multirow{2}{*}{\quad basis\quad} & \multicolumn{2}{c}{source} & \multicolumn{2}{c}{detection} \\   
\cline{2-5}
 & ideal & real & ideal & real \\
\hline
\multirow{2}{*}{$Z$} & $\ket{0}\ket{m}$ & $\ket{0}\ket{\sqrt{\mu}e^{i\varphi_a}}$ & \multirow{2}{*}{$\hat{\Pi}(q_1,q_2)$, \eqref{Eq::homo2fock}} & \multirow{2}{*}{$\hat{\Pi}(q_1,q_2)$} \\
 & $\ket{m}\ket{0}$ & $\ket{\sqrt{\mu}e^{i\varphi_a}}\ket{0}$ & & \\
 \hline
 \multirow{2}{*}{$X$} & \multirow{2}{*}{$\frac{1}{\sqrt{2}}(\ket{0}\ket{m}\pm\ket{m}\ket{0})$} & \multirow{2}{*}{$\quad \ket{\sqrt{\frac{\mu}{2}}e^{i\varphi_a^1}}\ket{\sqrt{\frac{\mu}{2}}e^{i\varphi_a^2}}\quad$} & \multirow{2}{*}{$\frac{1}{2}\left(\ket{0}\ket{m}\pm\ket{m}\ket{0}\right)\left(\bra{0}\bra{m}\pm\bra{m}\bra{0}\right)$} & \quad$\hat{Q}_{\varphi_1}\otimes\hat{Q}_{\varphi_2}$, \eqref{eq:QuadratureObs}, \\ & & & & \quad estimation via \eqref{eq:HomoTomoEst}\\
\hline
\hline
\end{tabular}
\label{tab:Realization}
\end{table*}

We briefly remark on the performance of homodyne detection. In key decoding, one may consider the homodyne detection as ill-performed single-photon detectors that introduce an inevitable bit-error rate. On the other hand, homodyne detection allows for more efficient parameter estimation than single-photon detection. As we shall discuss later, the set of all quadrature operators spans the underlying mode, thus allowing one to express any linear operator in terms of the quadrature operators. Therefore, with proper transformation of the quadrature measurement results, homodyne detection allows one to obtain an unbiased estimation of linear operator expectations. This is the reason for accurately estimating phase-error rates with repeated homodyne measurements, including those of the multi-photon components. In comparison, as the single-photon detection is not information-complete, estimation of multi-photon observables requires more complex setups such as sequential beam splitting~\cite{kumazawa2019rigorous}, and one can only obtain upper and lower bounds rather than an unbiased estimation.

\section{Security analysis}
\label{Sec::security}
\begin{figure*}[htbp!]
    \centering
    \includegraphics[width = 0.9\linewidth]{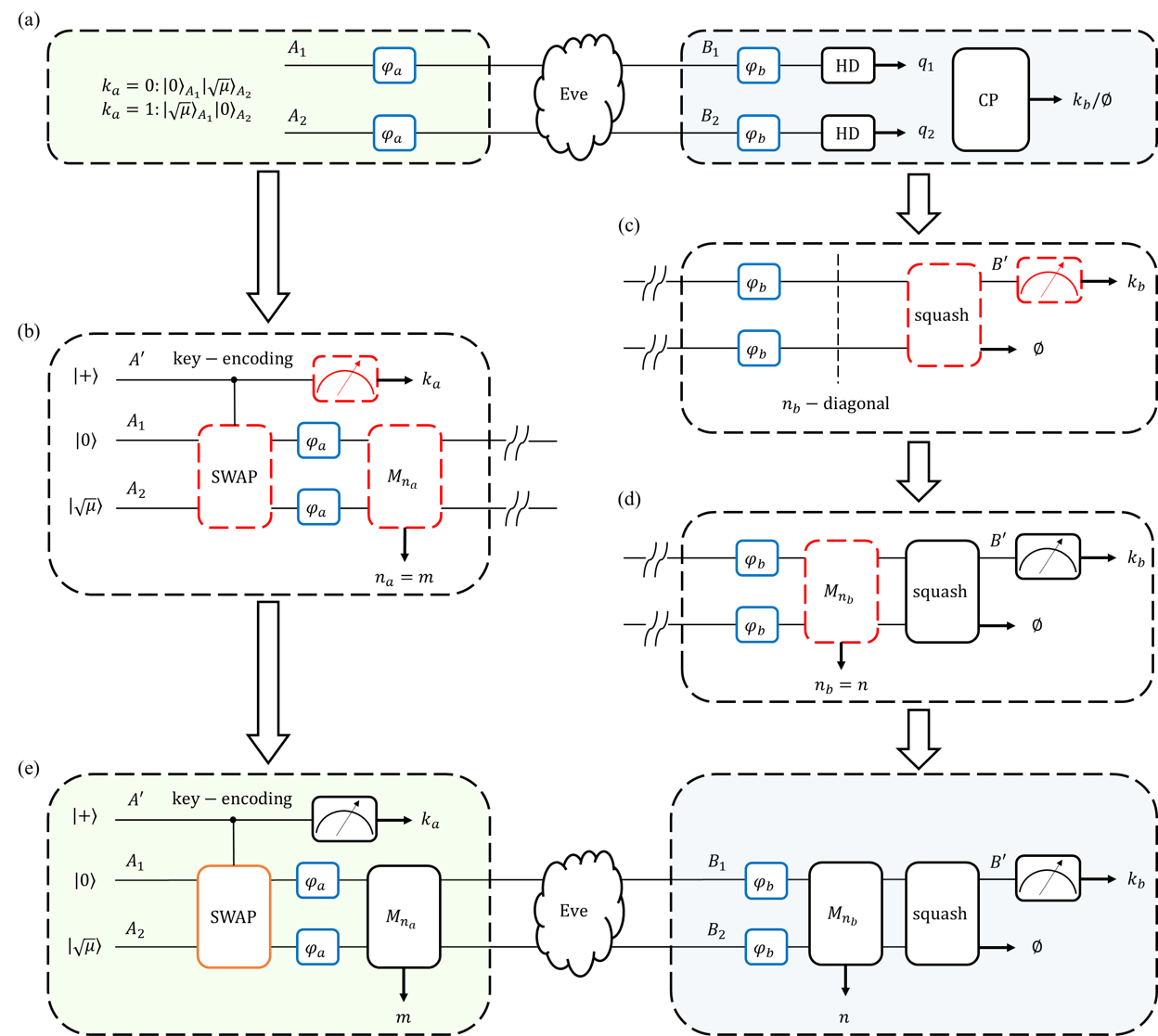}
    \captionsetup{justification = raggedright}
    \caption{Equivalent quantum circuits in key generation rounds. Reductions in each step are plotted with red dashed boxes. (a) The realistic implementation. The operations on Alice's and Bob's sides are shaded in green and blue, respectively. Alice prepares weak coherent states on two modes, which depend on the basis choice and the raw key value. On Bob's side, Bob measures the two modes with homodyne detectors (HD) and obtains quadratures $q_1$ and $q_2$. Afterward, Bob performs classical post-processing (CP) on the data and obtains a raw key $k_b$ probabilistically, where the key decoding may fail due to the key mapping threshold, denoted as $\emptyset$. The blue rounded boxes represent phase randomization processes in state preparation or for the LOs in homodyne detection.
    (b) Equivalent entanglement-based state preparation. Key encoding can be interpreted as a qubit control operation on two modes where the control qubit measurement gives Alice's raw key $k_a$. The joint state on the two modes is diagonal on the Fock basis after phase randomization. One can insert a photon-number measurement, $\hat{M}_{n_a}$, and read out the total photon number, $m$, without changing the state. (c) Equivalent key-decoding measurement. The joint state of the two modes becomes diagonal on the Fock basis due to detector phase randomization. In key decoding, the modes are first squashed into a qubit probabilistically, where the failure gives the abort signal $\emptyset$. Upon successful squashing into a qubit, the computational-basis measurement gives the raw key bit. (d) Due to detector phase randomization, one can insert a photon-number measurement, $\hat{M}_{n_b}$, and read out the total photon number, $n$, without changing the state. (e) Equivalent circuit for security analysis. After the above reductions, the key generation measurements can be equivalently defined on a pair of (sub-normalized) qubit states.}
    \label{Fig::setup}
\end{figure*}

We analyze the security of our phase-randomized time-bin-encoding CV QKD protocol along the complementarity approach~\cite{shor2000Simple,GLLP2004,koashi2009simple}. As outlined in Fig.~\ref{Fig::setup}, we shall set up a series of equivalent protocols of the realistic implementation that do not change the statistics of any observer, with which we define the phase-error observable and estimate the key privacy. In Sec.~\ref{Subsec::EBsquashing}, we shall prove that raw key generation can be effectively regarded as qubit measurements on a pair of entangled qubits, which allows us to borrow the mature complementarity-based security analysis in the DV regime. In brief, on the source side, we transform the preparation of key states to an entanglement-based protocol~\cite{lo1999Unconditional,shor2000Simple}, where a qubit measurement controls the key-encoding process, as shown in Fig.~\ref{Fig::setup}(b). On the detection side, we prove that the homodyne measurement can be squashed into an effective qubit measurement, as shown in Fig.~\ref{Fig::setup}(c). Moreover, in Sec.~\ref{Sec::Photon number tagging}, we shall rigorously prove that phase randomization twirls the photonic modes into diagonal states on the Fock basis and explain how to apply the tagging idea of the GLLP framework~\cite{GLLP2004,Ma2008PhD}. We also show how to estimate the phase-error rates for different photon-number components from Fock-basis observables in Sec.~\ref{Sec::PER calculation}. Later in Sec.~\ref{Sec::Parameter estimation}, we show that the estimation can be realized in the realistic implementation with coherent states and homodyne detection.

To focus on the essence of security analysis, we present the result in a single-round analysis in this section, where one can interpret it as the quantum Shannon limit under collective attacks. Namely, Eve applies the same attacking strategy to the quantum signal of Alice and Bob in each round. At the end of the protocol, Eve applies an optimal joint operation to guess the users' final key. Note that a collective attack is stronger than an individual attack, where Eve must measure her side information before the classical post-processing in an individual attack.
Nevertheless, the complementarity-based security analysis is inherently adapted to the most general case, namely the coherent attack, where the statistics over the rounds may not be independent and identically distributed (i.i.d.)~\cite{xu2020secure}. Unlike the collective attack, Eve may vary her attacking strategies in different rounds. Readers may refer to Ref.~\cite{scarani2009security} for a detailed discussion on the various types of attacks.
We will discuss the parameter estimation with non-i.i.d.~finite statistics in Sec.~\ref{Sec:CoherentAttack}.

\subsection{Entanglement-based squashing protocol}\label{Subsec::EBsquashing}
Here, we show the equivalence of the time-bin CV QKD protocol to a qubit-based entanglement distribution protocol, where the protocols generate the same transmitted quantum states and measurement statistics. The latter protocol enables us to simplify the security analysis and estimate the information leakage from phase-error rates.

We first focus on the key-generation rounds in the protocol where both users choose the $Z$-basis, of which the whole procedure is depicted in Fig.~\ref{Fig::setup}(a). In the realistic implementation, Alice prepares phase-randomized coherent states,
%\begin{strip}
\begin{equation}
  \int_0^{2\pi} \frac{d\varphi_a}{2\pi}\ket{\Psi(k_a)_{\varphi_a}}_{A_1A_2}\bra{\Psi(k_a)_{\varphi_a}},
\end{equation}
where
\begin{eqnarray}
\ket{\Psi(k_a)_{\varphi_a}}_{A_1A_2} =
\left\{
\begin{tabular}{ll} $\ket{0}_{A_1}\ket{\sqrt{\mu}e^{i\varphi_a}}_{A_2}$, &
if $k_a=0$, \\
$\ket{\sqrt{\mu}e^{i\varphi_a}}_{A_1}\ket{0}_{A_2}$, & if $k_a=1$.
\end{tabular}
\right.
\end{eqnarray}
% \end{strip}
We denote the optical modes sent to Bob as $A_1$ and $A_2$, which are CV systems. Throughout this paper, we treat the phase of optical modes, $\varphi_a$, as fully randomized over $[0,2\pi)$. Finite phase randomization, $\varphi_a\in\{2j\pi/D\}_{j\in[D]}$, suffices for a practical implementation, where its difference from the full phase randomization is negligible when $D$ is sufficiently large~\cite{Cao2015discrete}. This is also the case in later discussions on the detector phase randomization. Alice's key-state preparation can be effectively seen as an entanglement-based protocol~\cite{lo1999Unconditional,shor2000Simple}. Given the phase value, $\varphi_a$, Alice first prepares the following entangled state,
\begin{equation}
\label{Eq::DCCEnt}
\begin{split}
    \ket{\Psi_{\varphi_a}}_{A'A_1A_2} =&\frac{1}{\sqrt{2}}\big(\ket{0}_{A'}\ket{\Psi(k_a=0)_{\varphi_a}}_{A_1A_2} \\ &+ \ket{1}_{A'}\ket{\Psi(k_a=1)_{\varphi_a}}_{A_1A_2}\big),
\end{split}
\end{equation}
where system $A'$ is a qubit system that superposes the two possible key states. The entangled state can be prepared by the quantum circuit in Fig.~\ref{Fig::setup}(b). Up to phase randomization, systems $A'$ and $A_1A_2$ are initialized in $\ket{+}$ and $\ket{0}\ket{\sqrt{\mu}}$, and a control-swap operation is then applied from the qubit system to the optical modes. Alice obtains raw key bit $k_a$ by measuring system $A'$ on the computational basis, and the optical modes are prepared into the corresponding key state, $\ket{\Psi(k_a)_{\varphi_a}}$. The complementary observable of Alice's key-generation measurement can thus be defined over qubit system $A'$, which measures the complementary basis of  $\{\ket{+},\ket{-}\}:=\{(\ket{0}\pm\ket{1})/\sqrt{2}\}$.

At the detection side in Fig.~\ref{Fig::setup}(a), Bob receives two optical modes $B_1$ and $B_2$, takes homodyne measurements, and maps the quadratures to a raw key or an abort signal. This process can be described by a trace-non-preserving completely positive map,
\begin{equation}
\label{Eq::SquashPR}
\begin{split}
  \mathcal{F}_{\text{rand}}^{B_1B_2\to B'}(\hat{\rho}_{B_1B_2}) =&\int_0^{2\pi}\frac{d\varphi_b}{2\pi}\int_{\mathbf{R_0}}dq_1 dq_2  \\ &\quad\hat{K}^{(q_1,q_2,\varphi_b)}\hat{\rho}_{B_1B_2} \hat{K}^{(q_1,q_2,\varphi_b)\dagger},
\end{split}
\end{equation}
where
\begin{equation}
\begin{split}
  \hat{K}^{(q_1,q_2,\varphi_b)} :=& \ket{0}_{B'}\bra{q_1(\varphi_b),q_2(\varphi_b)}_{B_1,B_2} \\ &+ \ket{1}_{B'}\bra{q_2(\varphi_b),q_1(\varphi_b)}_{B_1,B_2},
\end{split}
\end{equation}
$\ket{q(\varphi)}$ is the rotated position eigenstate of quadrature observable
\begin{equation}\label{eq:QuadratureObs}
  \hat{Q}_\varphi = \hat{a} e^{-i\varphi} + \hat{a}^\dagger e^{i\varphi},
\end{equation}
with $\hat{a}$ and $\hat{a}^\dagger$ denoting the annihilation and creation operators, respectively, and $\mathbf{R_0}\in\mathbb{R}^2$ records the region that decodes the real-valued tuple, $(q_1,q_2)$, as $k_b=0$. Note that in our protocol, $\mathbf{R_0} = \{|q_1|<\tau\} \times \{|q_2|>\tau\}$, and the region decodes the quadratures to $k_b=1$ under the mapping $(q_1,q_2) \mapsto (q_2,q_1)$, which we denote as $\mathbf{R_1} = \{|q_1|>\tau\} \times \{|q_2|<\tau\}$. The LOs of homodyne measurements are synchronically randomized, as denoted by $\varphi_b$ in \eqref{Eq::SquashPR}. As the key-decoding region does not cover the entire parameter space, $\mathcal{F}_{\text{rand}}^{B_1B_2\to B'}$ is hence not trace-preserving, where $\tr[\mathcal{F}_{\text{rand}}^{B_1B_2\to B'}(\hat{\rho}_{B_1B_2})]$ gives the probability of obtaining raw key bit $k_b\in\{0,1\}$. Bob's raw key can be equivalently seen as obtained by measuring the squashed sub-normalized qubit on the computational basis, and the probabilities are given by
\begin{equation}
\begin{aligned}
&\Pr(k_b=0) = \bra{0}\mathcal{F}_{\text{rand}}^{B_1B_2\to B'}(\hat{\rho}_{B_1B_2})\ket{0} \\ =& \int_0^{2\pi}\frac{d\varphi_b}{2\pi}\int_{\mathbf{R_0}}dq_1 dq_2 \bra{q_1(\varphi_b),q_2(\varphi_b)}\hat{\rho}_{B_1B_2}\ket{q_1(\varphi_b),q_2(\varphi_b)},\\
&\Pr(k_b=1) =\bra{1}\mathcal{F}_{\text{rand}}^{B_1B_2\to B'}(\hat{\rho}_{B_1B_2})\ket{1} \\
=&\int_0^{2\pi}\frac{d\varphi_b}{2\pi}\int_{\mathbf{R_1}}dq_1 dq_2 \bra{q_1(\varphi_b),q_2(\varphi_b)}\hat{\rho}_{B_1B_2}\ket{q_1(\varphi_b),q_2(\varphi_b)}.
\end{aligned}
\end{equation}
Similar to the treatment to $A'$, we can define the complementary observable of Bob's key generation measurement on qubit system $B'$.

\subsection{Photon-number tagging of the source and receiver} 
\label{Sec::Photon number tagging}
In the last section, we have shown that raw keys can be equivalently seen as generated from qubit measurements on $A'$ and $B'$. Should Alice and Bob instead measure the qubit system on the complementary bases, the probability they obtain different results, or the phase-error rate, $e^X$, could be used to upper-bound the average privacy amplification cost per round as $h(e^X)$, where $h(p)=-p\log{p}-(1-p)\log(1-p)$ is the binary entropy function. Nevertheless, the actual privacy leakage may be less than the direct calculation. Note that the above privacy leakage estimation is averaged over the overall quantum state transmitted from Alice to Bob. The contribution to the privacy leakage of different components in quantum signals can differ. For instance, Eve can apply the photon-number-splitting (PNS) attack in the rounds in which Alice transmits two photons and Bob receives only a single photon~\cite{Brassard2000limitations,Ltkenhaus2000security}; hence no privacy should be expected, rendering the phase-error probability to be $1/2$ in these rounds. If Alice and Bob can distinguish such rounds from the others, they can simply discard them in privacy amplification. The GLLP framework makes the above statement rigorous~\cite{GLLP2004,Ma2008PhD}. Suppose Alice and Bob can categorize the transmitted quantum signals into different groups, or tags, and evaluate phase-error probabilities separately. The privacy amplification cost can be evaluated by $\sum_iQ_ih(e^X_i)$, where $Q_i$ is the probability that a signal in the $i$'th group is transmitted and detected, namely the gain, and $e^X_i$ is the phase-error probability of the group. Due to the concavity of the entropy function, this estimation is no larger than $h(\sum_iQ_ie^X_i)$.

In DV QKD, the tagging idea has been well practiced. In the coherent-state-based BB84 protocol, phase randomization on the source side diagonalizes the quantum signals on the Fock basis~\cite{Lo2005Decoy,wang2005decoy}, and an ideal single-photon detector naturally distinguishes the single photon components from other detected Fock components, allowing Alice and Bob to tag the quantum states with respect to the photon number~\cite{Ma2005practical}. Similarly, we now prove that the photon-number tag can also be applied to the phase-randomized CV QKD protocol in Table~\ref{table:protocol}. On the source side, the phase randomization diagonalizes the state on the joint Fock basis,
\begin{equation}
\label{Eq::Zsource}
\begin{split}
\hat{\rho}^Z=&\int_0^{2\pi} \frac{d\varphi_a}{2\pi} \quad \frac{1}{2}\ket{0}_{A_1}\bra{0} \otimes \ket{\sqrt{\mu}e^{i\varphi_a}}_{A_2}\bra{\sqrt{\mu}e^{i\varphi_a}} \\ &+ \frac{1}{2}\ket{\sqrt{\mu}e^{i\varphi_a}}_{A_1}\bra{\sqrt{\mu}e^{i\varphi_a}} \otimes \ket{0}_{A_2}\bra{0}\\
= &\sum_{m=0}^{\infty} \text{Pr}_{\mu}(m)\frac{1}{2}\left(\ket{0m}_{A_1A_2}\bra{0m} + \ket{m0}_{A_1A_2}\bra{m0}\right),
\end{split}
\end{equation}
where $\text{Pr}_\mu(m) = e^{-\mu}\mu^m/m!$ is the Poisson distribution. Consequently, one can virtually insert a photon-number measurement after phase randomization to measure the total photon number on the two modes without changing the state, as shown in Fig.~\ref{Fig::setup}(b). On the detection side, when Bob takes the $Z$-basis measurement, the phase-randomized homodyne detector POVM elements can be expanded on the Fock basis~\cite{Primaatmaja2022Discrete},
\begin{equation}
\label{Eq::homo2fock}
\begin{aligned}
& \hat{\Pi}(q_1,q_2) \\ 
=& \int_0^{2\pi} \frac{d\varphi_b}{2\pi} \ket{q_1(\varphi_b)}_{B_1}\bra{q_1(\varphi_b)} \otimes \ket{q_2(\varphi_b)}_{B_2}\bra{q_2(\varphi_b)}\\
=& \sum_{n=0}^\infty\sum_{k_0=0}^n\sum_{l_0=0}^n \psi_{k_0}(q_1)\psi_{l_0}(q_1)\psi_{n-k_0}(q_2)\psi_{n-l_0}(q_2) \\ &\quad\ket{k_0,n-k_0}_{B_1B_2}\bra{l_0,n-l_0},
\end{aligned}
\end{equation}
where
\begin{equation}
\psi_{n}(q_j) = \frac{1}{\sqrt{2^nn!\sqrt{2\pi}}}H_n(q_j/\sqrt{2})e^{-q_j^2/4}
\end{equation}
is the coordinate representation of Fock state $\ket{n}$, with $H_n$ being the $n$-th Hermite polynomial. Therefore, one can virtually insert another photon-number measurement after phase randomization before the squashing channel \eqref{Eq::SquashPR} on the detection side to measure the total photon number of the received state, as shown in Fig.~\ref{Fig::setup}(d).

Based on the above results, we depict a virtual quantum circuit of the protocol when both Alice and Bob chooses the $Z$-basis in Fig.~\ref{Fig::setup}(e). We denote the photon-number measurement results on the source side and the detection side as $m$ and $n$, respectively. Alice and Bob can thus distill secrete keys separately based on the photon-number tag of $(m,n)$.

A lower bound on the key rate can then be given by~\cite{GLLP2004,Ma2008PhD}
\begin{equation}
\label{Eq::Keyforward}
  r \geq \sum_{m=0}^{\infty} Q_{m,m}[1-h(e_{m,m}^X)] - fQ^Z h(e^Z),
\end{equation}
where $Q_{m,m}$ and $e_{m,m}^X$ denote the gain and the phase-error rate in the rounds where $m$ photons are sent and $m$ photons are accepted, $Q^Z$ is the $Z$-basis gain, $e^Z$ is the bit-error rate, and $f$ is the efficiency of information reconciliation. Note not to confuse the gains with quadrature observables. In addition, since Bob's key decoding succeeds probabilistically where he only accepts quadratures above the threshold, we use the term ``accepting'' to represent receiving a certain state and passing the post-selection. All the gains and error rates in the key-rate formula are restricted to the rounds with light intensity $\mu_a=\mu$. We discard the rounds where the total photon number decreases after state transmission, as the photons that are lost may come from Eve's interception, with which Eve can apply a PNS attack. The corresponding phase-error probability is $1/2$; hence these rounds do not contribute to key generation. In addition, as the transmission channel is naturally lossy in a usual setting, we do not account for the terms where the total photon number increases.

Note that the key-rate formula in \eqref{Eq::Keyforward} assumes forward reconciliation, where Bob reconciles his raw keys to Alice's, $k_a$, and then the users perform privacy amplification. The rounds where Alice sends a non-vacuum state while Bob receives a vacuum state are hence insecure, since the information carriers are lost through the channel. Instead, if reverse reconciliation is used, where Alice reconciles her raw keys to Bob's, the rounds where Bob receives a vacuum state become secure. One can interpret Bob's raw keys in these rounds as generated from local random numbers, and no information is known {\emph{a priori}} in transmission. 
This is a common practice in usual CV QKD and in accordance with the observation in Ref.~\cite{Qi2021BB84}. 
The fact that the vacuum component can also contribute to key rate formula is first observed in Ref.~\cite{lo2005getting}.
We present as Theorem~\ref{Thm:reversekeyrate} the key-rate lower bound with reverse reconciliation as the main key-rate formula to be used throughout this paper.

\begin{theorem}
\label{Thm:reversekeyrate}
For the time-bin CV QKD protocol in Table~\ref{table:protocol}  with reverse reconciliation, in the asymptotic limit of an infinite data size, the distillable secure key rate $r$ is lower bounded by $r_{\mathrm{rev}}$,  
\begin{equation}
\label{Eq::Keyrev}
  r\geq r_{\mathrm{rev}} = Q_{*,0} + \sum_{m=1}^{\infty} Q_{m,m}[1-h(e_{m,m}^X)] - fQ^Z h(e^Z),
\end{equation}
where $Q_{m,m}$ and $e_{m,m}^X$ denote the gain and the phase-error rate in the rounds where $m$ photons are sent and $m$ photons are accepted, $Q^Z$ is the $Z$-basis gain, $e^Z$ is the bit-error rate, and $f$ is the efficiency of information reconciliation. $Q_{*,0}$ represents the gain of the rounds where Bob accepts a vacuum state for whatever state sent by Alice.
\end{theorem}

\subsection{Phase-error probability calculation}
\label{Sec::PER calculation}
We now evaluate the key-rate formula in \eqref{Eq::Keyrev} with Fock-basis observables~\cite{Matsuura2021Finite}. The bit-error rate $e^Z$ can be directly measured, as the $Z$-measurement statistics in the entanglement-based squashing model are the same as the realistic statistics. To evaluate the gains and phase-error probabilities, we first determine the state before the phase-error measurement under each photon-number tag. Define $\hat{P}_m$ as the projector onto the $m$-photon state on modes $A_1$ and $A_2$. When sending $m$ photons, the source in Fig.~\ref{Fig::setup}(b) collapses to
\begin{equation}
\label{Eq::photonsource}
\begin{split}
  \hat{P}_m^{A_1A_2}\hat{\rho}_{A'A_1A_2}\hat{P}_m^{A_1A_2} &= \text{Pr}_\mu(m)\ket{\Psi_m}_{A'A_1A_2}\bra{\Psi_m},
\end{split}
\end{equation}
where
\begin{equation}
\label{eq:Poisson}
\begin{split}
  \ket{\Psi_m}_{A'A_1A_2} &= \frac{1}{\sqrt{2}}(\ket{0}_{A'}\ket{0m}_{A_1A_2} + \ket{1}_{A'}\ket{m0}_{A_1A_2}), \\
  \text{Pr}_\mu(m) &= \frac{e^{-\mu}\mu^m}{m!}.
\end{split}
\end{equation}
%is the probability that the source emits $m$ photons.
Upon transmitting the $m$-photon state, $\ket{\Psi_m}_{A'A_1A_2}$, the $n$-photon state is selected on the detection side after the squashing channel,
\begin{equation}
\begin{aligned}
\label{Eq::squashnorm}
\mathcal{F}_{\text{rand}}^{B_1B_2\to B'} & \left\{\hat{P}_n^{B_1B_2}\mathcal{N}_E^{A_1A_2\to B_1B_2}\left[\text{Pr}_\mu(m)\ket{\Psi_m}_{A'A_1A_2}\bra{\Psi_m}\right]\hat{P}_n^{B_1B_2}\right\} \\ 
&= Q_{m,n}\hat{\rho}_{A'B'}^{(m,n)},
\end{aligned}
\end{equation}
where $\mathcal{N}_E^{A_1A_2\to B_1B_2}$ represents Eve's channel, and  $Q_{m,n}$ denotes the probability of sending an $m$-photon state and accepting an $n$-photon state, namely the gain for the states tagged by the photon-number tuple, $(m,n)$. Note the probability that Bob aborts the signal is reflected in $Q_{m,n}$. The normalized state, $\hat{\rho}_{A'B'}^{(m,n)}$, is a bipartite qubit, with which we evaluate the phase-error probability,
\begin{equation}
  e_{m,n}^X = \tr\left[\hat{\rho}_{A'B'}^{(m,n)}\left(\ket{+-}_{A'B'}\bra{+-}+\ket{-+}_{A'B'}\bra{-+}\right)\right].
\end{equation}
With respect to the complementary-basis measurement result on the qubit $A'$, $+$ or $-$, the state on modes $A_1$ and $A_2$ collapses to
\begin{equation}
\label{Eq:psipm}
  \ket{\Psi_m^\pm}_{A_1A_2} = \frac{1}{\sqrt{2}}(\ket{0m}\pm\ket{m0})_{A_1A_2}
\end{equation}
with equal probabilities. For the state on Bob's systems $B_1$ and $B_2$ under tag $(m,n)$, $\hat{\rho}_{B_1B_2}^{(m,n)}$,
% i.e. the state processed by the phase-randomized squashing channel $\mathcal{F}^{\text{rand}}$ in \eqref{Eq::squashnorm},
the statistics of the complementary measurement are given by
\begin{equation}
\begin{aligned}
\label{Eq::XonB}
_{B'}\bra{\pm}\mathcal{F}_{\text{rand}}^{B_1B_2\to B'}[\hat{\rho}_{B_1B_2}^{(m,n)}]\ket{\pm}_{B'}  &= \tr\left[\hat{\rho}_{B_1B_2}^{(m,n)}\hat{M}_{\pm}\right] \\ &= \tr\left[\hat{\rho}_{B_1B_2}^{(m,n)}\hat{P}_n\hat{M}_{\pm}\hat{P}_n\right],
\end{aligned}
\end{equation}
where
\begin{equation}\label{Eq:Mpm}
\begin{aligned}
\hat{M}_{\pm} &= \frac{1}{2}\int_{\mathbf{R_0}}dq_1 dq_2\int_0^{2\pi}\frac{d\varphi_b}{2\pi} \big[\ket{q_1(\varphi_b),q_2(\varphi_b)}\pm \\
& \ket{q_2(\varphi_b),q_1(\varphi_b)} \big] \left[\bra{q_1(\varphi_b),q_2(\varphi_b)}\pm\bra{q_2(\varphi_b),q_1(\varphi_b)}\right].
\end{aligned}
\end{equation}
In the last equation in \eqref{Eq::XonB}, we utilize the fact that $\hat{\rho}_{B_1B_2}^{(m,n)}$ acts on the $n$-photon space of system $B_1B_2$. Combining the above results, we can express the phase-error rate for each tag with observables on optical modes:
\begin{proposition} \label{prop:phase_error}
The phase-error rate $e^X_{m,n}$ of the rounds where $m$ photons are sent and $n$ photons are accepted can be calculated by:
\begin{equation}
\label{Eq::PERinfock}
\begin{split}
\frac{Q_{m,n}e^{X}_{m,n}}{\Pr_\mu(m)} =& \frac{1}{2}\tr\big[\mathcal{N}_E^{A_1A_2\to B_1B_2}(\ket{\Psi_m^+}_{A_1A_2}\bra{\Psi_m^+})\hat{P}_n\hat{M}_{-}\hat{P}_n \\ 
& + \mathcal{N}_E^{A_1A_2\to B_1B_2}(\ket{\Psi_m^-}_{A_1A_2}\bra{\Psi_m^-})\hat{P}_n\hat{M}_{+}\hat{P}_n\big],
\end{split}
\end{equation}
where $Q_{m,n}$ denotes the probability of sending an $m$-photon state and accepting an $n$-photon state, and $\Pr_\mu(m)$ is the probability of the source emitting $m$ photons. $\mathcal{N}_E$ denotes Eve's channel on the two optical modes and $\hat{P}_n$ denotes the projector onto the $n$-photon subspace. $\ket{\Psi_m^\pm}$ and $\hat{M}_{\pm}$ defined in \eqref{Eq:psipm} and~\eqref{Eq:Mpm} respectively.
\end{proposition} 

We can write $\hat{P}_n\hat{M}_{+(-)}\hat{P}_n$ on the Fock basis using \eqref{Eq::homo2fock}, expressing the phase-error rate as quantities on optical modes. Here, we list the final results for a protocol that utilizes up to the two-photon components. The detailed calculation is placed in Appendix~\ref{App:PERcal}. For the single-photon component,
\begin{strip}
\begin{equation}
\label{Eq::ex11}
\begin{split}
 \frac{Q_{1,1}e^X_{1,1}}{\Pr_\mu(1)} = \frac{c_1}{2}\bigg\{\tr\left[\mathcal{N}_E^{A_1A_2\to B_1B_2}(\ket{\Psi_1^+}_{A_1A_2}\bra{\Psi_1^+})\frac{1}{2}(\ket{01}_{B_1B_2}-\ket{10}_{B_1B_2})(\bra{01}_{B_1B_2}-\bra{10}_{B_1B_2})\right]\\
+ \tr\left[\mathcal{N}_E^{A_1A_2\to B_1B_2}(\ket{\Psi_1^-}_{A_1A_2}\bra{\Psi_1^-})\frac{1}{2}(\ket{01}_{B_1B_2}+\ket{10}_{B_1B_2})(\bra{01}_{B_1B_2}+\bra{10}_{B_1B_2})\right]\bigg\},
\end{split}
\end{equation}
where
\begin{equation}
\label{eq:SinglePhotonAccept}
c_1 = \int_{\mathbf{R_0}}dq_1 dq_2 [\psi_0^2(q_1)\psi_1^2(q_2) + \psi_0^2(q_2)\psi_1^2(q_1)],
\end{equation}
and gain $Q_{1,1}$ is given by
\begin{equation} \label{Eq::Q11}
\begin{aligned}
\frac{Q_{1,1}}{\Pr_\mu(1)} &= c_1\tr\big\{\mathcal{N}_E\left[\tr_{A'}(\ket{\Psi_1}_{A'A_1A_2}\bra{\Psi_1})\right] \\
& (\ket{01}_{B_1B_2}\bra{01}+\ket{10}_{B_1B_2}\bra{10})\big\}.
\end{aligned}
\end{equation}
Up to the less-than-unity factor $c_1$ that arises from the data post-selection in key mapping, the formulae are the same as the complementary-basis result in the coherent-state-based BB84 protocol~\cite{Marand1995QKD}. It can also be seen that the phase-error rate of the rounds where Alice transmits two photons and Bob accepts one photon involves in the probability where Alice transmits $(\ket{02}\pm\ket{20})/\sqrt{2}$ and Bob receives $(\ket{01}\mp\ket{10})/\sqrt{2}$. In a pure-loss channel, the superimposed two-photon state $(\ket{02}\pm\ket{20})/\sqrt{2}$ would lose coherence if one photon is lost during the channel, thus giving 50\% phase-error rate. This observation validates the intuition of the PNS attack. For the two-photon subspace, based on \eqref{Eq::homo2fock} and \eqref{Eq::PERinfock}, we have,

\begin{equation}
\label{Eq::eX22}
\begin{split}
\frac{Q_{2,2}e_{2,2}^X}{\Pr_\mu(2)}=& \frac{1}{2}c^{-}_2\tr\left[\mathcal{N}_E^{A_1A_2\to B_1B_2}(\ket{\Psi_2^+}_{A_1A_2}\bra{\Psi_2^+})\frac{1}{2}(\ket{02}_{B_1B_2}-\ket{20}_{B_1B_2})(\bra{02}_{B_1B_2}-\bra{20}_{B_1B_2})\right]\\
&+\frac{1}{2}c^{+}_2\tr\left[\mathcal{N}_E^{A_1A_2\to B_1B_2}(\ket{\Psi_2^-}_{A_1A_2}\bra{\Psi_2^-})\frac{1}{2}(\ket{02}_{B_1B_2}+\ket{20}_{B_1B_2})(\bra{02}_{B_1B_2}+\bra{20}_{B_1B_2})\right]\\
&+c_2^{11}\tr\left[\mathcal{N}_E^{A_1A_2\to B_1B_2}(\ket{\Psi_2^-}_{A_1A_2}\bra{\Psi_2^-})\ket{11}_{B_1B_2}\bra{11}\right],
\end{split}
\end{equation}

where
\begin{equation}\label{eq:TwoPhotonCoeff}
\begin{split}
 &c_{2}^{+} = \int_{\mathbf{R_0}}dq_1 dq_2 [\psi_0(q_1)\psi_2(q_2) + \psi_2(q_1)\psi_0(q_2)]^2,\\
 &c_{2}^{-} = \int_{\mathbf{R_0}}dq_1 dq_2 [\psi_0(q_1)\psi_2(q_2) - \psi_2(q_1)\psi_0(q_2)]^2,\\
 &c_2^{11} = \int_{\mathbf{R_0}}dq_1 dq_2 [2\psi_1^2(q_1)\psi_1^2(q_2)],
\end{split}
\end{equation}
and the two-photon gain is given by

\begin{equation}
\label{Eq::Q22}
\begin{split}
  \frac{Q_{2,2}}{\Pr_\mu(2)} =& c_2^+\tr\left\{\mathcal{N}_E^{A_1A_2\to B_1B_2}[\tr_{A'}(\ket{\Psi_2}_{A'A_1A_2}\bra{\Psi_2})]\frac{1}{2}(\ket{02}_{B_1B_2}+\ket{20}_{B_1B_2})(\bra{02}_{B_1B_2}+\bra{20}_{B_1B_2})\right\} \\ 
  &+c_2^-\tr\left\{\mathcal{N}_E^{A_1A_2\to B_1B_2}[\tr_{A'}(\ket{\Psi_2}_{A'A_1A_2}\bra{\Psi_2})]\frac{1}{2}(\ket{02}_{B_1B_2}-\ket{20}_{B_1B_2})(\bra{02}_{B_1B_2}-\bra{20}_{B_1B_2})\right\}\\
  &+2c_2^{11}\tr\left\{\mathcal{N}_E^{A_1A_2\to B_1B_2}[\tr_{A'}(\ket{\Psi_2}_{A'A_1A_2}\bra{\Psi_2})]\ket{11}_{B_1B_2}\bra{11}\right\}.
\end{split}
\end{equation}
\end{strip}
The probability of
accepting a vacuum state when employing reverse reconciliation is given by
\begin{equation}
\label{eq:ReverseGain}
Q_{*,0} = c_0\tr\left[\hat{P}_0^{B_1B_2}\mathcal{N}_E^{A_1A_2\to B_1B_2}\left(\hat{\rho}^Z\right)\hat{P}_0^{B_1B_2}\right],
\end{equation}
where
\begin{equation}\label{eq:ZeroPhotonCoeff}
    c_0 = \int_{\mathbf{R_0}} 2\psi_0^2(q_1)\psi_0^2(q_2) dq_1 dq_2,
\end{equation}
and $\hat{\rho}^Z$ is the $Z$-basis state sent by the source given in \eqref{Eq::Zsource}; hence $Q_{*,0}$ is given by the product of the probability of receiving a vacuum-state in the $Z$-basis rounds and a post-selection-related integration factor. Note that the former value is independent of the post-selection.

\section{Parameter estimation and practical protocol}
\label{Sec::Parameter estimation}
We briefly show how to estimate the parameters derived in Sec.~\ref{Sec::PER calculation} with a practical setup. In the actual protocol, we do not have photon-number-resolving detectors, with which one can directly measure the above parameters. In addition, the phase-error probabilities and gains are defined by particular Fock-basis states, yet the actual photon source emits coherent states. Nevertheless, we can construct unbiased estimators with the available states and detection settings to evaluate these values. On the detection side, we apply the homodyne tomography technique to evaluate the photon-number observables~\cite{vogel1989determination,smithey1993measurement,d1994detection,d1995homodyne,leonhardt1995measuring,d1995tomographic}. The homodyne tomography allows unbiased estimation of the expected value of a variety of observables, including the photon-number observables, of measuring an unknown quantum state. On the source side, we extend the decoy-state method~\cite{Lo2005Decoy,wang2005decoy,zhang2017improved} to evaluate the statistics defined by the non-classical Fock states with the use of the coherent states at hand. We will give a practical version of the protocol at the end of this section. A fully-detailed discussion is placed in the Appendix~\ref{app:paraest} on how the specific parameters related to key rate calculation can be practically estimated. 

\subsection{Effective photon-number resolving via homodyne tomography}\label{Sec:Homotomo}
Since the eigenstates of the quadrature observables, $\ket{q(\varphi)}$, form a complete basis on an optical mode, one can reconstruct a general observable on an optical mode with homodyne measurements. In our study, the parameters to be estimated involve photon-number measurements on two modes in the form of $\hat{O}_1\otimes \hat{O}_2=\ket{n_1}_{B_1}\bra{m_1}\otimes\ket{n_2}_{B_2}\bra{m_2}$. Their measurements on an arbitrary state, $\hat{\rho}$, can be obtained from two independent homodyne measurements with randomized LO phases, 
\begin{equation}\label{eq:HomoTomoEst}
\begin{aligned}
    & \langle\hat{O}_1\otimes\hat{O}_2\rangle = \tr[(\hat{O}_1\otimes\hat{O}_2)\hat{\rho}] \\
    & \quad :=\int_0^\pi\frac{d\varphi_1}{\pi}\int_{-\infty}^{\infty}dq_1\int_0^\pi\frac{d\varphi_2}{\pi}\int_{-\infty}^{\infty}dq_2 \\&\quad\mathcal{R}[\hat{O}_1](q_1,\varphi_1)\mathcal{R}[\hat{O}_2](q_2,\varphi_2)p(q_1,q_2|\varphi_1,\varphi_2), \\
\end{aligned}
\end{equation}
where $p(q_1,q_2|\varphi_1,\varphi_2)$ is the joint probability of the quadrature measurements on the two modes conditioned on phases $\varphi_1$ and $\varphi_2$. The estimators, $\mathcal{R}[\hat{O}_1](q_1,\varphi_1)$ and $\mathcal{R}[\hat{O}_2](q_2,\varphi_2)$, link the quadrature measurement statistics with $\langle\hat{O}_1\otimes\hat{O}_2\rangle$. For a homodyne detector with efficiency $\eta$, the estimator for observable $\ket{n}\bra{n+d}$ is given by
\begin{equation}
\begin{aligned}
\label{eq:homotomoPN}
    &\mathcal{R}_\eta[\ket{n}\bra{n+d}](q,\varphi) = e^{id(\varphi+\frac{\pi}{2})}\sqrt{\frac{n!}{(n+d)!}} \\ & \quad \int_{-\infty}^{\infty}dk|k|\exp\left(\frac{1-2\eta}{2\eta}k^2-ikq\right)k^dL_n^d(k^2),
\end{aligned}
\end{equation}
where $L_n^d$ is the generalized Laguerre polynomial. The estimator is shown to be bounded for detector efficiency $\eta > 1/2$~\cite{d1995homodyne,d1995tomographic}, a mild requirement for current technologies~\cite{Hansen2001Ultra,Grandi2017Experimental}. Consequently, repeated measurements allow the users to obtain an unbiased estimation of the photon-number observables that converges in probability. Note that the detector imperfection does not need to be trusted. The homodyne tomography is valid as long as the detector is well-calibrated so that the quadrature measurement is genuine. In Appendix~\ref{App::homotomo}, we shall provide more details of the homodyne tomography techniques.

\subsection{Generalized decoy-state method}
\label{Subsec::decoy}
To effectively realize the non-classical states on the source side, we extend the standard decoy-state method~\cite{Lo2005Decoy,wang2005decoy}. We take advantage of two-mode coherent states with simultaneous phase randomization on the two modes. We denote the state with phase difference $\varphi$ as

\begin{equation}
\begin{split}
 &\hat{\rho}_\mu^\varphi = \int_0^{2\pi}\frac{d\theta}{2\pi}\ket{\sqrt{\frac{\mu}{2}}e^{i\theta}}\bra{\sqrt{\frac{\mu}{2}}e^{i\theta}} \otimes \ket{\sqrt{\frac{\mu}{2}}e^{i(\theta+\varphi)}}\bra{\sqrt{\frac{\mu}{2}}e^{i(\theta+\varphi)}}\\
 &= \sum_{m=0}^\infty\sum_{k=0,l=0}^{m} \frac{e^{-\mu}\left(\frac{\mu}{2}\right)^me^{i(l-k)\varphi}}{\sqrt{k!l!(m-k)!(m-l)!}}\ket{k,m-k}\bra{l,m-l},
\end{split}
\end{equation}

where we specify the light intensity with the subscript, $\mu$. With proper linear combination of these states, we can effectively construct the photon-number-cat states that we are interested in. It is well-known that $(\ket{01}\pm\ket{10})/\sqrt{2}$ is the single-photon component of $\hat{\rho}_\mu^{0(\pi)}$,
\begin{equation}
\label{Eq::decoy1}
\text{Pr}_\mu(1)\ket{\Psi_1^{+(-)}}\bra{\Psi_1^{+(-)}} = \hat{P}_1 \hat{\rho}_\mu^{0(\pi)} \hat{P}_1,
\end{equation}
where $\Pr_{\mu}$ represents the Poisson distribution determined by light intensity $\mu$, as given in \eqref{eq:Poisson}.
Thus, the estimation problem is transformed into the estimation of the single-photon yields of $\hat{\rho}_\mu^0$ and $\hat{\rho}_\mu^\pi$. For the multi-photon components $(\ket{0m}\pm\ket{m0})/\sqrt{2}$, a direct calculation shows

\begin{equation}
\label{Eq::decoyPsi+}
\begin{split}
\text{Pr}_\mu(m)&\ket{\Psi_m^{+}}\bra{\Psi_m^{+}} = \\
&\hat{P}_m\big(\hat{\rho}^Z_\mu + \frac{2^{m-2}}{m}\sum_{k=0}^{m-1} \hat{\rho}^{\frac{2\pi k}{m}}_\mu - \frac{2^{m-2}}{m}\sum_{k=0}^{m-1} \hat{\rho}^{\frac{2\pi k}{m}+ \delta}_\mu\big)\hat{P}_m,
\end{split}
\end{equation}
\begin{equation}
\label{Eq::decoyPsi-}
\begin{split}
\text{Pr}_\mu(m)&\ket{\Psi_m^{-}}\bra{\Psi_m^{-}} = \\
&\hat{P}_m\big(\hat{\rho}^Z_\mu -\frac{2^{m-2}}{m}\sum_{k=0}^{m-1} \hat{\rho}^{\frac{2\pi k}{m}}_\mu+ \frac{2^{m-2}}{m}\sum_{k=0}^{m-1} \hat{\rho}^{\frac{2\pi k}{m} + \delta}_\mu\big)\hat{P}_m,
\end{split}
\end{equation}

where $\delta = \pi$ for odd $m$ and $\pi/2$ for even $m$, and $\hat{\rho}_\mu^Z$ is the state emitted from the source in a key generation round. Consequently, the terms that define $e_{m,m}^X$ and $Q_{m,m}$ can be constructed from the statistics when emitting the states of $\hat{\rho}_\mu^Z$ and $\hat{\rho}_\mu^\varphi$ with $\varphi\in\{2\pi k/m,2\pi k/m+\delta\}_{k=0}^{m-1}$. Notably, the extended decoy method allows estimating the gains with the number of parameters increasing only linearly in the photon number. In later discussions, we shall utilize up to the two-photon components. Specifically, for $m=2$,
\begin{equation}
\begin{split}
\label{Eq::decoy2}
\text{Pr}_\mu(2)\ket{\Psi_2^{\pm}}\bra{\Psi_2^{\pm}} = &\hat{P}_2\big[\hat{\rho}_\mu^Z \pm\left(\frac{1}{2}\hat{\rho}_\mu^0 + \frac{1}{2}\hat{\rho}_\mu^\pi\right) \\ &\mp \left(\frac{1}{2}\hat{\rho}_\mu^{\frac{\pi}{2}} + \frac{1}{2}\hat{\rho}_\mu^{\frac{3\pi}{2}}\right)\big]\hat{P}_2.
\end{split}
\end{equation}

One may notice in \eqref{Eq::decoy2} there are states outside the encoding subspace $\ket{02}$ and $\ket{20}$ being introduced, which gives Eve possibility to distinguish the $X$-basis states. In fact, the $X$ basis is comprised of the mixture of $(\ket{02}\pm\ket{20})/\sqrt{2}$ and $\ket{11}$. As a result, it is not possible for Eve to distinguish between the $Z$-basis states and the $(\ket{02}\pm\ket{20})/\sqrt{2}$ states of the $X$ basis. It is possible for Eve to distinguish the $\ket{11}$ state, yet it does not yield knowledge on the encoded key information since it is orthogonal to the $\ket{02}$ and $\ket{20}$ space. Hence, the standard decoy argument still applies even if the parameter-estimation space consists of a direct sum of the key-encoding space and some orthogonal spaces.

\subsection{General parameter estimation under the coherent attack}\label{Sec:CoherentAttack}
In this section, we discuss the security analysis and parameter estimation in the most general case. In the most general adversarial scenario, namely the coherent attack, Eve can apply a joint quantum operation over the rounds for eavesdropping, which may correlate or even entangle the states transmitted to Bob. Eve collects all the side information leaked to her in the protocol and then guesses the legitimate users' keys. Under such an attack, the measurement statistics obtained by Bob are generally correlated over the rounds~\cite{xu2020secure}. 

The complementarity-based security analysis remains valid with finite statistics under a coherent attack~\cite{koashi2009simple}. The information leakage is quantified via the number of phase errors, while the occurrence of a phase error in each round may be non-i.i.d. That is, one should interpret the gains and phase-error rates in \eqref{Eq::Keyrev} as frequencies in non-i.i.d.~statistics. To avoid any misunderstanding, throughout this work, the term ``frequency'' is used in the context of probability theory and refers to the number of occurrences of an event.
For instance, $Q_{1,1}$ should be regarded as the frequency of the events that Alice sends a single-photon state, and Bob accepts a single-photon state among key generation rounds in the virtual experiment. The remaining problem is to estimate these parameters via observed statistics. 

To tackle the non-i.i.d.~parameter estimation problem, we can apply a martingale-based analysis. We shall present the details in Appendix~\ref{app:martingale}. Here, we explain its basic idea. As the starting point, in the $i$'th round, the users can evaluate the probability of choosing some experimental setting and observing a particular event conditioned on the experimental history, including the events of sending an $m$-photon state and accepting an $n$-photon state and the occurrence of a phase error if they choose the key generation setting, and observing a particular homodyne detection result if they choose to perform the parameter estimation operations. The events' correlations with the experimental history are inherently taken into account in the definitions of conditional probabilities. 
We can then set up martingales for a series of events, such as the occurrence of phase errors in each round of the virtual protocol, and link their frequencies with the associated conditional probabilities via concentration results like Azuma's inequality~\cite{azuma1967weighted}. 
Note that such concentration results work for general non-i.i.d.~correlations. To tighten the estimation, we shall apply a new variant of Azuma's inequality, Kato's inequality~\cite{kato2020concentration}. Furthermore, the setting choices randomly chosen by Alice and Bob are independent of the experimental history and unknown to Eve. Therefore, conditioned on the experimental history, the probabilities of different possible events in a round are linked. For instance, the probability that the users take key generation measurements and a phase error occurs in a round is measurable via the probability that they instead take parameter estimation measurements and observe certain statistics. The relation is in the form of \eqref{Eq::PERinfock}, while now the probabilities are interpreted as conditional ones that cover the correlations. The relations between conditional probabilities then link the martingales for the parameter estimation measurement with the ones for the gains and phase-error rates, completing the parameter estimation. In the end, the total number of keys that can be securely distilled from finite statistics under the coherent attack is given by a formula of the following form:

\begin{theorem}[Informal]
For the CV QKD protocol with $M^{Z,s}$ rounds for key generation with the signal intensity $s$, given the failure probability in parameter estimation $\varepsilon_{\mathrm{pe}}$, suppose the quantum bit error rate is $e^{Z,s}$, the number of key generation rounds accepting a vacuum state is lower-bounded by $\underline{M_{*,0}^{Z,s}}$, the number of key generation rounds sending and accepting an $m$-photon state is lower-bounded by $\underline{M_{m,m}^{Z,s}}$, and the $m$-photon phase-error rate is upper-bounded by $\overline{e_{m,m}^{X}}$. 
Then, given the failure probability in privacy amplification $\varepsilon_{\mathrm{pa}}$, conditioned on the success of information reconciliation with efficiency $f$, except a total failure probability $\varepsilon=\varepsilon_{\mathrm{pe}}+\varepsilon_{\mathrm{pa}}$, the finite-size key length is lower bounded by:
\begin{equation}
\begin{split}
  n\geq& \underline{M_{*,0}^{Z,s}} + \sum_{m=1}^{\infty} \underline{M_{m,m}^{Z,s}}\left[1-h\left(\overline{e_{m,m}^{X}}\right)\right] \\& - fM^{Z,s} h(e^{Z,s}) -\log{\frac{1}{\varepsilon_{\mathrm{pa}}}},
\end{split}
\end{equation}
and the key rate is lower-bounded by $r\geq n/M^{Z,s}$.
\end{theorem}

The term $\log{\varepsilon_{\mathrm{pa}}}$ in the key-rate formula originates from the failure probability in privacy amplification~\cite{koashi2009simple,fung2010practical}. The parameter estimation failure probability $\varepsilon_{\mathrm{pe}}$ comes from the use of concentration results to link frequencies and probabilities. In the asymptotic limit of infinite data size, $\varepsilon_{\mathrm{pe}}$ converges to zero, and the effect of $\varepsilon_{\mathrm{pa}}$ on the key rate becomes negligible; hence the key rate formula degenerates to that in \eqref{Eq::Keyrev}, where $\underline{M_{*,0}^{Z,s}}\sim M^{Z,s}Q_{*,0}$, and $\underline{M_{m,m}^{Z,s}}\sim M^{Z,s}Q_{m,m}$. In Appendix~\ref{app:paraest}, we provide the details of non-i.i.d. parameter estimation and the formal description of the key-rate formula.

\subsection{Practical protocol}
Combining the above ingredients, we provide a practical protocol that utilizes up to the two-photon components in Table~\ref{table:protocolPractial}. In parameter estimation, Bob applies homodyne tomography to estimate the statistics of measuring photon-number observables, including $\ket{00}$, $(\ket{01}\pm\ket{10})/\sqrt{2}$, $(\ket{02}\pm\ket{20})/\sqrt{2}$, and $\ket{11}$, on various states transmitted from the source, originally $\hat{\rho}_{\mu_a}^Z$ and $\hat{\rho}_{\mu}^{\varphi_a}$. Afterward, the users can obtain upper and lower bounds on the gains and phase-error rates by applying the extended decoy-state method. 

In the end, we make some remarks on the protocol. Notice that in contrast to the conventional BB84-type protocols, our protocol also uses for parameter estimation the signals where Alice chooses $Z$ basis and Bob chooses $X$ basis. Alice's announcement of the relative phase does not reveal key information since the key is encoded in the relative intensity between the two modes. We assume Alice and Bob apply continuous phase randomization, although it is only practical to use discrete random phases. The effect of the discretization requires further investigation. In addition, in the $X$ basis, Alice only transmits coherent states with relative phases in $\{0,\pi/2,\pi,3\pi/2\}$. These relative phases are enough to estimate the phase-error rate of an up-to-two-photon protocol according to \eqref{Eq::decoy2}.

\begin{table*}[htbp!] 
  \caption{Practical time-bin CV QKD with decoy states using up to two photons}\label{table:protocolPractial}
  \centering
  \begin{tabular}{p{0.95\linewidth}}%{p{20cm}}
  \hline
  \hline
  \begin{enumerate}
  \item On Alice's side (source):
  \begin{itemize}
  \item \emph{$Z$-basis}:
  \begin{enumerate}
    \item Randomly select a key bit $k_a \in \{0,1\}$, a phase factor $\varphi_a\in [0,2\pi)$, and a light intensity $\mu_a\in\{\mu,\nu_1,\nu_2,0\}$.
    \item Prepare a coherent state of $\ket{0}_{A1} \ket{\sqrt{\mu_a}e^{i\varphi_a}}_{A2}$ for $k_a=0$ or $\ket{\sqrt{\mu_a}e^{i\varphi_a}}_{A1} \ket{0}_{A2}$ for $k_a=1$.
  \end{enumerate}
  \item \emph{$X$-basis}:
  \begin{enumerate}
    \item Randomly select a phase factor $\varphi_a^1 \in [0,2\pi)$ and another phase factor with relative phase $\varphi_a$ randomly in $\varphi_a^2 - \varphi_a^1 \in \{0,\pi/2,\pi,3\pi/2\}$. Randomly select a light intensity $\mu_a\in\{\mu,\nu_1,\nu_2,0\}$.
    \item Prepare a coherent state of $\ket{\sqrt{\mu_a/2}e^{i\varphi_a^1}}\ket{\sqrt{\mu_a/2}e^{i\varphi_a^2}}$.
  \end{enumerate}
\end{itemize}
  \item Alice sends the state through an authenticated channel to Bob.
  \item On Bob's side (detection):
  \begin{itemize}
  \item \emph{$Z$-basis}:
  \begin{enumerate}
    \item Randomly select a phase factor $\varphi_b\in [0,2\pi)$.
    \item Use homodyne detectors with LO phases $\varphi_b$ to measure the modes and obtain quadratures $q_1$ and $q_2$.
    \item Decode the key bit as $0$ if $|q_1|<\tau\wedge|q_2|>\tau$, $1$ if $|q_1|>\tau\wedge|q_2|<\tau$, and $\emptyset$ otherwise.
  \end{enumerate}
  \item \emph{$X$-basis}:
  \begin{enumerate}
    \item Randomly select two phases $\varphi_b^1,\varphi_b^2\in[0,\pi)$.
    \item Use homodyne detectors with LO phases $\varphi_b^1$ and $\varphi_b^2$ to measure the modes and obtain quadratures $q_1$ and $q_2$.
  \end{enumerate}
\end{itemize}
\item Alice announces the light intensity in each round and relative phase between the two modes in $X$-basis states $\varphi_a=\varphi^2_a-\varphi^1_a$.
\item Alice and Bob perform basis sifting, where they obtain raw keys in the rounds they both choose $Z$-basis with light intensity $\mu_a=\mu$ and $k_b\neq\emptyset$.
\item Bob estimates the gains and phase-error rates from the statistics in the rounds where Alice sends the $Z$-basis states $\hat{\rho}_{\mu_a}^Z$ or $X$-basis states $\hat{\rho}_{\mu_a}^{\varphi_a}$ with $\varphi_a\in\{0,\pi/2,\pi,3\pi/2\}$.
\item Alice and Bob perform reverse information reconciliation and privacy amplification to obtain final keys.
\end{enumerate} \\
\hline
\hline
  \end{tabular}
\end{table*}

\section{Performances and comparison}
\label{Sec::results}

\begin{figure*}
\centering
\includegraphics[width=\linewidth]{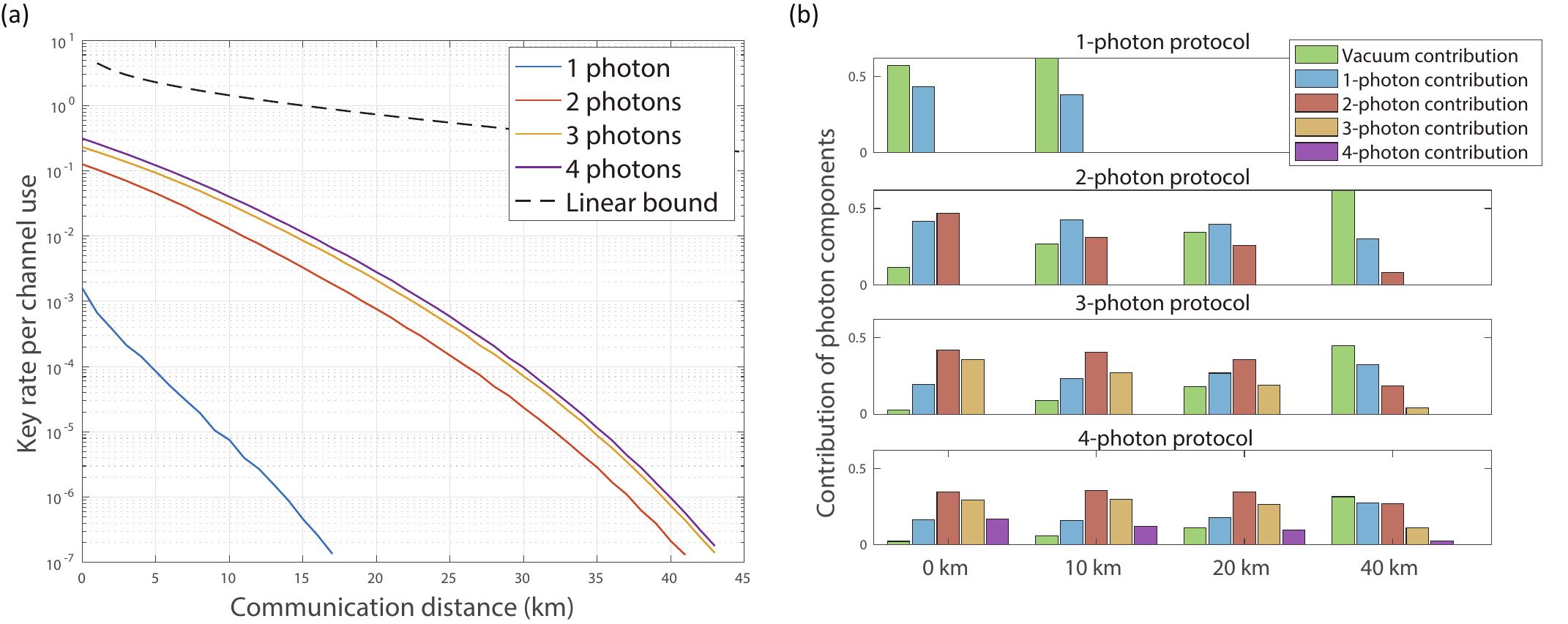}
\caption{\textbf{(a)} The solid lines illustrate the asymptotic key rates of protocols using maximal one, two, three and four photons to generate keys. The dotted line is the linear key rate bound~\cite{Takeoka2014fundamental,Pirandola2017fundamental}. We plot the PLOB bound here. The channel and devices are assumed to be ideal with no excess noise and inefficiency. \textbf{(b)} The relative contribution of $Q_{m,m}$, i.e. the gain of the rounds where $m$ photons are sent and $m$ photons are received. The $m$-photon contribution of the $i$-photon protocol is relative to the raw key rate. Each group of bars illustrate the contribution of vacuum, one, two, three and four-photon components of the protocol at a certain distance.}
\label{Fig:234Results}
\end{figure*}

We demonstrate in this section the asymptotic and finite-size key rate-distance performances of the time-bin CV QKD protocol. We consider a thermal noise channel with a unit-efficiency homodyne detector. An inefficient detector with thermal electronic noise can be equated to a fiber section with transmittance equal to the detector efficiency, and the electronic noise absorbed into the channel excess noise (\eqref{Eq:thermalconcatenate}). In practice, the excess noise may originate from issues including an imperfect modulation, Raman scattering in the fiber, phase fluctuation, etc.
The fiber attenuation is 0.2 dB/km, and the error-correction efficiency $f$ is taken to be 1. The simulation formulae can be found in Appendix~\ref{Sec:simufomula}. 
According to the key rate formula \eqref{Eq::Keyrev}, the rounds where Alice sends $m$ photons and Bob receives $m$ photons can assure to generate secure keys. We plot in Fig.~\ref{Fig:234Results}(a) the asymptotic key rate of the $i$-photon protocol assuming perfect decoy estimation and no excess noise, where in an $i$-photon protocol we only extract secure keys from a maximal $i$-photon components. In this ideal case, the phase error rates of all the protocols are zero. The optimized source intensities $\mu$ and the post-selection thresholds $\tau$ are listed in Table~\ref{Tab::234Para}, as well as the resulted $Z$-basis error rate. Notice that the two-photon-protocol key rate derived from our DV method is similar to that from Ref.~\cite{Primaatmaja2022Discrete} using CV method, both reversely reconciled. This implies the connection between DV and CV security analysis, as well as the validity of the DV reverse reconciliation idea in Section~\ref{Sec::Photon number tagging}. To facilitate the discussion, we also plot in Fig.~\ref{Fig:234Results}(b) the contribution of each photon components to the key rate at different distances. In each group of bars, the relative contribution of the vacuum, one, two, three, four-photon components are plotted respectively, where the $m$-photon contribution of the $i$-photon protocol is defined to be $Q_{m,m}/(Q_{*,0}+\sum_{m=1}^i Q_{m,m})$ and the vacuum contribution is $Q_{*,0}/(Q_{*,0}+\sum_{m=1}^i Q_{m,m})$, i.e., the relative contribution to the raw key rate.

It can be seen that the key rate improves as we make use of the multi-photon components. The improvement is most remarkable between the one and two-photon protocols. This is reasonable since in the one-photon protocol, the multi-photon components are considered insecure, thus limiting the source intensity. The low source intensity would result in severe bit error rate and higher post-selection thresholds, which in turn suppress the key rate. Whilst in the two-photon protocol where the two-photon components are considered secure, the limit on the source intensity can be lifted, and the bit error rate would drop, resulting in higher key rates. This is manifested in Fig.~\ref{Fig:234Results}(b), where the single-photon protocol sees significant vacuum contribution, whilst the two-photon protocol, at short distances, does not. Since the vacuum component would yield 50\% bit error rate, we see the lower bit error rate of the two-photon protocol than the single-photon protocol as in Table~\ref{Tab::234Para}.    

When we further make use of the three-photon components, the key rate as well as the source intensity still increase, yet less obviously. This is mainly because the fraction of the rounds where three photons are sent and three photons are received, decaying cubically with the channel transmittance, are not dominating, especially at longer distances. For example, we see in Fig.~\ref{Fig:234Results}(b) that at 20 km, the contribution of the three-photon component is less than that of the single and two-photon components, and at 40 km the three-photon component rarely has contribution to the key rate. This trend is justified further in the four-photon protocol, where in Fig.~\ref{Fig:234Results}(b) we see the four-photon-component contribution is quite small for longer distances, and in turn the key rate of the four-photon protocol only improves marginally than that of the three-photon protocol. Simulation shows that resorting to higher-than-four photon-number components has negligible increase to the key rate. Hence, If we consider the protocol with infinite photon-number components, the $0$-km key rate is around 0.31 bit/channel, and the BB84 protocol with currently the best single-photon detector of $80\%$ efficiency~\cite{Grnenfelder2020performance,Li2023highrate} has $0$-km key rate 0.29 bit/channel, based on the model from Ref.~\cite{Ma2005practical}. This key rate advantage lasts for about 1 km. Our key rate thus matches the best BB84 key rate with practically favorable devices.

\begin{table*}[htbp!] 
\centering
\captionsetup{justification = raggedright}
\caption{The optimized intensities and post-selection thresholds of protocols using one to four photons respectively at different distances. $\mu_i$, $\tau_i$ and $e_Z^{i}$ denote the optimized intensity, post-selection threshold and the $Z$-basis error rate of the $i$-photon protocol. These parameters generate the four key rate plots in Fig.~\ref{Fig:234Results}(a), assuming infinite decoy levels.}
\begin{tabular}
{c||c|c|c|c||c|c|c|c||c|c|c|c}
\hline
\hline
 & $\mu_1$ & $\mu_2$ & $\mu_3$ & $\mu_4$ & $\tau_1$ & $\tau_2$ & $\tau_3$ & $\tau_4$ & $e_Z^{1}$ & $e_Z^{2}$ & $e_Z^{3}$ & $e_Z^{4}$\\
\hline
0 km & 0.356 
&1.487&2.395&2.395&1.437&1.641&1.845&1.845&30.95\%&10.52\%&5.31\%&5.31\%\\
10 km & 0.137 &0.924&1.887&2.395&3.476&2.253&2.457&2.457&29.80\%&14.84\%&5.66\%&4.17\%\\
20 km & --- &0.728&1.487&1.887& --- &3.068&3.068&3.272&---&15.48\%&6.91\%&3.85\%\\
40 km & --- &0.356&0.728&1.172&---&4.495&4.495&4.699&---&28.52\%&17.07\%&8.81\%\\
\hline
\end{tabular}
\label{Tab::234Para}
\end{table*}

\begin{figure*}
\centering
\includegraphics[width=\linewidth]{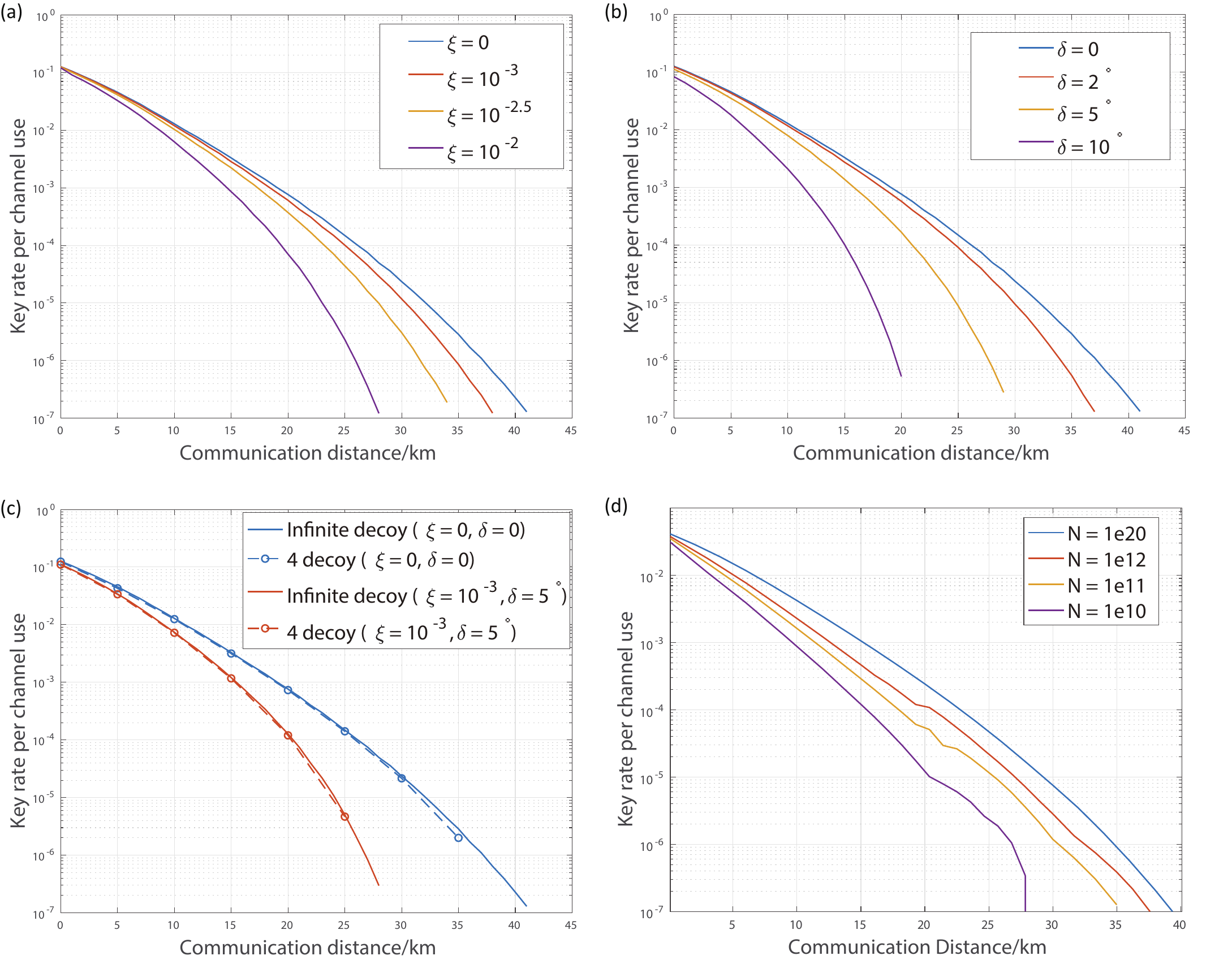}
\caption{Practical performances of the two-photon protocol. \textbf{(a)} Asymptotic key rate against excess noise with respect to channel output, assuming infinite decoy levels. \textbf{(b)} Asymptotic key rate against misalignment, i.e. the phase-reference difference between the two optical modes generating the time-bin qubit, assuming infinite decoy levels. \textbf{(c)} Asymptotic Key rate derived using decoy methods. The noiseless setup (blue curves) uses fixed decoy levels at $1.2\times 10^{-4}$, $1 \times 10^{-4}$ and vacuum, and the optimized protocol parameters of the practical setup (red curves) are listed in Table~\ref{Tab::Para}. \textbf{(d)} Finite-size key rates with various data sizes derived using the decoy-state method under no excess noise and misalignment. The security parameter is at most $5\times 10^{-10}$, and the source $Z/X$ basis setting is 1:2.}
\label{Fig:Results}
\end{figure*}

The practical performances of the two-photon time-bin CV QKD protocol are illustrated in Fig.~\ref{Fig:Results}. For a reasonable range of excess noise $\xi$ from $10^{-3}$ to $10^{-2}$ with respect to channel output, the key rate decays mildly as shown in Fig.~\ref{Fig:Results}(a). Notice that the key rate is almost unaffected at 0 km since no noise photon is introduced to give phase error, and the bit error is almost unchanged for a negligible increase in the shot-noise variance. This demonstrates the robustness of the phase-error analysis to the excess noise.

Fig.~\ref{Fig:Results}(b) illustrates the key rate against the mode reference misalignment, where the two optical modes generating the time-bin qubit differ by $\delta$ in the reference phases intrinsically. The misalignment in relative phases does not affect the $Z$ basis as we encode the key bits into the relative intensities, and it only affects the $X$ basis where the phase error is defined as the flips in relative phases. Our protocol thus has robustness against misalignment.

Fig.~\ref{Fig:Results}(c) illustrates the key rates of the decoy-state protocol in Sec.~\ref{Subsec::decoy}. We set one decoy level at vacuum, and heuristically optimize the two decoy intensities $\nu_1$ and $\nu_2$ and the signal intensity $\mu$. The decoy estimations are done by linear programming with a cutoff photon number 20~\cite{Ma2012Statistical,Xu2014protocol}. A detailed treatment of the decoy estimation of the two-photon protocol is placed in Appendix~\ref{app:decoy}. We see for both the noiseless setup, with no excess noise and misalignment, and the practical setup, with $10^{-3}$ excess noise and $5^{\circ}$ misalignment, the 4-level decoy estimation is almost exact. Especially on the two-photon components, the decoy estimation has only a 1\% discrepancy for yield and no discrepancy for phase error rate. This clearly surpassed the decoy performance of the protocol in Ref.~\cite{Primaatmaja2022Discrete}, since our protocol uses simpler estimation of the phase error by identifying the principal components in key generation. The optimized parameters of the practical setup are listed in Table~\ref{Tab::Para}.  

Table~\ref{Tab::CompareProtocols} below demonstrates the performance comparison of our protocol with other discrete-modulated CV protocols. Compared with the protocol in Ref.~\cite{Primaatmaja2022Discrete} with similar $Z$-basis operations, our protocol generates higher key rate and reaches longer distance with 4-intensity-level decoy estimation. This manifests the effectiveness of our parameter estimation process since our extended decoy method imposes more accurate restriction to Eve's attack identified by the phase-error rate. Compared with the binary-modulated single-mode CV protocol in Ref.~\cite{Matsuura2021Finite}, under $10^{-3}$ excess noise, our protocol generates about a quarter of their key rate yet reaches longer distance without pilot-reference transmission. This could be attributed to the noise-robustness of the two-mode encoding.

Fig.~\ref{Fig:Results}(d) illustrates the finite-size key rate when there is no channel excess noise or misalignment. The security parameter reflected by the failure probability~\cite{koashi2009simple} is set to be smaller than $5\times 10^{-10}$ in this simulation. The ratio between the $Z$- and $X$-basis setting choices on the source side is fixed at 1:2. 
For this reason, the asymptotic key rate drops to a third of that, where the basis ratio is also optimized. Again, we use four decoy levels. We optimize their intensities and ratio setting as well as the basis setting from the detection side with respect to the distance. Given this huge parameter space, we apply efficient searching algorithms such as particle swarming with random seeding. The searching is rapid yet not guaranteed to reach the global maximum. This explains the non-smoothness in the key rate plot. It can be seen that with a data size of $10^{20}$ rounds, the key rate approaches the asymptotic key rate. A high short-distance key rate can be reached with a reasonable data size of $10^{10}$ to $10^{11}$. Moreover, with a data size of $10^{12}$, the communication distance can almost reach the maximal distance of the asymptotic case. Our time-bin CV protocol thus enjoys good finite-size performance and is one of the very few CV schemes that admits robust finite-size analysis~\cite{Leverrier2017security,Matsuura2021Finite}.

\begin{table*}[hbtp!]
\centering
\captionsetup{justification = raggedright}
\caption{Optimal protocol parameters in generating the key rate plot with $10^{-3}$ excess noise and $5^\circ$ misalignment, using 4 decoy levels, as in Fig.~\ref{Fig:Results}(c). The heuristically optimized signal intensity, post-selection threshold and the decoy intensities are as given. There is one more decoy intensity set to be vacuum. The decoy estimation is done by linear programming with cutoff photon number 20.
}
\begin{tabular}{c c c c c}
\hline
\hline
Distance (km) & Signal intensity $\mu$ & Threshold $\tau$ & Decoy intensity $\nu_1$ & Decoy intensity $\nu_2$ \\ 
\hline
0 & 1.487 & 1.641 & $1.737 \times 10^{-1}$ & $1.000\times 10^{-4}$ \\
5 & 1.172 & 2.049 & $3.406 \times 10^{-3}$ & $2.740 \times 10^{-4}$\\
10 & 0.924 & 2.457 & $2.993 \times 10^{-2}$ & $1.000\times 10^{-4}$ \\
15 & 0.924 & 3.068 & $1.861 \times 10^{-2}$ & $1.000\times 10^{-4}$ \\
20 & 0.728 & 3.476 & $1.355 \times 10^{-2}$ & $2.441\times 10^{-4}$ \\
25 & 0.728 & 4.291 & $1.355 \times 10^{-2}$ & $1.562\times 10^{-4}$ \\
\hline
\hline
\end{tabular}
\label{Tab::Para}
\end{table*}

\begin{table*}[hbtp!]
\centering
\captionsetup{justification = raggedright}
\caption{Comparison between protocols. Our protocol gives higher key rate and longer distance than Ref.~\cite{Primaatmaja2022Discrete} under 4-intensity decoy estimation. Our protocol generates slightly lower key rate than the binary-modulated CV protocol in Ref.~\cite{Matsuura2021Finite}, yet reaches longer distance.}
\begin{tabular}{c c c c c}
\hline
\hline
Distance (km) & 0 km  & 10 km & 20 km & 30 km \\ 
\hline
Ref.~\cite{Primaatmaja2022Discrete} (4-decoy and no noise) & $8.8\times 10^{-2}$ &$9.8\times 10^{-3}$ & Below $1\times 10^{-4}$ & No key rate \\
This work (4-decoy and no noise) & $1.2\times 10^{-1}$& $1.2\times 10^{-2}$ & $7.3 \times 10^{-4}$ & $2.2 \times 10^{-5}$\\
\hline
Ref.~\cite{Matsuura2021Finite} ($10^{-3}$ excess noise) & $4.1\times 10^{-1}$ &$6.2\times 10^{-2}$ & $2.1\times 10^{-3} $ & No key rate \\
This work (4-decoy and $10^{-3}$ excess noise) & $1.2\times 10^{-1}$& $1.2\times 10^{-2}$ & $5.8 \times 10^{-4}$ & $1.2 \times 10^{-5}$\\
\hline
\hline
\end{tabular}
\label{Tab::CompareProtocols}
\end{table*}

\section{Conclusion and outlook}
In summary, we present the time-bin-encoding CV QKD protocol with a phase-error-based security analysis. Similar to the ideas in DV protocols~\cite{Beaudry2008squashing} and other CV protocols~\cite{Matsuura2021Finite}, we introduce a squashing channel to ``squash'' the original privacy-estimation problem on two optical modes to a single qubit, enabling the definition of phase-error rate. Phase randomization on both the source and detector enables the introduction of the photon-number-tagging method, identifying the central components for key generation. Combined with the decoy-state estimation, the parameter estimation is made simple and efficient. We are thus able to obtain a finite-size analysis with decent performances under a practical setting. We expect our methods of constructing squashing models and applying phase randomization can be applied to many other CV protocols. 

In the protocol design, one of our major observations is that coherent detectors can be used to estimate the privacy of multi-photon signals. This is also pointed out in Ref.~\cite{Primaatmaja2022Discrete}. Such detectors may also be helpful to the DV protocols. In fact, we may consider a hybrid protocol: single-photon detectors for key generation and homodyne detectors for parameter estimation. The multi-photon components in this protocol can contribute to key generation compared with the single-photon BB84 protocol.

To evaluate the finite-size performance of this CV protocol, we provide a detailed analysis based on martingale theory, which is valid against coherent attacks. Notably, combined with the photon-number tagging method, the phase-error approach greatly simplifies the finite-size analysis. From the numerical simulation under practical security parameters, it can be seen that our protocol enjoys a robust finite-size performance with a reasonable data size requirement for practical usage.
% A direct follow-up of this work is to complete the finite-size analysis, encompassing the effects on the distillable key rate, the decoy-method accuracy, and the deviation of the homodyne tomography. In the literature, variants of Azuma's inequality have been applied for faster convergence of parameter estimation in quantum key distribution~\cite{curras2021tight,zhang2023quantum}, such as Kato's inequality~\cite{kato2020concentration}. One can bring such techniques to the protocol in this work for better practicality.}

It is tempting to further enhance the key rate and the maximal distance of this protocol. We may consider the high-dimensional time-bin encoding, which is relatively easy to implement experimentally~\cite{islam2017provably,Islam2019scalable,Vagniluca2020efficient}. The high-dimensional complementarity security analysis~\cite{Jin2021reference} can be invoked, and the squashing channel should map the optical modes to a qudit. We can also apply the trusted-noise model to alleviate the effect of the detector noise~\cite{Usenko2016trusted,Qi2018noise}. The model requires the modification of the detector POVM, which is still block-diagonal on the Fock basis~\cite{Primaatmaja2022Discrete}. One may also consider using squeezed states as the light source to reduce the shot noise in one quadrature and use the other only for parameter estimation. This may tackle the large bit error rate due to the shot noise, the issue that renders the 0-km performance of our protocol not as good as the usual CV QKD scheme. We can also examine the variations of our protocol based on the combination with new DV QKD schemes such as the measurement-device-independent-type schemes~\cite{Lo2012MDI,Ma2012alternative} and their extensions, including the twin-field-type~\cite{Lucamarini2018overcoming,Ma2018PM,Wang2018twin} and the mode-pairing schemes~\cite{Zeng2022mode,Xie2022breaking}.

\begin{backmatter}
%\bmsection{Funding}
%Content in the funding section will be generated entirely from details submitted to Prism. Authors may add placeholder text in this section to assess length, but any text added to this section will be replaced during production and will display official funder names along with any grant numbers provided. If additional details about a funder are required, they may be added to the Acknowledgment, even if this duplicates some information in the funding section. For preprint submissions, please include funder names and grant numbers in the manuscript.

\bmsection{Acknowledgment}
We acknowledge insightful discussions with Hoi-Kwong Lo and Xiongfeng Ma. A.J. and R.V.P. acknowledge support from the UK EPSRC Quantum Communications Hub, project EP/T001011/1. A.J. acknowledges funding from Cambridge Trust. 
X.~Z. acknowledges support from Shenzhen-Hong Kong cooperation zone for technology and innovation (Contract NO. HZQB-KCZYB-2020050) and Hong Kong Research Grant Council through grant number R7035-21 of the Research Impact Fund and No.~27300823. 
P.~Z.~ and L.~J.~acknowledge support from the ARO(W911NF-23-1-0077), ARO MURI (W911NF-21-1-0325), AFOSR MURI (FA9550-19-1-0399, FA9550-21-1-0209), NSF (OMA-1936118, ERC-1941583, OMA-2137642), NTT Research, Packard Foundation (2020-71479), and the Marshall and Arlene Bennett Family Research Program.

A.~J. and X.~Z. contributed equally to this work. 

\bmsection{Disclosures}
The authors declare no conflicts of interest.

\bmsection{Data availability} Data underlying the results presented in this paper may be obtained from the authors upon reasonable request.

\end{backmatter}

%%%%%%%%%%%%%%%%%%%%%%% References %%%%%%%%%%%%%%%%%%%%%%%%%

%%%%%%%%%% If using BibTeX:
\bibliography{bibCVwDV}

\onecolumn
\appendix
\renewcommand{\thesubsection}{\thesection.\arabic{subsection}}
\section{Refined key mapping scheme}
\label{Sec:refinedkeymapping}

\begin{figure}[htbp!] 
\centering
\includegraphics[width=0.7\linewidth]{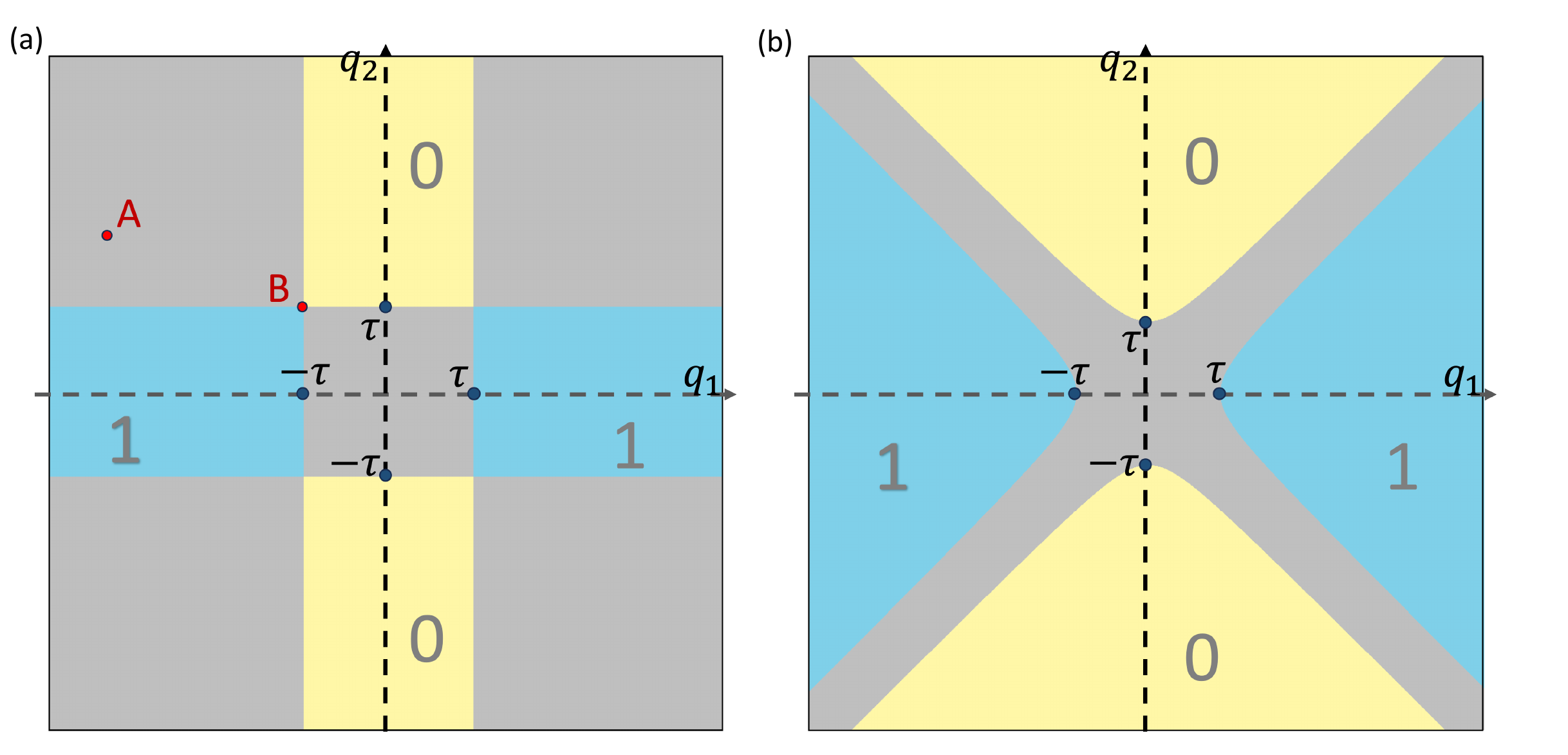}
\caption{The simple and the maximum-likelihood-based key mapping. The two axes represent the two homodyne measurement results $(q_1,q_2)$. The yellow region is decoded as 0 and the blue decoded as 1. The grey region is discarded. \textbf{(a)} The simple key mapping scheme used throughout this paper, being rectangular. Point $A$ has key mapping error 0.02$\%$ and point $B$ 50$\%$, yet point $A$ is discarded whilst $B$ is accepted. \textbf{(b)} The maximum-likelihood key mapping scheme, being approximately triangular. The post-selection is done for coherent light amplitude $\alpha=1$, where only the region with likelihood ratio greater than 10 is kept in this showcase. In this refined key mapping, point $A$ is accepted and point $B$ is discarded.}
\label{Fig:refinedregions}
\end{figure}

In the time-bin-encoded CV QKD protocol in Table~\ref{table:protocol}, we consider a simple key mapping strategy with a threshold value $\tau$ illustrated in Fig.~\ref{Fig:refinedregions}(a). It can be seen that the security analysis in Sec.~\ref{Sec::security} does not rely on the specific shapes of the key mapping regions $\mathbf{R_0}$ and $\mathbf{R_1}$, as long as they differ by a swap of the two optical modes. As a result, we can optimize $\mathbf{R_0}$ and $\mathbf{R_1}$ for a higher key-rate performance indicated by $r$ in \eqref{Eq::Keyrev}. 

To get a higher key rate $r$, we want to avoid the post-selection of the detected signal region $(q_1, q_2)$ as long as the bit error rate $e^Z$ can be ensured low. To this end, we introduce a maximum-likelihood-based key mapping and analyze its performance. Corresponding to bit 0 or 1, Bob would receive coherent states $\ket{0}\ket{\alpha e^{i\varphi_a}}$ or $\ket{\alpha e^{i\varphi_a}}\ket{0}$ for some randomized phase $\varphi_a$ and $\alpha$ after attenuation in a pure-loss channel. Corresponding to bit value 0 and 1, the probability distributions $f_0$ and $f_1$ of the homodyne measurement results $(q_1,q_2)$ are given by
\begin{equation}
\label{Eq:likelihood0}
f_0(q_1,q_2) = \int_0^{2\pi} \frac{d\varphi}{4\pi^2}\exp\frac{\left[-q_1^2-(q_2-2|\alpha| \cos\varphi)^2\right]}{2}, 
\end{equation}
\begin{equation}
\label{Eq:likelihood1}
f_1(q_1,q_2) = \int_0^{2\pi} \frac{d\varphi}{4\pi^2}\exp\frac{\left[-q_2^2-(q_1-2|\alpha| \cos\varphi)^2\right]}{2},   
\end{equation}
where $\varphi$ is the difference between the source and detector phase randomization. 

Upon detecting a specific pair of quadratures $(q_1,q_2)$, the maximum-likelihood key mapping scheme requires to decode bit 0 if $f_0(q_1,q_2) > f_1(q_1,q_2)$ and bit 1 if $f_1(q_1,q_2) > f_0(q_1,q_2)$. According to \eqref{Eq:likelihood0} and~\eqref{Eq:likelihood1}, the maximum-likelihood key mapping is equivalent to the decoding of bit 0 if $q_1^2 < q_2^2$ and bit 1 if $q_1^2 > q_2^2$. This refined key mapping takes in detection results such as point $A$ in Fig.~\ref{Fig:refinedregions}(a) where $q_1^2$ is significantly different from $q_2^2$, thus reducing the post-selection loss of the gain. However, in the region where $q_1^2$ and $q_2^2$ are comparable, the key mapping error will be large, making a large contribution to the final bit error rate $e^Z$. Conditioned on the detection outcome $(q_1,q_2)$, the key mapping error $e^Z_0(q_1,q_2)$ in $\mathbf{R_0}$ is related to the likelihood function by
\begin{equation}
    e^Z_0(q_1,q_2) = \mathrm{Pr}((q_1,q_2)\in \mathbf{R_0}|k_a=1) = \frac{f_1(q_1,q_2)}{Q(q_1,q_2)},
\end{equation}
with the key mapping gain $Q(q_1,q_2) = f_0(q_1,q_2) + f_1(q_1,q_2)$.
We can similarly define $e^Z_1(q_1,q_2)$ for the $\mathbf{R_1}$ region. The final bit error rate $e^Z$ is
\begin{equation}
\begin{aligned}
    e^Z = \frac{1}{2} \int_{\mathbf{R_0}} dq_1 dq_2\,  e^Z_0(q_1,q_2) \, Q(q_1,q_2) \\
    + \frac{1}{2} \int_{\mathbf{R_1}} dq_1 dq_2\,  e^Z_1(q_1,q_2) \, Q(q_1,q_2). 
\end{aligned}
\end{equation}

For example, at point $B$ in Fig.~\ref{Fig:refinedregions}(a), the key mapping error is 50$\%$. To discard the erroneous results, we can set a threshold $t > 1$ and require to decode bit 0 if $f_0(q_1,q_2) > t f_1(q_1,q_2)$ and bit 1 if $f_1(q_1,q_2) > t f_0(q_1,q_2)$. The region in between will be discarded. Fig.~\ref{Fig:refinedregions}(b) below illustrates the numerically plotted key mapping regions given the light amplitude $\alpha = 1$ at $t = 10$, where the grey region is discarded. It can be seen that the key mapping regions are nearly isosceles right triangles with non-zero intercepts $\pm \tau$ with the quadrature axes. We thus present the approximately maximum-likelihood key mapping scheme as the following:
\begin{equation}
 k_b = 
\begin{cases}
0 \quad &\text{ if } (q_2 - \tau > \pm q_1) \vee (q_2 + \tau < \pm q_1),\\
1 \quad &\text{ if } (q_1 - \tau > \pm q_2) \vee (q_1 + \tau < \pm q_2),\\
\emptyset \quad &\text{otherwise}.
\end{cases} 
\end{equation}   

As in Sec.~\ref{Sec::results}, we regard the intercept $\tau$ as a protocol parameter. We optimize it and the source intensity $\mu$ for each distance to yield the optimal key rate. However, numerical optimization shows that this refined key mapping would only improve the key-rate performances marginally. It increases the key rate at 0 km of the ideal 2-photon protocol from 0.1261 bit/channel to 0.1314 bit/channel and that of the ideal 4-photon protocol from 0.3131 bit/channel to 0.3242 bit/channel, with almost no increase to the maximal transmission distance. This is due to the low gain of the newly-accepted region such as point $A$ in Fig.~\ref{Fig:refinedregions}(a). Considering the experimental cost as well, we thus suggest that using the simple threshold key mapping as in Sec.~\ref{Sec:protocol} is good enough in practice.

\section{Simulation formulae under thermal-noise channel}
\label{Sec:simufomula}
We present the simulation formulae of the asymptotic time-bin CV QKD under a thermal-noise channel with excess noise $\xi$ from the output. We set $\hbar = 2$ so that the vacuum variance is 1. A thermal noise channel is characterized as a Gaussian completely positive map transforming the first and second moment $(\bar{r},V)$, representing the mean vector and covariance matrix of the quadrature operators, of the input state as~\cite{weedbrook2012gaussian}:
\begin{equation}
\begin{split}
&\bar{r} \mapsto \sqrt{\eta} \bar{r},\\
&V \mapsto \eta V + (1-\eta)\mathbb{I} + \xi \mathbb{I},
\end{split}
\end{equation}
where $\eta$ is the channel transmittance. Two thermal channels with transmittance $\eta$ and $\eta'$ and excess noise $\xi$ and $\xi'$ concatenate to another thermal channel with transmittance $\eta\eta'$ and excess noise $(\eta'\xi + \xi')$ since
\begin{equation}
\label{Eq:thermalconcatenate}
\begin{split}
&\bar{r} \mapsto \sqrt{\eta'\eta} \bar{r},\\
&V \mapsto \eta'(\eta V + (1-\eta)\mathbb{I} + \xi \mathbb{I}) + (1-\eta')\mathbb{I} + \xi' \mathbb{I}\\
&= \eta'\eta V + (1-\eta'\eta)\mathbb{I} + (\eta'\xi+\xi')\mathbb{I}. 
\end{split}
\end{equation}

On the bit-error side, the thermal noise can be seen as adding $\xi$ to the unity variance of the coherent states. Hence, if Alice transmits a coherent state $\ket{\sqrt{\mu}e^{i\theta}}$ through a thermal-noise channel with transmittance $\eta$ and excess noise $\xi$, and Bob applies homodyne detection with LO phase $\varphi$, the detection result $q$ will follow a distribution
\begin{equation}
\text{Pr}(q|\mu,\theta-\varphi) = \frac{1}{\sqrt{2\pi}}\exp\left\{-\frac{[q-2\sqrt{\eta\mu}\cos(\theta-\varphi)]^2}{2(1+\xi)}\right\}.  
\end{equation}
Since both the signal states and the receiver LO are uniformly phase randomized, $(\theta - \varphi)$ is also uniformly randomized with $[0,2\pi)$ in a cyclic manner. The bit error rate $e^Z_{\mu}$ and the $Z$-basis gain $Q^Z_{\mu}$ can thus be calculated according to the post-selection threshold $\tau$, uniformly randomizing over $[0,2\pi)$.

The calculation of the vacuum gain $Q_{*,0}$, according to \eqref{eq:ReverseGain}, requires the probability of sending the $Z$-basis state whilst receiving vacuum. This can be calculated via the Wigner function for Gaussian state, and in specific
\begin{equation}
\tr\left[\hat{P}_0^{B_1B_2}\mathcal{N}_E^{A_1A_2\to B_1B_2}\left(\hat{\rho}^Z\right)\hat{P}_0^{B_1B_2}\right] = \left(\frac{2}{2+\xi}\right)^2\exp\left(-\frac{2\eta\mu}{2+\xi}\right).
\end{equation}

The single- and two-photon gains, $Q_{1,1}$ and $Q_{2,2}$, and phase-error rates, $e_{1,1}^X$ and $e_{2,2}^X$, are more complicated in calculation. In the infinite-decoy setup, we calculate the photon gains directly. We decompose the thermal noise $\hat{\rho}_{\mathrm{th}}$ into Fock states,
\begin{equation}
\label{Eq:thermalphoton}
\hat{\rho}_{\mathrm{th}} = \sum_{k=0}^{\infty} \frac{\bar{k}^k}{(\bar{k}+1)^{k+1}}\ket{k}\bra{k},
\end{equation}
where $\bar{k} = \xi/2(1-\eta)$ is the average photon number of the thermal noise. The optical mode from Alice can be seen as mixing with the thermal noise through an $\eta$-transmittance beam splitter. We calculate the effect of the thermal noise in an ensemble manner, that is, we calculate the case where the channel injects $k$ and $l$ noise photons to the two consecutive optical modes respectively, and mix the results according to the noise photon-number distribution in \eqref{Eq:thermalphoton}. We set a cutoff photon-number at $N_c = 3$ since the thermal noise is relatively low. Simulation shows that higher cutoffs have negligible effects on the key rate. We also account for the effects of the misalignment angle $\delta$, which introduces a $\sin^2(m\delta/2)$ error to the $m$-photon phase error rate. We ignore the correlation between the misalignment and thermal noise photon as a second-order small quantity.

The calculations of the quantities of interest are listed below. The notation of $(k,l)$ represents the conditional probability that the thermal sources emit $k$ and $l$ photons respectively to the two optical modes:
\begin{enumerate}
\item The probability of sending $(\ket{01}\bra{01}+\ket{01}\bra{01})/2$ whilst accepting one photon in total (\eqref{Eq::Q11}):
\begin{equation}
Q_{1,1}(k,l) = c_1{\Pr(1)}\eta^{k+l-1} \left\{[(k+1)\eta-k]^2+l(k+1)(1-\eta)^2\right\},
\end{equation}
\begin{equation}
Q_{1,1} = \sum_{k=0,l=0}^{N_c}\text{P}_{\mathrm{th}}(k)\text{P}_{\mathrm{th}}(l)Q_{1,1}(k,l),
\end{equation}
where 
\begin{equation}
\text{P}_{\mathrm{th}}(k) = \frac{\bar{k}^k}{(\bar{k}+1)^{k+1}}\text{ with }\bar{k} = \frac{\xi}{2(1-\eta)}.
\end{equation}

\item The probability of sending $(\ket{01}\pm\ket{10})/\sqrt{2}$ whilst receiving $(\ket{01}\mp\ket{10})/\sqrt{2}$ (\eqref{Eq::ex11}):
\begin{equation}
\frac{e^X_{1,1}(k,l)Q_{1,1}}{\text{Pr}(1)} = \frac{c_1}{4}\eta^{k+l-1}(1-\eta)^2(k^2+l^2+k+l)+c_1\sin^2\left(\frac{\delta}{2}\right),
\end{equation}
\begin{equation}
e_{1,1}^X = \sum_{k=0,l=0}^{N_c}\text{P}_{\mathrm{th}}(k)\text{P}_{\mathrm{th}}(l) e^X_{1,1}(k,l).
\end{equation}

\item The probability of sending $(\ket{02}\bra{02}+\ket{20}\bra{20})/2$ whilst accepting within the $(\ket{02}\bra{02}+\ket{20}\bra{20})$ and $\ket{11}\bra{11}$  subspace (\eqref{Eq::Q22}):
\begin{equation}
Q^{02}_{2,2}(k,l) = \frac{1}{2}c_2^{02}\Pr(2)\eta^{k+l-2}\left\{\left[\eta^2-2k\eta(1-\eta)+\frac{1}{2}k(k-1)(1-\eta)^2\right]^2 + \frac{1}{4}l^2(l-1)^2(1-\eta)^4\right\} + \{k\leftrightarrow l\},
\end{equation}
\begin{equation}
Q^{11}_{2,2}(k,l) =  \frac{1}{2}c_2^{11}\Pr(2)\eta^{k+l-2}\left[\sqrt{2(k+1)l}\eta(1-\eta)-\sqrt{\frac{1}{2}kl(k+1)}(1-\eta)^2\right]^2 + \{k\leftrightarrow l\},   
\end{equation}
\begin{equation}
Q_{2,2} = \sum_{k=0,l=0}^{N_c}\text{P}_{\mathrm{th}}(k)\text{P}_{\mathrm{th}}(l)\left[Q^{02}_{2,2}(k,l) + Q^{11}_{2,2}(k,l)\right],   
\end{equation}
where the expression $\{k\leftrightarrow l\}$ denotes exchanging the $k$'s and $l$'s in the term ahead.

\item The probability of sending $(\ket{02}\pm\ket{20})/\sqrt{2}$ whilst receiving $(\ket{02}\mp\ket{20})/\sqrt{2}$ and $\ket{11}$ (\eqref{Eq::eX22}).
\begin{equation}
\frac{e^{02,X}_{2,2}(k,l)Q_{2,2}}{\Pr(2)} = \frac{c_2^{02}}{4}\eta^{k+l-2}\left[2(k-l)(1-\eta)\eta + (k^2-k-l^2-l)(1-\eta)^2\right]^2+ c_2^{02}\sin^2(\delta),
\end{equation}
\begin{equation}
\frac{e^{11,X}_{2,2}(k,l)Q_{2,2}}{\Pr(2)} = c_2^{11}\eta^{k+l-2}(1-\eta)^2\left\{l(k+1)\left[\eta-\frac{1}{2}k(1-\eta)\right]^2+k(l+1)\left[\eta-\frac{1}{2}l(1-\eta)\right]^2\right\},
\end{equation}
\begin{equation}
\frac{e_{2,2}^XQ_{2,2}}{\text{Pr}(2)} = \sum_{k=0,l=0}^{N_c}\text{P}_{\mathrm{th}}(k)\text{P}_{\mathrm{th}}(l)\left[e_{2,2}^{02,X}(k,l) + e_{2,2}^{11,X}(k,l)\right].    
\end{equation}
\end{enumerate}

In the finite-decoy setup, the estimations of photon gains are derived from the statistics of coherent-state gains. We need to calculate the probability of transmitting certain coherent states whilst receiving certain photon states. This can be also be done by the Gaussian-state Wigner function. Let $\kappa = 2/(2+\xi)$. Denote the output of a thermal noise channel when transmitting the coherent state $\ket{\alpha}$ as $\rho_\alpha$.  Its Fock-basis matrix elements are:
\begin{equation}
\bra{0}\rho_\alpha\ket{0} = \kappa\exp(-\kappa |\alpha|^2),
\end{equation}
\begin{equation}
\bra{1}\rho_\alpha\ket{1} = \kappa (\kappa^2|\alpha|^2 + 1-\kappa)\exp(-\kappa |\alpha|^2),  
\end{equation}
\begin{equation}
\bra{0}\rho_\alpha\ket{1} = -\kappa^2\alpha^{*}\exp(-\kappa |\alpha|^2), 
\end{equation}
\begin{equation}
\bra{2}\rho_\alpha\ket{2} = \kappa(\frac{1}{2}\kappa^4|\alpha|^4 + 2(\kappa^2-\kappa^3)|\alpha|^2+(1-\kappa^2))\exp(-\kappa |\alpha|^2),
\end{equation}
\begin{equation}
\bra{0}\rho_\alpha\ket{2} = \frac{1}{\sqrt{2}}\kappa^3(\alpha^{*})^2 \exp(-\kappa |\alpha|^2).
\end{equation}
The statistics required by the decoy method are all based on the gains of separable coherent states. For example, the probability of sending $\ket{\alpha}\otimes\ket{\beta}$ whilst receiving $(\ket{02}+\ket{20})/\sqrt{2}$ can be computed by
\begin{equation}
\begin{split}
&\quad\frac{1}{2}(\bra{02}+\bra{20})\rho_{\alpha}\otimes\rho_{\beta}(\ket{02}+\ket{20})\\ 
&= \frac{1}{2} \left(\bra{0}\rho_\alpha\ket{0}\bra{2}\rho_\beta\ket{2} + \bra{2}\rho_\alpha\ket{2}\bra{0}\rho_\beta\ket{0} + \bra{0}\rho_\alpha\ket{2}\bra{2}\rho_\beta\ket{0} +\bra{2}\rho_\alpha\ket{0}\bra{0}\rho_\beta\ket{2}\right).
\end{split}
\end{equation}

\section{Phase-error calculation details}
\label{App:PERcal}
We give the detailed derivation of the phase-error probability expressions \eqref{Eq::ex11} to~\eqref{eq:ReverseGain} for the zero, one and two-photon components. According to \eqref{Eq::PERinfock}, the calculation involves in expanding the $X$-basis measurement operator $M_{\pm}$ based on \eqref{Eq::homo2fock}. For the single-photon subspace, we have
\begin{equation}
\begin{split}
\hat{P}_1\left(\int_0^{2\pi} \frac{d\varphi}{2\pi} \ket{q_1(\varphi)}\bra{q_1(\varphi)} \otimes \ket{q_2(\varphi)}\bra{q_2(\varphi)}\right)\hat{P}_1 = \psi_0^2(q_1)\psi_1^2(q_2)\ket{01}\bra{01} + \psi_1^2(q_1)\psi_0^2(q_2)\ket{10}\bra{10} +\\
\psi_0(q_1)\psi_0(q_2)\psi_1(q_1)\psi_1(q_2)\ket{01}\bra{10} + \psi_0(q_1)\psi_0(q_2)\psi_1(q_1)\psi_1(q_2)\ket{10}\bra{01}
\end{split}
\end{equation}

\begin{equation}
\begin{split}
\hat{P}_1\left(\int_0^{2\pi} \frac{d\varphi}{2\pi} \ket{q_2(\varphi)}\bra{q_2(\varphi)} \otimes \ket{q_1(\varphi)}\bra{q_1(\varphi)}\right)\hat{P}_1 =
\psi_1^2(q_1)\psi_0^2(q_2)\ket{01}\bra{01} + \psi_0^2(q_1)\psi_1^2(q_2)\ket{10}\bra{10} +\\
\psi_0(q_1)\psi_0(q_2)\psi_1(q_1)\psi_1(q_2)\ket{01}\bra{10} + \psi_0(q_1)\psi_0(q_2)\psi_1(q_1)\psi_1(q_2)\ket{10}\bra{01}
\end{split}
\end{equation}

\begin{equation}
\begin{split}
\hat{P}_1\left(\int_0^{2\pi} \frac{d\varphi}{2\pi} \ket{q_1(\varphi)}\bra{q_2(\varphi)} \otimes \ket{q_2(\varphi)}\bra{q_1(\varphi)}\right)\hat{P}_1 =
\psi_0^2(q_1)\psi_1^2(q_2)\ket{01}\bra{10} + \psi_1^2(q_1)\psi_0^2(q_2)\ket{10}\bra{01} +\\
\psi_0(q_1)\psi_0(q_2)\psi_1(q_1)\psi_1(q_2)\ket{01}\bra{01} + \psi_0(q_1)\psi_0(q_2)\psi_1(q_1)\psi_1(q_2)\ket{10}\bra{10}
\end{split}
\end{equation}

\begin{equation}
\begin{split}
\hat{P}_1\left(\int_0^{2\pi} \frac{d\varphi}{2\pi} \ket{q_1(\varphi)}\bra{q_2(\varphi)} \otimes \ket{q_2(\varphi)}\bra{q_1(\varphi)}\right)\hat{P}_1 =
\psi_1^2(q_1)\psi_0^2(q_2)\ket{01}\bra{10} + \psi_0^2(q_1)\psi_1^2(q_2)\ket{10}\bra{01} +\\
\psi_0(q_1)\psi_0(q_2)\psi_1(q_1)\psi_1(q_2)\ket{01}\bra{01} + \psi_0(q_1)\psi_0(q_2)\psi_1(q_1)\psi_1(q_2)\ket{10}\bra{10}
\end{split}
\end{equation}

Hence the single-photon phase-error operator is given by
\begin{equation}
\begin{split}
\label{Eq::fockX1}
\hat{P}_1\hat{M}_{\pm}\hat{P}_1 &= \int_{\mathbf{R_0}}dq_1 dq_2  [\psi_0(q_1)\psi_1(q_2) \pm \psi_0(q_2)\psi_1(q_1)]^2\quad\frac{1}{2}(\ket{01}\pm\ket{10})(\bra{01}\pm\bra{10})\\
&= \int_{\mathbf{R_0}}dq_1 dq_2 [\psi_0^2(q_1)\psi_1^2(q_2)+ \psi_1^2(q_1)\psi_0^2(q_2)]\quad\frac{1}{2}(\ket{01}\pm\ket{10})(\bra{01}\pm\bra{10}),
\end{split}
\end{equation}
Note that the second equality is deduced as the cross terms are odd functions with respect to $q_1$ and $q_2$, and the key-mapping region is symmetrical. This gives \eqref{Eq::ex11}. The gain $Q_{1,1}$ involves in measuring $\hat{P}_1(\hat{M}_{+}+\hat{M}_{-})\hat{P}_1$ which is clearly in the form of \eqref{Eq::Q11}.

The two-photon case involves in more terms, but we can make use of the symmetry of the key mapping region $\mathbf{R_0}$ to eliminate the odd terms. The calculation goes by:
\begin{equation}
\begin{split}
\label{Eq::fockX2}
&\hat{P}_2\left(\int_0^{2\pi} \frac{d\varphi}{2\pi} \ket{q_1(\varphi)}\bra{q_1(\varphi)} \otimes \ket{q_2(\varphi)}\bra{q_2(\varphi)}\right)\hat{P}_2 = \psi_0^2(q_1)\psi_2^2(q_2)\ket{02}\bra{02}+\psi_0(q_1)\psi_0(q_2)\psi_2(q_1)\psi_2(q_2)\ket{02}\bra{20}\\
&\quad+\psi_0(q_1)\psi_0(q_2)\psi_2(q_1)\psi_2(q_2)\ket{20}\bra{02} + \psi_0^2(q_2)\psi_2^2(q_1)\ket{20}\bra{20} + \psi_1^2(q_1)\psi_1^2(q_2)\ket{11}\bra{11} + \text{ odd terms}
\end{split}    
\end{equation}
The $X$-basis measurement is thus given by:
\begin{equation}
\begin{split}
\hat{P}_2\hat{M}_{+}\hat{P}_2 &= \int_{\mathbf{R_0}}dq_1dq_2 [\psi_0(q_1)\psi_2(q_2) + \psi_2(q_1)\psi_0(q_2)]^2\quad \frac{1}{2}(\ket{02}+\ket{20})(\bra{02}+\bra{20})\\  
&+\int_{\mathbf{R_0}}dq_1dq_2 2\psi_1^2(q_1)\psi_1^2(q_2)\ket{11}\bra{11}
\end{split}
\end{equation}
\begin{equation}
\hat{P}_2\hat{M}_{-}\hat{P}_2 = \int_{\mathbf{R_0}}dq_1dq_2 [\psi_0(q_1)\psi_2(q_2) - \psi_2(q_1)\psi_0(q_2)]^2\quad \frac{1}{2}(\ket{02}-\ket{20})(\bra{02}-\bra{20})
\end{equation}
This recovers \eqref{Eq::eX22}, and adding the two equations together gives the expression for the gain $Q_{2,2}$ as in \eqref{Eq::Q22}. Expanding the vacuum subspace according to \eqref{Eq::homo2fock} gives the coefficients as in \eqref{eq:ReverseGain}.

\section{Preliminaries for parameter estimation}
\subsection{Homodyne tomography}\label{App::homotomo}
We provide the in Supplementary material the complete finite-size analysis with detailed parameter estimation. The first issue to tackle is the estimation of photon-number statistics. Due to the lack of photon-number-resolving detectors, these operators are not directly measurable. Nevertheless, we can apply homodyne tomography and obtain unbiased estimation~\cite{vogel1989determination,smithey1993measurement,d1994detection,d1995homodyne,leonhardt1995measuring,d1995tomographic}. For a systematic review, we recommend the tutorial textbook of Ref.~\cite{DAriano2007HomoTomo}.

We start with a single-mode system. Consider the displacement operators given by
\begin{equation}
\begin{split}
    \hat{D}(\alpha)=&\exp(\alpha \hat{a}^\dag-\alpha^*\hat{a}) \\
    =&\exp\left[(-ik)\left(\hat{a}^\dag e^{i\varphi}+\hat{a}e^{-i\varphi}\right)\right] \\
    :=&\exp(-ik\hat{Q}_\varphi),
\end{split}
\end{equation}
where $\hat{a}$ and $\hat{a}^\dag$ are the annihilation and creation operators of the mode, respectively, $\alpha$ is a complex scalar, and $\alpha^*$ denotes the complex conjugate of $\alpha$. In the second equation, we use the polar variables to represent $\alpha$, $\alpha=-ike^{i\varphi}$. We call $\hat{Q}_\varphi$ the quadrature operator. Measuring $\hat{Q}_\varphi$ corresponds to the homodyne measurement, where the phase of the LO is $\varphi$. By definition, $\hat{D}(\alpha)$ is a Hermitian operator. The set of all displacement operators forms an orthogonal and complete function basis on a mode; hence any linear operator on a mode, $\hat{O}$, can be expanded with displacement operators,
\begin{equation}\label{eq:QuorumExp}
    \hat{O}=\int_0^\pi\frac{d\varphi}{\pi}\int_{-\infty}^{\infty}\frac{dk|k|}{4}\tr(\hat{O}e^{ik\hat{Q}_\varphi})e^{-ik\hat{Q}_\varphi}.
\end{equation}
When measuring $\hat{O}$ on a state, $\hat{\rho}$, the expected value is given by
\begin{equation}
\label{eq:homodynetomo}
\begin{split}
    \langle \hat{O}\rangle =& \tr(\hat{O}\hat{\rho}) \\
    =&\int_0^\pi\frac{d\varphi}{\pi}\int_{-\infty}^{\infty}\frac{dk|k|}{4}\tr(\hat{O}e^{ik\hat{Q}_\varphi})\tr(\hat{O}e^{-ik\hat{Q}_\varphi}) \\
    :=&\int_0^\pi\frac{d\varphi}{\pi}\int_{-\infty}^{\infty}dq\tr[\hat{O}K(\hat{Q}_\varphi-q)]p(q|\varphi) \\
    :=&\int_0^\pi\frac{d\varphi}{\pi}\int_{-\infty}^{\infty}dq\mathcal{R}[\hat{O}](q,\varphi)p(q|\varphi),
\end{split}
\end{equation}
where the value of the term
\begin{equation}
    K(q):=\int_{-\infty}^{\infty}\frac{dk}{4}|k|e^{ikq}
\end{equation}
should be determined by the Cauchy principal value, $p(q|\varphi)$ is the conditional probability of obtaining quadrature $q$ when the phase of the homodyne measurement is $\varphi$, and $\mathcal{R}[\hat{O}](q,\varphi)$ is the kernel function of $\hat{O}$ with respect to the homodyne measurement. Eq.~\eqref{eq:homodynetomo} gives a sampling procedure to estimate $\langle \hat{O}\rangle$ for a general unknown system using homodyne measurements~\cite{d1994detection,d1995homodyne,d1995tomographic}. Namely,
\begin{enumerate}
\item Repeat the following process for $N$ times:
\begin{enumerate}
    \item Choose the LO phase of the homodyne measurement, $\varphi_i\in[0,\pi]$, uniformly at random.
    \item Measure the system and record the result, $q_i$.
\end{enumerate}
\item Calculate the average value of the kernel function with respect to the observed statistics, $\sum_{i=1}^N\mathcal{R}[\hat{O}](q_i,\varphi_i)/N$.
\end{enumerate}
When the kernel function is bounded, the law of large numbers guarantees the convergence,
\begin{equation}
    \langle \hat{O}\rangle=\lim_{N\rightarrow\infty}\frac{1}{N}\sum_{i=1}^N\mathcal{R}[\hat{O}](q_i,\varphi_i).
\end{equation}
Note that, when $\hat{O}$ is an Hermitian operator or we know that $\mathrm{Im}\left( \langle \hat{O}\rangle \right)=0$, we have
\begin{equation}
\begin{aligned}
    \langle \hat{O}\rangle &= \mathrm{Re}\left( \langle \hat{O}\rangle \right) \\
    &= \int_0^\pi\frac{d\varphi}{\pi}\int_{-\infty}^{\infty}dq \, 
 \mathrm{Re}\left( \mathcal{R}[\hat{O}](q,\varphi) \right) p(q|\varphi).
\end{aligned}
\end{equation}
That is, we can redefine the estimator to be the real part of $\mathcal{R}[\hat{O}](q,\varphi)$, which is still unbiased.

In our protocol, we are interested in the photon-number operators, $\ket{n}\bra{n+d}$, where $\ket{n}$ is the eigenstate of the $n$-photon eigenstate. For a single mode, the estimator of this operator is given by
\begin{equation}\label{eq:Fockkernel}
\begin{split}
    \mathcal{R}_\eta[\ket{n}\bra{n+d}](q,\varphi)=&e^{id(\varphi+\frac{\pi}{2})}\sqrt{\frac{n!}{(n+d)!}}\int_{-\infty}^{\infty}dk|k| \exp\left(\frac{1-2\eta}{2\eta}k^2-ikq\right)k^dL_n^d(k^2),
\end{split}
\end{equation}
where $\eta$ is the detector efficiency, and $L_n^d$ is the generalized Laguerre polynomial~\cite{d1994detection,d1995homodyne,leonhardt1995tomographic}. The kernel function is bounded when $\eta>1/2$, allowing for a converging tomography result by increasing samples~\cite{d1995homodyne,d1995tomographic}.homodyne tomography can be generalized to estimate the statistics of a multiple-mode observable. In our case, one needs to estimate separable observables on two modes, $\hat{O}_1\otimes\hat{O}_2$, where we specify the modes with subscripts. One can apply independent homodyne measurements to each mode for the estimation. Notably, as $\hat{D}(\alpha_1)\otimes\hat{D}(\alpha_2)$ forms a complete basis on the joint system, then
\begin{equation}
\begin{split}
    \hat{O}_1\otimes \hat{O}_2=\int_0^\pi\frac{d\varphi_1}{\pi}\int_{-\infty}^{\infty}\frac{dk_1|k_1|}{4}\tr(\hat{O}_1e^{ik_1\hat{Q}_{\varphi_1}})\int_0^\pi\frac{d\varphi_2}{\pi}\int_{-\infty}^{\infty}\frac{dk_2|k_2|}{4}\tr(\hat{O}_2e^{ik_2\hat{Q}_{\varphi_2}})(e^{-ik_1\hat{Q}_{\varphi_1}}\otimes e^{-ik_2\hat{Q}_{\varphi_2}}).
\end{split}
\end{equation}
Consequently,
\begin{equation}\label{eq:twomodekernel}
\begin{split}
    \langle \hat{O}_1\otimes \hat{O}_2\rangle &= \tr[(\hat{O}_1\otimes \hat{O}_2)\hat{\rho}] \\
    &=\int_0^\pi\frac{d\varphi_1}{\pi}\int_{-\infty}^{\infty}dq_1\int_0^\pi\frac{d\varphi_2}{\pi}\int_{-\infty}^{\infty}dq_2\mathcal{R}[\hat{O}_1](q_1,\varphi_1)\mathcal{R}[\hat{O}_2](q_2,\varphi_2)p(q_1,q_2|\varphi_1,\varphi_2).
\end{split}
\end{equation}
As a remark, note that the quantum state of the two modes, $\hat{\rho}$, can generally be entangled. In the experiment, the users simply need two independently phase-randomized homodyne detectors and record the joint conditional probability distribution of quadratures $(q_1,q_2)$ given the LO phases $(\varphi_1,\varphi_2)$.

In practical homodyne measurements, due to the usage of analog-to-digital converters (ADC), the measured quadrature value $q$ will be discrete with a finite resolution $\Delta$. In the former CV QKD protocols like Ref.~\cite{Primaatmaja2022Discrete}, the parameter estimation is done by linear programming~\cite{lavie2021estimating} instead of optical homodyne tomography; as a result, the finite resolution $\Delta$ will not affect the estimation accuracy as long as $\Delta$ is small. 

To characterize how the finite resolution $\Delta$ will disturb the parameter estimation by optical homodyne tomography, we numerically estimate the bias caused by $\Delta$. We estimate the expectation values of the (non-Hermitian) observables $\hat{O}_{01}:=\ket{0}\bra{1}$ and $\hat{O}_{02}:=\ket{0}\bra{2}$ with a coherent state $\ket{\sqrt{\mu}}$ input ($\mu=0.5$). We choose $\hat{O}_{01}$ and $\hat{O}_{02}$ since they are directly related to the phase error operator in the Proposition 1 of the main text. 

When the homodyne detection results is quantized with the bin width $\Delta$, if we use the observable estimator $\mc{R}_{\eta}[\ket{n}\bra{n+d}](q,\phi)$ given by Eq.~\eqref{eq:Fockkernel}, the expectation value becomes
\begin{equation}
    \mathbb{E}\left(\mc{R}[\hat{O}](q,\phi)\right) = \int_0^\pi \frac{d\phi}{\pi} \sum_{t=-\infty}^{\infty} \Pr(q=t\Delta|\phi) \, \mc{R}[\hat{O}](q=t\Delta,\phi),
\end{equation}
where
\begin{equation}
    \Pr(q=t\Delta|\phi) = \int_{t\Delta}^{(t+1)\Delta} p(q|\phi),
\end{equation}
is the quantized version of the probabilistic distribution of the homodyne detection result $q$. 
The finite-bin effect is then characterized by the estimation bias $\left| \mathbb{E}\left(\mc{R}[\hat{O}](q,\phi)\right) - \hat{O} \right|$.

\begin{figure*}[htbp!]
    \centering
    \includegraphics[width = 0.55\textwidth]{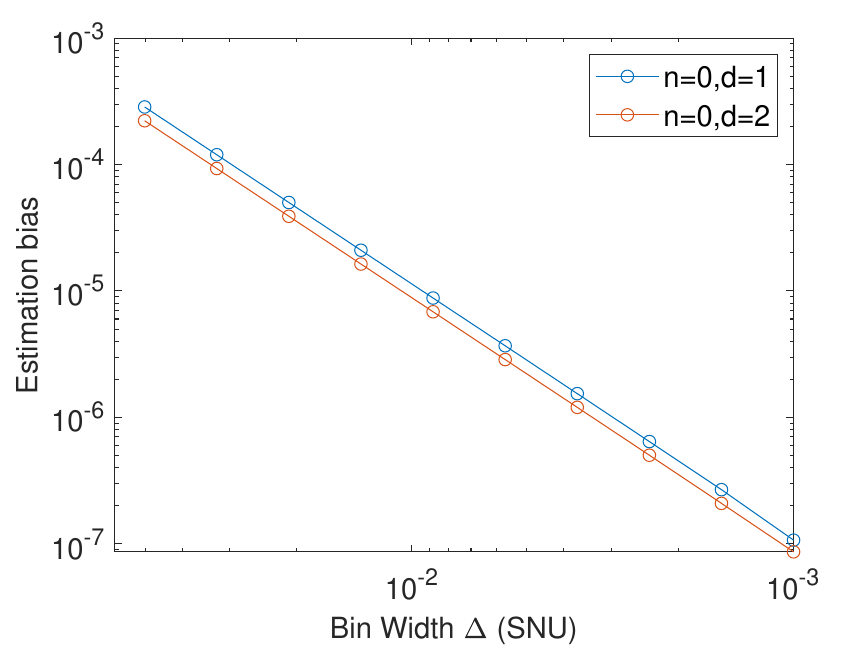}
    % \captionsetup{justification = raggedright}
    \caption{Estimation bias of the observable $\ket{n}\bra{n+d}$ with respect to the homodyne bin width $\Delta$. The input state is a coherent state $\ket{\sqrt{\mu}}$ with $\mu=0.5$. The detection efficiency $\eta_{det}=1$.}
    \label{Fig::HDdiscrete}
\end{figure*}

We plot the estimator bias of $\hat{O}_{01}$ and $\hat{O}_{02}$ with respect to the bin width $\Delta$ in Fig.~\ref{Fig::HDdiscrete}. We see that the bias is almost proportional to $\Delta^{2}$, i.e., it decays rapidly if we reduce the bin width $\Delta$. With a width $\Delta=0.01$ shot noise unit which is easily achievable in the experiment, the bias of $\hat{O}_{01}$ and $\hat{O}_{02}$ are smaller than $10^{-5}$. As a result, the estimation bias is negligible with practical homodyne measurement devices.

\subsection{Kato's inequality}
To tackle the most general attack, namely, a coherent attack, we apply a martingale-based approach, which naturally adapts to a sequential setting. For a martingale, the summation of its composed random variables converges to the expected value in probability, as shown by Azuma's inequality~\cite{azuma1967weighted} and its variants. In this work, we apply the variant, Kato's inequality~\cite{kato2020concentration}, for a better performance.

\begin{theorem}[Kato's Inequality~\cite{kato2020concentration}]
Let $\{X_i\}$ be a list of real-valued random variables satisfying $0\leq X_i\leq1$ for any $i$, and $\{\mathcal{F}_i\}$ be a filtration that identifies these random variables, i.e., $\mathcal{F}_i$ is a $\sigma$-algebra that satisfies $\mathcal{F}_i\subseteq\mathcal{F}_{i+1}$ and $E(X_{i'}|\mathcal{F}_i)=X_{i'}$ for $i'\leq i$. Then, for any $N\in\mathbb{N},\delta\in[0,\infty)$ and $\delta'\in\mathbb{R}$, the following inequalities hold:
  \begin{equation}%\label{eq:Kato}
  \begin{split}
    \Pr\left\{\sum_{i=1}^N \mathbb{E}(X_i|\mathcal{F}_{i-1}) \geq \left(1+\frac{2\delta'}{N}\right)\sum_{i=1}^NX_i+\delta-\delta'\right\} &\leq \exp\left\{-\frac{2(\delta^2-\delta'^2)}{N\left(1+\frac{4\delta'}{3N}\right)^2}\right\}, \\
    \Pr\left\{\sum_{i=1}^N \mathbb{E}(X_i|\mathcal{F}_{i-1}) \leq \left(1+\frac{2\delta'}{N}\right)\sum_{i=1}^NX_i-\delta-\delta'\right\} &\leq \exp\left\{-\frac{2(\delta^2-\delta'^2)}{N\left(1+\frac{4\delta'}{3N}\right)^2}\right\}.
  \end{split}
  \end{equation}
\label{Lemma:Kato}
\end{theorem}

In our discussions, we shall encounter variables $X_i$ that may take complex values. Nevertheless, the target quantities to be estimated are real values. Therefore, we apply Kato's inequality to variables $\mathrm{Re}(X_i)$, which still yields an unbiased estimation. In addition, we shall encounter cases where the maximum value of $\mathrm{Re}(X_i)$, denoted as $c$, exceeds $1$. For this purpose, we can normalize $\mathrm{Re}(X_i)$ to $Y_i=\mathrm{Re}(X_i)/c$, which takes the maximum value to be $1$, and apply Kato's inequality to variables $Y_i$. Then we have the following estimation result:
\begin{equation}\label{eq:Kato}
  \begin{split}
    \Pr\left\{\sum_{i=1}^N \mathbb{E}(X_i|\mathcal{F}_{i-1}) \geq\left(1+\frac{2\delta'}{cN}\right) \sum_{i=1}^N\mathrm{Re}(X_i) +\delta-\delta'\right\}&\leq \exp\left\{-\frac{2(\delta^2-\delta'^2)}{Nc^2\left(1+\frac{4\delta'}{3Nc}\right)^2}\right\}\equiv\varepsilon(\delta,\delta'), \\
    \Pr\left\{\sum_{i=1}^N \mathbb{E}(X_i|\mathcal{F}_{i-1}) \leq\left(1+\frac{2\delta'}{cN}\right) \sum_{i=1}^N\mathrm{Re}(X_i) -\delta-\delta'\right\}&\leq \exp\left\{-\frac{2(\delta^2-\delta'^2)}{Nc^2\left(1+\frac{4\delta'}{3Nc}\right)^2}\right\}\equiv\varepsilon(\delta,\delta').
  \end{split}
  \end{equation}

\section{Finite-size analysis against coherent attacks}
% \label{App::martingale}
\label{app:paraest}
In this section, we show how to estimate the quantities in the key-rate formula with realistic devices. 
% We first state parameter estimation in terms of probabilities in a single round, and one can interpret the results as obtained from sufficiently many rounds under a collective attack. In the next section, we will generalize the results to the finite-size regime under a coherent attack. 
We first review the terms to be estimated in the virtual protocol that utilizes up to the two-photon component. 
Throughout the finite-size key rate analysis, we consider working under the equivalent protocol where Alice and Bob would perform a global photon number measurement upon emitting and receiving the signals, respectively.
For convenience here, we state the parameters as (conditional) probabilities of random variables, as given in Sec.~3.C of the main text. Through subsequent discussions in this section, we shall estimate their rates or observed frequencies in the virtual protocol. As a reminder, note that we distinguish the terms ``receiving'' and ``accepting.'' That is, when accepting a signal, the pulse passes the filtering and is decoded as a conclusive key bit.
\begin{enumerate}
\item $Q_{*,0}$: The probability of accepting a vacuum state, given by Eq.~(27) in the main text.
\item $Q_{1,1}$: The probability of sending $\ket{\Psi_1}_{A'A_1A_2}$ and accepting a single-photon state, given by Eq.~(23) in the main text.
%together with the integration factor $c_1$ gives  (Eq.~\eqref{Eq::Q11}).
\item $Q_{2,2}$: The probability of sending $\ket{\Psi_2}_{A'A_1A_2}$ and accepting a two-photon state, given by Eq.~(26) in the main text.
%together with the integration factors $c_2^{02}$ and $c_2^{11}$ gives  ().
\item $e_{1,1}^X$: the phase-error probability when sending a single-photon state and accepting a single-photon state, determined by the probabilities of sending $(\ket{01}\pm\ket{10})/\sqrt{2}$ and accepting $(\ket{01}\mp\ket{10})/\sqrt{2}$, given by Eq.~(21) in the main text.
\item $e_{2,2}^X$: the phase-error probability when sending a two-photon state and accepting a two-photon state, determined by the probabilities of sending $(\ket{02}\pm\ket{20})/\sqrt{2}$ and accepting $(\ket{02}\mp\ket{20})/\sqrt{2}$ and sending $(\ket{02}-\ket{20})/\sqrt{2}$ and accepting $\ket{11}$, given by Eq.~(24) in the main text.
\end{enumerate}

\subsection{Parameter estimation procedure}\label{app:EstProcedure}
\begin{figure*}[htbp]
    \centering
    \includegraphics[width = 0.95\textwidth]{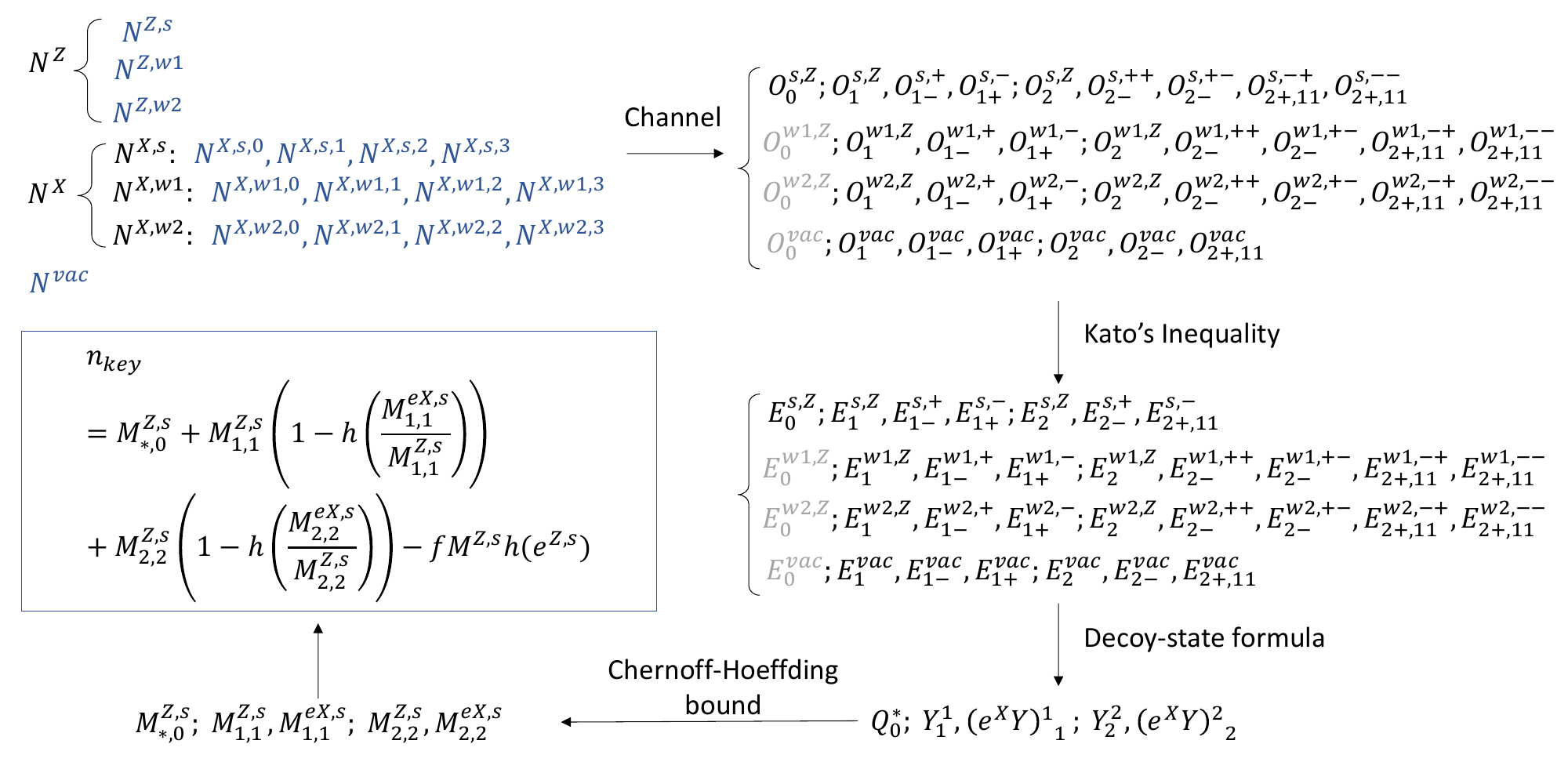}
    \caption{Parameters used in the two-photon time-bin CV QKD protocol. We omit the lower and upper bound notation in all the estimated parameters. The gray observables are not required.}
    \label{Fig::finitePara}
\end{figure*}

First, we summarize the parameter estimation procedure to evaluate the finite-size key rate in Fig.~\ref{Fig::finitePara}. 

\begin{enumerate}
    \item Data categorization on Alice's side: We first categorize the rounds with respect to the settings on Alice's side. For all the $N$ rounds, we first divide them into three settings: $Z$-basis data $N^{Z}$, $X$-basis data $N^{X}$ and vacuum data $N^{vac}$. For the $Z$- and $X$-basis data, we divide the data with respect to the light intensity, where the light intensity $\mu$ is denoted as the signal pulse $s$, and the light intensities $\nu_1$ and $\nu_2$ are denoted as weak pulses $w1$ and $w2$, respectively. The number of sending rounds in $Z$-basis with intensity $\mathrm{int}\in\{s,w1,w2\}$ is denoted as $N^{Z,\mathrm{int}}$. For the $X$-basis data, we further determine the relative phase between two pulses $\{\frac{\pi}{2} j\}_{j=0}^3$. The number of sending rounds in $X$-basis with intensity $\mathrm{int}\in\{s,w1,w2\}$ and relative phase $\frac{\pi}{2}j$ is denoted as $N^{X,\mathrm{int},j}$.
    
    \item Data categorization on Bob's side: From now on, we use $Z \to X$ to denote the case where Alice emits $Z$-basis signals and Bob measures on the $X$-basis, and similarly for other settings. Alice and Bob use $Z\rightarrow Z$ data for key generation, among which they calculate the number of successful detection rounds $M^{Z,s}$ where a key bit is generated and the quantum bit error rate $e^Z$. Alice and Bob use $Z\rightarrow X$ and $X\rightarrow X$ data for parameter estimation.
    
    \item Homodyne tomography estimation: Based on the homodyne measurement results, Bob performs homodyne tomography and calculates the average ``observed statistics'' for the quantities of interest, which we denote as $O$-values. In the next step, we introduce the quantities to be calculated. Their calculation shall be given in Appendix~\ref{app:martingale}.
    
    \item Kato inequality: By applying Kato's inequality to the ``observed statistics,'' Alice and Bob obtain the average underlying values, which we denote as $E$-values. The $O$-values and $E$-values with the same subscripts and superscripts correspond with each other. Here, we describe the terms in a single round. In Appendix~\ref{app:martingale}, we shall explain the exact meaning of their averages over the rounds.
    \begin{itemize}
        \item Proportion of $0,1$ and $2$-photon component. In the rounds that emit a signal or a weak pulse,
        \begin{equation}
        \begin{aligned}
            E^{\mathrm{int},Z}_0 &:= c_0 \tr[ \mc{N}_E(\hat{\rho}_{\mathrm{int}}^{Z}) \ket{00}\bra{00} ], \\ 
            E^{\mathrm{int},Z}_1 &:= c_1 \tr[ \mc{N}_E(\hat{\rho}_{\mathrm{int}}^{Z}) (\ket{01}\bra{01}+\ket{10}\bra{10}) ], \\ 
            E^{\mathrm{int},Z}_2 &:= \frac{c_2^+}{2} \tr[ \mc{N}_E(\hat{\rho}_{\mathrm{int}}^{Z})(\ket{02}+\ket{20})(\bra{02}+\bra{20}) ] \\
            &\quad + \frac{c_2^-}{2} \tr[ \mc{N}_E(\hat{\rho}_{\mathrm{int}}^{Z}) (\ket{02}-\ket{20})(\bra{02}-\bra{20}) ] + 2c_2^{11} \tr[ \mc{N}_E(\hat{\rho}_{\mathrm{int}}^{Z}) \ket{11}\bra{11} ], \\ 
        \end{aligned}
        \end{equation}
        where $\hat{\rho}_{\mathrm{int}}^{Z}$ represents the uniformly mixed $Z$-basis emitted state over all the relative phases with light intensity $\mathrm{int}$, and $\mathcal{N}_E$ represents the quantum channel controlled by Eve in this round. The coefficients of $c_0,c_1,c_2^+,c_2^-$ are listed in Eq.~(22),~(25), and~(28) of the main text.
        
        In the rounds that emit a vacuum pulse,
        \begin{equation}
        \begin{aligned}
            E^{vac}_0 &:= c_0 \tr[ \mc{N}_E(\ket{0}\bra{0}) \ket{00}\bra{00} ], \\ 
            E^{vac}_1 &:= c_1 \tr[ \mc{N}_E(\ket{0}\bra{0}) (\ket{01}\bra{01}+\ket{10}\bra{10}) ], \\ 
            E^{vac}_2 &:= \frac{c_2^+}{2} \tr[ \mc{N}_E(\ket{0}\bra{0}) (\ket{02}+\ket{20})(\bra{02}+\bra{20}) ] \\
            &\quad + \frac{c_2^-}{2} \tr[ \mc{N}_E(\ket{0}\bra{0})(\ket{02}-\ket{20})(\bra{02}-\bra{20}) ] + 2c_2^{11} \tr[ \mc{N}_E(\ket{0}\bra{0}) \ket{11}\bra{11} ]. \\ 
        \end{aligned}
        \end{equation}
        
        \item Phase error rate when Alice emits $+$ signals in the virtual protocol. In the rounds that emit a single or a weak pulse,
        \begin{equation}\label{eq:phaseerror}
        \begin{aligned}
            E^{\mathrm{int},+}_{1-} &:= \frac{c_1}{4} \tr[ \mc{N}_E(\hat{\rho}_{\mathrm{int}}^{0}) (\ket{01}-\ket{10})(\bra{01} - \bra{10})], \\
            E^{\mathrm{int},++}_{2-} &:= \frac{c_2^-}{4} \tr[ \mc{N}_E(\hat{\rho}_{\mathrm{int}}^{++}) (\ket{02}-\ket{20})(\bra{02} -\bra{20})], \\
            E^{\mathrm{int},+-}_{2-} &:= \frac{c_2^-}{4} \tr[ \mc{N}_E(\hat{\rho}_{\mathrm{int}}^{+-}) (\ket{02}-\ket{20})(\bra{02} -\bra{20})], \\
        \end{aligned}
        \end{equation}
        where the effective source states are given by
        \begin{equation}\label{eq:+state}
        \begin{aligned}
            \hat{\rho}_{\mathrm{int}}^{++} &:= \frac{1}{2} \hat{\rho}_{\mathrm{int}}^{Z} + \frac{1}{4}(\hat{\rho}_{\mathrm{int}}^{0}+\hat{\rho}_{\mathrm{int}}^{\pi}), \\
            \hat{\rho}_{\mathrm{int}}^{+-} &:= \frac{1}{2}(\hat{\rho}_{\mathrm{int}}^{\pi/2} + \hat{\rho}_{\mathrm{int}}^{3\pi/2}). \\
            % \rho^{s,+}  &:= 2\rho^{s,++} - \rho^{s,+-}.
        \end{aligned}
        \end{equation}
        In the subscripts of this set of $E$-terms, the number denotes the photon number, i.e., a single-photon or a two-photon component, and the minus sign represents that Bob's measurement result in the virtual protocol is $-$. We split the two-photon phase-error rate estimation into two terms, $E^{\mathrm{int},++}_{2-}$ and $E^{\mathrm{int},+-}_{2-}$. This processing guarantees that $\hat{\rho}_{\mathrm{int}}^{++}$ and $\hat{\rho}_{\mathrm{int}}^{+-}$ are effective quantum density operators, which is necessary for the subsequent decoy-state analysis. 
        % The Table~\ref{table:protocolPractial}.

        In the rounds that emit a vacuum pulse,
        \begin{equation}
        \begin{aligned}
            E^{vac}_{1-} &:= \frac{c_1}{4} \tr[ \mc{N}_E(\ket{0}\bra{0}) (\ket{01}-\ket{10})(\bra{01} - \bra{10})], \\
            E^{vac}_{2-} &:= \frac{c_2^-}{4} \tr[ \mc{N}_E(\ket{0}\bra{0}) (\ket{02}-\ket{20})(\bra{02} -\bra{20})]. \\
        \end{aligned}
        \end{equation}
        
        \item Phase error rate when Alice emit $-$ signals in the virtual protocol. In the rounds that emit a single or a weak pulse,
        \begin{equation}
        \begin{aligned}
            E^{\mathrm{int},-}_{1+} &:= \frac{c_1}{4} \tr[ \mc{N}_E(\hat{\rho}_{\mathrm{int}}^{2}) (\ket{01} + \ket{10})(\bra{01} + \bra{10})], \\
            E^{\mathrm{int},-+}_{2+,11} &:= \frac{c_2^+}{4} \tr[ \mc{N}_E(\hat{\rho}_{\mathrm{int}}^{-+})(\ket{02}+\ket{20})(\bra{02}+\bra{20})] + c_2^{11} \tr[ \mc{N}_E(\hat{\rho}_{\mathrm{int}}^{-+}) \ket{11}\bra{11}], \\
            E^{\mathrm{int},--}_{2+,11} &:= \frac{c_2^+}{4} \tr[ \mc{N}_E(\hat{\rho}_{\mathrm{int}}^{--}) (\ket{02}+\ket{20})(\bra{02}+\bra{20})] + c_2^{11} \tr[ \mc{N}_E(\hat{\rho}_{\mathrm{int}}^{--}) \ket{11}\bra{11}], \\
        \end{aligned}
        \end{equation}
        where the effective source states are given by
        \begin{equation}\label{eq:-state}
        \begin{aligned}
            \hat{\rho}_{\mathrm{int}}^{-+} &:= \frac{1}{2} \hat{\rho}_{\mathrm{int}}^{Z} + \frac{1}{4}(\hat{\rho}_{\mathrm{int}}^{\pi/2}+\hat{\rho}_{\mathrm{int}}^{3\pi/2}), \\
            \hat{\rho}_{\mathrm{int}}^{--} &:= \frac{1}{2}(\hat{\rho}_{\mathrm{int}}^{0} + \hat{\rho}_{\mathrm{int}}^{\pi}). \\
            % \rho^{s,-}  &:= 2\rho^{s,-+} - \rho^{s,--}.
        \end{aligned}
        \end{equation}
    \end{itemize}
    The subscripts follow a similar convention as above.

    In the rounds that emit a vacuum pulse,
        \begin{equation}
        \begin{aligned}
            E^{vac}_{1+} &:= \frac{c_1}{4} \tr[ \mc{N}_E(\ket{0}\bra{0}) (\ket{01} + \ket{10})(\bra{01} + \bra{10})], \\
            E^{vac}_{2+,11} &:= \frac{c_2^+}{4} \tr[ \mc{N}_E(\ket{0}\bra{0})(\ket{02}+\ket{20})(\bra{02}+\bra{20})] + c_2^{11} \tr[ \mc{N}_E(\ket{0}\bra{0}) \ket{11}\bra{11}].
        \end{aligned}
        \end{equation}
    
    \item Decoy-state formula: On the source side, to evaluate the single-photon and two-photon subspaces in the gains and phase error rates from the estimated statistics, we apply the extended decoy-state formulae presented in Appendix~\ref{app:decoy}.

    \item Chernoff-Hoeffding bound: By the last step, we have obtained the probabilities for the target values. As gains and phase-error rates can be interpreted as frequencies of Bernoulli random variables, we can apply the Chernoff-Hoeffding bound for Bernoulli random variables in Ref.~\cite{zhang2017improved} to estimate their values in the virtual experiment, $M^Z_{*,0}, M^Z_{1,1}, M^Z_{2,2}, M^{eX}_{1,1}$ and $M^{eX}_{2,2}$. We present the derivations in this part in Appendix~\ref{app:Chernoff}.
\end{enumerate}

\subsection{Summary of data categorization}
In Table~\ref{tab:DataCat}, we summarize the data categorization and processing in homodyne tomography relevant to each term's estimation. Terms labelled with the same variable value of $\mathrm{int}$ share the same light intensity. On Bob's side, he maps his homodyne measurement results, $(\varphi_b^1,\varphi_b^2,q_1,q_2)$, to proper kernel functions $\mathcal{R}_{\eta}[\ket{n_1}\bra{n_1+d_1}]\mathcal{R}_{\eta}[\ket{n_2}\bra{n_2+d_2}]$ shown in Eq.~\eqref{eq:Fockkernel} for the estimation. Note that the same data set can be re-used to form various effective source states or calculate various kernel functions and hence estimate multiple terms. It is worth mentioning that Bob's $Z$-basis data can also be used for homodyne tomography, where the relative phase between the LOs of the two modes is zero. For instance, we can use $Z\rightarrow Z$ data to estimate $E_0^{\mathrm{int},Z}$. Nevertheless, for simplicity in our statements and numerical simulation, we simply use Bob's $X$-basis data in our discussions, where the LOs of the two modes are independently randomized.

\begin{table*}[hbtp!]
\centering
\captionsetup{justification = raggedright}
\caption{Data categorization. The first column lists the terms to be estimated. The second column lists the source states for the estimation, which come from Alice's relevant data listed in the third column. The measurement operators to be estimated from homodyne tomography are listed in the fourth column.}
\begin{tabular}{cccc}
\hline
\hline
 Estimation term & \quad Source states & \quad Alice's relevant data & \quad Operators in homodyne tomography \\
\hline
$E^{\mathrm{int},Z}_0$ & $\hat{\rho}_{\mathrm{int}}^{Z}$ & $Z$-basis & $\ket{00}\bra{00}$ \\
$E^{\mathrm{int},Z}_1$ & $\hat{\rho}_{\mathrm{int}}^{Z}$ & $Z$-basis & $\ket{01}\bra{01},\ket{10}\bra{10}$ \\
$E^{\mathrm{int},Z}_2$ & $\hat{\rho}_{\mathrm{int}}^{Z}$ & $Z$-basis & $\ket{02}\bra{02},\ket{02}\bra{20},\ket{20}\bra{02},\ket{20}\bra{20},\ket{11}\bra{11}$ \\
$E^{vac}_0$ & $\ket{0}$ & vacuum & $\ket{00}\bra{00}$ \\
$E^{vac}_1$ & $\ket{0}$ & vacuum & $\ket{01}\bra{01},\ket{10}\bra{10}$ \\
$E^{vac}_2$ & $\ket{0}$ & vacuum & $\ket{02}\bra{02},\ket{02}\bra{20},\ket{20}\bra{02},\ket{20}\bra{20},\ket{11}\bra{11}$ \\
$E^{\mathrm{int},+}_{1-}$ & $\hat{\rho}_{\mathrm{int}}^{0}$ & $X$-basis & $\ket{01}\bra{01},\ket{01}\bra{10},\ket{10}\bra{01},\ket{10}\bra{10}$ \\
$E^{\mathrm{int},++}_{2-}$ & $\hat{\rho}_{\mathrm{int}}^{Z},\hat{\rho}_{\mathrm{int}}^{0},\hat{\rho}_{\mathrm{int}}^{\pi}$ & $X$- and $Z$-basis & $\ket{02}\bra{02},\ket{02}\bra{20},\ket{20}\bra{02},\ket{20}\bra{20}$ \\
$E^{\mathrm{int},+-}_{2-}$ & $\hat{\rho}_{\mathrm{int}}^{\pi/2},\hat{\rho}_{\mathrm{int}}^{3\pi/2}$ & $X$-basis & $\ket{02}\bra{02},\ket{02}\bra{20},\ket{20}\bra{02},\ket{20}\bra{20}$ \\
$E^{vac}_{1-}$ & $\ket{0}$ & vacuum & $\ket{01}\bra{01},\ket{01}\bra{10},\ket{10}\bra{01},\ket{10}\bra{10}$ \\
$E^{vac}_{2-}$ & $\ket{0}$ & vacuum & $\ket{02}\bra{02},\ket{02}\bra{20},\ket{20}\bra{02},\ket{20}\bra{20}$ \\
$E^{\mathrm{int},-}_{1+}$ & $\hat{\rho}_{\mathrm{int}}^{2}$ & $X$-basis & $\ket{01}\bra{01},\ket{01}\bra{10},\ket{10}\bra{01},\ket{10}\bra{10}$ \\
$E^{\mathrm{int},-+}_{2+,11}$ & $\hat{\rho}_{\mathrm{int}}^{Z},\hat{\rho}_{\mathrm{int}}^{\pi/2},\hat{\rho}_{\mathrm{int}}^{3\pi/2}$ & $X$- and $Z$-basis & $\ket{02}\bra{02},\ket{02}\bra{20},\ket{20}\bra{02},\ket{20}\bra{20},\ket{11}\bra{11}$ \\
$E^{\mathrm{int},--}_{2+,11}$ & $\hat{\rho}_{\mathrm{int}}^{0}, \hat{\rho}_{\mathrm{int}}^{\pi}$ & $X$-basis & $\ket{02}\bra{02},\ket{02}\bra{20},\ket{20}\bra{02},\ket{20}\bra{20},\ket{11}\bra{11}$ \\
$E^{vac}_{1+}$ & $\ket{0}$ & vacuum & $\ket{01}\bra{01},\ket{01}\bra{10},\ket{10}\bra{01},\ket{10}\bra{10}$ \\
$E^{vac}_{2+,11}$ & $\ket{0}$ & vacuum & $\ket{02}\bra{02},\ket{02}\bra{20},\ket{20}\bra{02},\ket{20}\bra{20},\ket{11}\bra{11}$ \\
\hline
\hline
\end{tabular}
\label{tab:DataCat}
\end{table*}

\subsection{Homodyne tomography and martingale-based parameter estimation}\label{app:martingale}
% {\color{red}ZXJ: Modify figure 8 and put it here (explain the meaning of estimating an E-value and its corresponding value in a single round)}

In the security analysis of our protocol, we hope to estimate the gains $Q_{m,n}$, defined as rates of sending an $m$-photon state and accepting an $n$-photon state, and the phase-error rates $e^X_{m,n}$, defined as frequencies of a phase error occurrence in a round of sending an $m$-photon state and accepting an $n$-photon state. 
While these variable values can be measured in the virtual protocol, we need to devise means for their evaluation from observed statistics in the realistic protocol. 
In this section and the next section, we show how to reach the above target, starting from the results on the expected values of relevant quantities through Eq.~(21) to Eq.~(27) of the main text. In this section, we show how to evaluate the photon-number measurement results from homodyne measurements, where the latter are directly observable in the experiment. We also exhibit a martingale-based analysis and show how to link frequencies and expected values. 

First, we describe the estimation target. Consider the $i$'th round in the experiment, where all the events occurred in the previous $i-1$ rounds form the underlying filtration $\mathcal{F}_{i-1}$. 
In the realistic experiment, Alice and Bob can only choose a specific experiment setting and perform the relevant operations, and their observations contribute to the filtration $\mathcal{F}_{i}$. Nevertheless, conditioned on $\mathcal{F}_{i-1}$, we can ask for the expected value of other measurable variables in this round. Moreover, we can consider a virtual experiment, where the users instead measured these measurable variables, and estimate the virtual measurement results.
For the key-rate analysis, we are interested in the following question: conditioned on the filtration $\{\mathcal{F}_i\}_{i=1}^N$ that originates from the real experiment, should Alice prepare the proper quantum states and Bob measure the operators that define the gains and phase errors in the key-generation rounds instead, what would be their measurement results?

To link the observed statistics in the real experiment with the quantities to be estimated, we apply the following general approach. Consider an observable $\hat{O}$ to be estimated. For its estimation, Alice and Bob measure a relevant observable $\hat{K}$ in a specific experiment setting denoted as $\mathcal{I}$, which is defined by a specific light intensity and basis choices. The observable $\hat{K}$ is linked with $\hat{O}$:
\begin{equation}
    \langle\hat{O}\rangle=\mathbb{E}_{\Pr(K=k)}[f(k)],
\end{equation}
where $K$ denotes the measurement result of $\hat{K}$, and $f(\cdot)$ is some kernel function. Then, in the $i$'th round, consider a counter variable $\zeta_{\{\mathcal{I},\hat{O}\}}^{(i)}$ such that
\begin{eqnarray}
  \zeta_{\{\mathcal{I},\hat{O}\}}^{(i)} =
  \left\{
  \begin{tabular}{ll} $f(k)$, & if setting $=\mathcal{I}$ and $K=k$, \\
    $0$, & otherwise. 
  \end{tabular}
\right.
\end{eqnarray}
By definition,
\begin{equation}
    \mathbb{E}\left[\zeta_{\{\mathcal{I},\hat{O}\}}^{(i)}|\mathcal{F}_{i-1}\right]=\Pr(\mathcal{I})\langle\hat{O}\rangle.
\end{equation}
Since the randomness in the measurement setting is embedded in the definition of $\zeta_{\{\mathcal{I},\hat{O}\}}^{(i)}$, we can hence link the observed statistics in the rounds where $\hat{K}$ is measured with the rest rounds. Specifically, suppose $f(k)$ is a non-negative value upper-bounded by $c$. By applying Kato's inequality, we have the following result: for any $\delta\in[0,\infty)$ and $\delta'\in\mathbb{R}$, except for a failure probability of
\begin{equation}\label{eq:martingalefail}
    \varepsilon=2\exp\left\{-\frac{2(\delta^2-\delta'^2)}{Nc^2\left(1+\frac{4\delta'}{3Nc}\right)^2}\right\},
\end{equation}
the following inequalities hold simultaneously,
\begin{equation}
\begin{split}
    \sum_{i=1}^{N}\mathbb{E}\left[\zeta_{\{\mathcal{I},\hat{O}\}}^{(i)}|\mathcal{F}_{i-1}\right] &\leq \left(1+\frac{2\delta'}{cN}\right) \left[\sum_{i=1}^N\zeta_{\{\mathcal{I},\hat{O}\}}^{(i)} +\delta-\delta'\right], \\
    \sum_{i=1}^{N}\mathbb{E}\left[\zeta_{\{\mathcal{I},\hat{O}\}}^{(i)}|\mathcal{F}_{i-1}\right] &\geq \left(1+\frac{2\delta'}{cN}\right) \left[\sum_{i=1}^N\zeta_{\{\mathcal{I},\hat{O}\}}^{(i)} -\delta-\delta'\right].
\end{split}
\end{equation}
Note that we apply the union bound in deriving the failure probability. In the above inequalities, the left-hand sides represent the quantities to be estimated, whose averaged values per round are equal to the estimation terms in Table~\ref{tab:DataCat} up to the setting choice probability, and the right-hand sides correspond to the quantities observed in the experiment.

After $\sum_{i=1}^{N}\mathbb{E}\left[\zeta_{\{\mathcal{I},\hat{O}\}}^{(i)}|\mathcal{F}_{i-1}\right]$ is estimated, we can apply Kato's inequality once more to obtain the final estimation, i.e., the estimation on the measurement results of $\hat{O}$ in a virtual experiment. Nevertheless, we shall apply the Chernoff-Hoeffding bound for a tighter analysis. We leave this part in Appendix~\ref{app:Chernoff}.

To summarize, in Fig.~\ref{Fig::martingale}, we list the counter variables and show the link between the observed statistics and the terms to be estimated. The blue boxes correspond to the experimental settings and observations in reality, and the orange boxes correspond to the settings and quantities in the virtual experiment. In the following subsections, we take the estimation of $E^{\mathrm{int},Z}_1$ and $E^{\mathrm{int},++}_{2-}$ as examples and present the thorough derivations. The estimation of other terms follows a similar approach.

\begin{figure*}[htbp!]
    \centering
    \includegraphics[width = \textwidth]{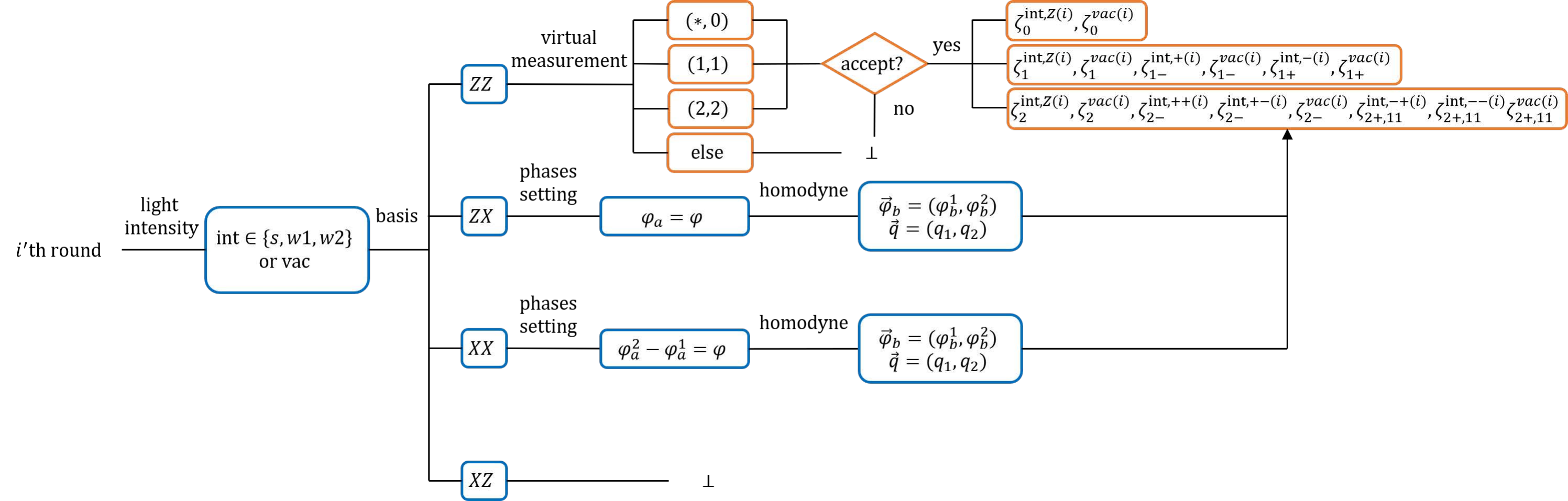}
    \captionsetup{justification = raggedright}
    \caption{The counter variables in the martingale-based parameter estimation. As a remark, we do not utilize the $X\rightarrow Z$ data in our protocol. These statistics might be useful with a clever protocol design.}
    \label{Fig::martingale}
\end{figure*}

%\subsubsection{Example 1: Estimation of $E^{\mathrm{int},Z}_1$}
\subsubsection{Example 1: Estimation of single-photon gains}\label{app:Q11}
As a relatively simple demonstration, in this section, we present a thorough analysis on the estimation of $E^{\mathrm{int},Z}_1$. We explain its definition in the general sequential setting against coherent attacks, where the statistics might be non-i.i.d. over the rounds, and show how to apply the martingale analysis to estimate its value. For each round, we construct a ``counter variable'' $\zeta_1^{\mathrm{int},Z(i)}$. In the round where the light intensity is $\mathrm{int}$, Alice chooses the $Z$-basis, Bob chooses the $X$-basis, and Bob's homodyne measurement result is given by $(q_1^{(i)},q_2^{(i)})=\vec{q},(\varphi_b^{1(i)},\varphi_b^{2(i)})=\vec{\varphi}$, the counter variable takes the following value:
\begin{equation}
    \zeta_{1}^{\mathrm{int},Z(i)}=c_1\left\{\mathcal{R}_{\eta}[\ket{01}\bra{01}](\vec{q},\vec{\varphi}_b) +\mathcal{R}_{\eta}[\ket{10}\bra{10}](\vec{q},\vec{\varphi}_b)\right\}.
\end{equation}
Otherwise, we set $\zeta_{1}^{\mathrm{int},Z(i)}=0$. Following the QKD protocol, we can measure the counter variable $\zeta_{1}^{\mathrm{int},Z(i)}$ in each round. The counter variable $\zeta_1^{\mathrm{int},Z(i)}$ is $\mathcal{F}_{i-1}$-measurable, conditioned on which the expected value is given by
\begin{equation}
    \mathbb{E}\left[\zeta_{1}^{\mathrm{int},Z(i)}|\mathcal{F}_{i-1}\right] = p_{\mathrm{int}}p_{zx} c_1 \tr[\mc{N}_E(\hat{\rho}_{\mathrm{int}}^{Z}) (\ket{01}\bra{01}+\ket{10}\bra{10})],
\end{equation}
where $p_{\mathrm{int}}$ represents the probability that Alice chooses the light intensity $\mathrm{int}$, and $p_{zx}$ represents the probability that Alice chooses the $Z$-basis and Bob chooses the $X$-basis. We follow the same notation convention in the following discussions. We denote 
\begin{equation}
    c_1^{Z}=\max\mathrm{Re}(\zeta_{1}^{\mathrm{int},Z(i)})=c_0\max_{\vec{q},\vec{\varphi}}{\mathrm{Re}\{\mathcal{R}_{\eta}[\ket{01}\bra{01}](\vec{q},\vec{\varphi})+\mathcal{R}_{\eta}[\ket{10}\bra{10}](\vec{q},\vec{\varphi})\}},
\end{equation}
which is finitely bounded when $\eta>1/2$. The value of $E^{\mathrm{int},Z}_1$ is defined as
\begin{equation}
    E^{\mathrm{int},Z}_1= \frac{1}{p_{\mathrm{int}}p_{zx}N}\sum_{i=1}^{N}\mathbb{E}\left[\zeta_{1}^{\mathrm{int},Z(i)}|\mathcal{F}_{i-1}\right].
\end{equation}
By applying Kato's inequality in Eq.~\eqref{eq:Kato}, we obtain its estimation: for any $\delta\in[0,\infty)$ and $\delta'\in\mathbb{R}$, except for a failure probability of
\begin{equation}
    \varepsilon=2\exp\left\{-\frac{2(\delta^2-\delta'^2)}{Nc^2\left(1+\frac{4\delta'}{3Nc}\right)^2}\right\},
\end{equation}
the value of $E^{\mathrm{int},Z}_1$ is bounded on both sides,
\begin{equation}
    \frac{1}{p_{\mathrm{int}}p_{zx}N}\left(1+\frac{2\delta'}{c_1^ZN}\right) \left[\sum_{i=1}^N\mathrm{Re}\left(\zeta_{1}^{\mathrm{int},Z(i)}\right) -\delta-\delta'\right] \leq E^{\mathrm{int},Z}_1\leq \frac{1}{p_{\mathrm{int}}p_{zx}N}\left(1+\frac{2\delta'}{c_1^ZN}\right) \left[\sum_{i=1}^N\mathrm{Re}\left(\zeta_{1}^{\mathrm{int},Z(i)}\right) +\delta-\delta'\right].
\end{equation}

\subsubsection{Example 2: Estimation of two-photon phase-error rate}\label{app:ep^{++}_{2-}}
As a more involved example, in this section, we present a thorough analysis on the estimation of $E^{\mathrm{int},++}_{2-}$, which takes more than one basis for the estimation. This example also demonstrates the parameter estimation for two-photon components, which are absent in the usual BB84 protocol. Similarly, for each round, we construct a counter variable $\zeta_{2-}^{\mathrm{int},++(i)}$. In the round where the light intensity is $\mathrm{int}$, Alice chooses the $Z$-basis or the $X$-basis with a relative phase of $\pi/2$ or $3\pi/2$, Bob chooses the $X$-basis, and Bob's homodyne measurement result is given by $(q_1^{(i)},q_2^{(i)})=\vec{q},(\varphi_b^{1(i)},\varphi_b^{2(i)})=\vec{\varphi}$, the counter variable takes the following value:
\begin{equation}
    \zeta_{2-}^{\mathrm{int},++(i)}=\frac{c_2^-}{4} 
\left\{\mathcal{R}_{\eta}[\ket{02}\bra{02}](\vec{q},\vec{\varphi}_b) -\mathcal{R}_{\eta}[\ket{02}\bra{20}](\vec{q},\vec{\varphi}_b) -\mathcal{R}_{\eta}[\ket{20}\bra{02}](\vec{q},\vec{\varphi}_b) +\mathcal{R}_{\eta}[\ket{20}\bra{20}](\vec{q},\vec{\varphi}_b)\right\}.
\end{equation}
Otherwise, we set $\zeta_{2-}^{\mathrm{int},++(i)}=0$. Following the QKD protocol, we can measure the counter variable $\zeta_{2-}^{\mathrm{int},++(i)}$ in each round. The counter variable $\zeta_{2-}^{\mathrm{int},++(i)}$ is $\mathcal{F}_{i-1}$-measurable, conditioned on which the expected value is given by

\begin{equation}
    \mathbb{E}\left[\zeta_{2-}^{\mathrm{int},++(i)}|\mathcal{F}_{i-1}\right] = p_{\mathrm{int}}\left(p_{zx}+\frac{1}{2}p_{xx}\right)\frac{c_2^-}{4} \tr[ \mc{N}_E(\hat{\rho}_{s}^{++}) (\ket{02}-\ket{20})(\bra{02} -\bra{20})].
\end{equation}
We denote 
\begin{equation}
\begin{split}
    c_{2-}^{\mathrm{int},++}&=\max\mathrm{Re}(\zeta_{2-}^{\mathrm{int},++(i)}) \\
    &=\frac{c_2^-}{4}\max_{\vec{q},\vec{\varphi}}\mathrm{Re}{\left\{\mathcal{R}_{\eta}[\ket{02}\bra{02}](\vec{q},\vec{\varphi}_b) -\mathcal{R}_{\eta}[\ket{02}\bra{20}](\vec{q},\vec{\varphi}_b) -\mathcal{R}_{\eta}[\ket{20}\bra{02}](\vec{q},\vec{\varphi}_b) +\mathcal{R}_{\eta}[\ket{20}\bra{20}](\vec{q},\vec{\varphi}_b)\right\}},
\end{split}
\end{equation}
which is finitely bounded when $\eta>1/2$. The value of $E^{\mathrm{int},Z}_1$ is defined as
\begin{equation}
    E^{\mathrm{int},++}_{2-}= \frac{1}{p_{\mathrm{int}}\left(p_{zx}+\frac{1}{2}p_{xx}\right)N}\sum_{i=1}^{N}\mathbb{E}\left[\zeta_{2-}^{\mathrm{int},++(i)}|\mathcal{F}_{i-1}\right].
\end{equation}
By applying Kato's inequality in Eq.~\eqref{eq:Kato}, we obtain its estimation: for any $\delta\in[0,\infty)$ and $\delta'\in\mathbb{R}$, except for a failure probability of
\begin{equation}
    \varepsilon=2\exp\left\{-\frac{2(\delta^2-\delta'^2)}{Nc^2\left(1+\frac{4\delta'}{3Nc}\right)^2}\right\},
\end{equation}
the value of $E^{\mathrm{int},++}_{2-}$ is bounded on both sides,
\begin{equation}
\begin{split}
    &\frac{1}{p_{\mathrm{int}}\left(p_{zx}+\frac{1}{2}p_{xx}\right)N}\left(1+\frac{2\delta'}{c_{2-}^{\mathrm{int},++}N}\right) \left[\sum_{i=1}^N\mathrm{Re}\left(\zeta_{2-}^{\mathrm{int},++(i)}\right) -\delta-\delta'\right] \\ 
    \leq &E^{\mathrm{int},++}_{2-} \\
    \leq &\frac{1}{p_{\mathrm{int}}\left(p_{zx}+\frac{1}{2}p_{xx}\right)N}\left(1+\frac{2\delta'}{c_{2-}^{\mathrm{int},++}N}\right) \left[\sum_{i=1}^N\mathrm{Re}\left(\zeta_{2-}^{\mathrm{int},++(i)}\right) +\delta-\delta'\right].
\end{split}
\end{equation}

\subsection{Extended decoy state method}\label{app:decoy}
In the previous section, we have established martingales to estimate the gains and phase-error rates from homodyne measurement results. A remaining issue is the source state, where we need to develop approaches to evaluating the single- and two-photon subspaces in the effective source states of signal pulses. For this purpose, we generalize the decoy-state method in Ref.~\cite{zhang2017improved}. Results in this section are given in terms of (conditional) expected values. As discussed in the last section, they should be understood as defined over the probability filtration embedding all the previous events in the experiment. For brevity, we omit the underlying filtration. Their transformation to frequencies in the virtual experiment shall be given in the next section.

% We start from the discussions of gains, which are relatively easier to deal with and sufficient to demonstrate how the decoy states come into effect. 

\subsubsection{Gains}
Among the gains $Q_{m,n}$, $Q_{*,0}$ can be directly estimated from the lower bound on $E^{s,Z}_0$; we simply focus on the discussions where $m,n\neq0$.
After phase randomization, the $Z$-basis state is given by
\begin{equation}\label{eq:decoy1}
    \hat{\rho}_{\mathrm{int}}^{Z} = \sum_{k=0}^{\infty} \Pr(k|\mathrm{int}) \hat{\rho}^Z_{[k]},
\end{equation}
where $\mathrm{int}\in\{s,w1,w2,vac\}$, and $\Pr(k|\mathrm{int})$ is the Poisson distribution given by
\begin{equation}\label{eq:poi}
    \Pr(k|\mathrm{int})=e^{-\mu_a}\frac{\mu_a^k}{k!},
\end{equation}
with $\mu_a$ the light intensity corresponding to the variable $\mathrm{int}$. Due to phase randomization, we have $\hat{\rho}^Z_{[k]} = \frac{1}{2}(\ket{k0}\bra{k0} + \ket{0k}\bra{0k})$. In each round, the light intensity is a random variable whose value is randomly chosen by Alice and unknown to Eve before post-processing. Besides, the light intensity also determines the Poisson distribution in Eq.~\eqref{eq:poi}. On the other hand, the photon number $k$ is a random variable that can be measured and fixed by Eve and unknown to Alice and Bob. As we have shown in Eq.~(9) in the main text, as Bob's homodyne POVM elements are photon-number-diagonalized, the most general operation Eve can perform on the signals is to change the photon number from $k$ to $n$. We can now define the yield values as
\begin{equation} \label{eq:Ymn}
\begin{split}
    Y^{m}_{n} &:= \Pr(n,\checkmark|m),
\end{split}
\end{equation}
which is the probability of accepting an $n$-photon state conditioned on sending an $m$-photon state.
Note that the yield value is independent of the basis and light intensity setting, as the setting choice can be postponed to the end in the entanglement-based protocol in Fig.~2(c) in the main text, i.e., after the photon numbers $m$ and $n$ are measured and the signal passes the filtering in the squashing procedure (indicated by $\checkmark$). Combined with Eq.~\eqref{eq:decoy1}, we have the following relation for the expected values:
\begin{equation}\label{eq:yieldsum}
    E^{\mathrm{int},Z}_{n} = \sum_{m=0}^{\infty}\Pr(m|\mathrm{int})Y^m_n.
\end{equation}
As analysed in Appendix~\ref{app:martingale}, we can bound $E^{\mathrm{int},Z}_{n}$ from observed statistics via Kato's inequality. Therefore, the target of Alice and Bob becomes estimating the lower bound on the yield subjected to Eq.~\eqref{eq:yieldsum}. In the protocol that utilizes up to two-photon components, we are interested in the values $Y^1_1$ and $Y^2_2$. 
The yield estimation is formalized as the following linear programming:
\begin{equation} \label{eq:decoy_Ynn}
\begin{split}
    \mr{min} &\quad Y^n_{n}, \\
    \mr{s.t.} &\quad \sum_{m=0}^{\infty} \Pr(m|\mathrm{int}) Y^m_{n} = E^{\mathrm{int},Z}_{n}, \\
    &\quad\mathrm{int}\in\{s,w1,w2,vac\}.
\end{split}
\end{equation}
Considering the Poisson distribution, we can take a numerical truncation $N_c$ in the summation over $m$ in Eq.~\eqref{eq:yieldsum}~\cite{Ma2008PhD}, which renders the equality constraints becoming inequality constraints. Taking in the Poisson distribution in Eq.~\eqref{eq:poi},
\begin{equation}
\begin{split}
\sum_{m=0}^{N_c} e^{-\mu_a}\frac{\mu_a^m}{m!} Y^m_{n} - \Delta(\mu_a, N_c) \leq E^{\mathrm{int},Z}_{n} \leq \sum_{m=0}^{N_c} e^{-\mu_a}\frac{\mu_a^m}{m!} Y^m_{n} + \Delta(\mu_a, N_c),\\
\end{split}
\end{equation}
where $\Delta(\mu_a,N_c) \leq 1 - \sum_{m=0}^{N_c} e^{-\mu_a}\frac{\mu_a^m}{m!}$ is the tail of the Poisson distribution, and it can be upper-bounded by the Chernoff bound,
\begin{equation}
    \Delta(\mu_a,N_c) \leq e^{-\mu_a} \left(\frac{e\mu_a}{N_c}\right)^{N_c}.
\end{equation}
For the two-photon protocol discussed in our work, $\mu_a<1.5$.
In practice, we set a cut-off photon number $N_c = 20$ in the linear programming. Then,
\begin{equation}
    \Delta(\mu_a,N_c) \leq e^{-1.5} \left(\frac{e\times 1.5}{20} \right)^{20} = 3.43\times 10^{-15},
\end{equation}
which is much smaller than the security parameter we set, $1\times 10^{-10}\leq \epsilon\leq 5\times 10^{-10}$. Therefore, the cut-off effect is negligible. In the following, we keep using this cut-off number. For brevity, we simply write the constraint as an equality, though it should be taken as inequalities during calculation.

\subsubsection{Phase error rates}
% The expected values of these random variables are given by
% \begin{equation}
%     \mathbb{E}[Q_{m,n}]=\mathbb{E}\left[\frac{}{}\right]
% \end{equation}
% To estimate the gain value $Q_{1,1}$ and $Q_{2,2}$, we use the following photon number component correspondence, Therefore, from the values of $\{E^{s,Z}_1\}$, we can estimate the lower bound $Q_{1,1}(L)$; from the values of $\{E^{s,Z}_2\}$, we can estimate $Q_{2,2}(L)$.

The estimation for the phase-error rates in signal pulses is more involved, as their estimation involves Alice's data on both the $Z$- and the $X$-basis. Nevertheless, for the use of decoy states, it follows a similar procedure as the gains. 
To estimate $e^X_{1,1}$ and $e^X_{2,2}$, we use the following photon number component correspondence, 
\begin{equation}\label{eq:decoy2}
\begin{split}
    \hat{\rho}_{\mathrm{int}}^{0} &= \sum_{k=0}^{\infty} \Pr(k|\mathrm{int}) \hat{\rho}^0_{[k]}, \\
    \hat{\rho}_{\mathrm{int}}^{\pi} &= \sum_{k=0}^{\infty} \Pr(k|\mathrm{int}) \hat{\rho}^{\pi}_{[k]}, \\
    \hat{\rho}_{\mathrm{int}}^{++} &= \sum_{k=0}^{\infty} \Pr(k|\mathrm{int}) \hat{\rho}^{++}_{[k]}, \\
    \hat{\rho}_{\mathrm{int}}^{+-} &= \sum_{k=0}^{\infty} \Pr(k|\mathrm{int}) \hat{\rho}^{+-}_{[k]}, \\
    \hat{\rho}_{\mathrm{int}}^{-+} &= \sum_{k=0}^{\infty} \Pr(k|\mathrm{int}) \hat{\rho}^{-+}_{[k]}, \\
    \hat{\rho}_{\mathrm{int}}^{--} &= \sum_{k=0}^{\infty} \Pr(k|\mathrm{int}) \hat{\rho}^{--}_{[k]}, \\
\end{split}
\end{equation}
% where Eq.~\eqref{eq:+state}
and the $X$-basis states are given by 
\begin{equation}
\begin{split}
    \ket{\Psi_1^+}\bra{\Psi_1^+}&=\hat{\rho}^0_{[1]}, \\
    \ket{\Psi_1^-}\bra{\Psi_1^-}&=\hat{\rho}^{\pi}_{[1]}, \\
    \ket{\Psi_2^+}\bra{\Psi_2^+}&=2\hat{\rho}^{++}_{[2]} - \hat{\rho}^{+-}_{[2]}, \\
    \ket{\Psi_2^-}\bra{\Psi_2^-}&=2\hat{\rho}^{-+}_{[2]} - \hat{\rho}^{--}_{[2]}.
\end{split}
\end{equation}
We remark that the two-photon states of $\hat{\rho}^{++}_{[2]}, \hat{\rho}^{+-}_{[2]}, \hat{\rho}^{-+}_{[2]}, \hat{\rho}^{--}_{[2]}$ correspond to the two-photon subspaces of $\hat{\rho}^{++}_{\mathrm{int}}, \hat{\rho}^{+-}_{\mathrm{int}}, \hat{\rho}^{-+}_{\mathrm{int}}, \hat{\rho}^{--}_{\mathrm{int}}$, which we defined in Eq.~\eqref{eq:+state} and~\eqref{eq:-state}.
To estimate the upper bounds on $e^X_{1,1} Q_{1,1}$ and $e^X_{2,2} Q_{2,2}$, we define the following yield values,
\begin{equation} \label{eq:YmnExpress}
\begin{aligned}
    (e^X Y)^{m}_{n} &:= \Pr(n,\checkmark,n,e^X|m), \\
    Y_{n-}^{m,+} &:= \Pr(n, \checkmark, B_{-}|m, A^+), \\ Y_{n+}^{m,-} &:= \Pr(n, \checkmark, B_{+}|m, A^-), \\
\end{aligned}
\end{equation}
where $e^X$ corresponds to the event that Alice and Bob perform $X$-basis measurement and obtain different results in the virtual protocol.
% We remark that the definition of these yield values is independent of the basis choice: in the entanglement-based protocol in Fig.~2(c) in the main text, the basis determination can be postponed after the photon number measurement of $m$ and $n$ and the squashing procedure indicated by $\checkmark$.  
Since the probability of Alice's $X$-basis measurement to obtain $+$ or $-$ is equal, we have the following yield correspondence,
\begin{equation}
    (e^X Y)^{m}_{n} = \frac{1}{2} Y_{n-}^{m,+} + \frac{1}{2} Y_{n+}^{m,-}.
\end{equation}
% However, for the case when $m>1$, we cannot directly estimate $Y_{n-}^{m,+}$ and $Y_{n+}^{m,-}$ since we are not able to prepare the $X$-basis state directly. As introduced in Sec.~IVB in the main text, we replace the $X$-basis source with the mixture of different sources: when $m=2$, we use the sources $\rho^{s,++}, \rho^{s,+-}, \rho^{s,-+}$ and $\rho^{s,--}$.
Following Eq.~\eqref{eq:decoy2}, we have the following relation:
\begin{equation}
\begin{split}
    E^{s,+}_{1-} &= \sum_{m} \Pr(m|s) Y^{m,+}_{1-}, \\  E^{s,-}_{1+} &= \sum_{m} \Pr(m|s) Y^{m,-}_{1+}, \\
    E^{s,++}_{2-} &= \sum_{m} \Pr(m|s) Y^{m,++}_{2-}, \\ E^{s,+-}_{2-} &= \sum_{m} \Pr(m|s) Y^{m,+-}_{2-}, \\
    E^{s,-+}_{2+,11} &= \sum_{m} \Pr(m|s) Y^{m,-+}_{2+,11}, \\
    E^{s,--}_{2+,11} &= \sum_{m} \Pr(m|s) Y^{m,--}_{2+,11}.
\end{split}
\end{equation}
% \blue{Note that all the definition given here is based on the entanglement-based scheme in Fig.~2(c).}
We can then perform the following linear programming to bound the yield values $Y_{n,n}^{+,-}$ and $Y_{n,n}^{-,+}$:

The linear program for $Y_{1-}^{1,+}$ is
    \begin{equation}
    \begin{aligned}
        \mr{max} &\quad Y_{1-}^{1,+} \\
        \mr{s.t.} &\quad \sum_{m=0}^{N_c} \Pr(m|\mathrm{int}) Y_{1-}^{m,+} = E^{\mathrm{int},+}_{1-}, \\
        &\quad \mathrm{int}\in\{s,w1,w2,vac\}.
    \end{aligned}
    \end{equation}

    The linear program for $Y_{1+}^{1,-}$ is
    \begin{equation}
    \begin{aligned}
        \mr{max} &\quad Y_{1+}^{1,-} \\
        \mr{s.t.} &\quad \sum_{m=0}^{N_c} \Pr(m|\mathrm{int}) Y_{1+}^{m,-} = E^{\mathrm{int},-}_{1+}, \\
        &\quad \mathrm{int}\in\{s,w1,w2,vac\}.
    \end{aligned}
    \end{equation}

    The linear program for $Y^{2,++}_{2-}$ is
    \begin{equation}
    \begin{aligned}
        \mr{max} &\quad Y^{2,++}_{2-} \\
        \mr{s.t.} &\quad \sum_{m=0}^{N_c} \Pr(m|\mathrm{int}) Y^{m,++}_{2-} = E^{\mathrm{int},++}_{2-}, \\
        &\quad \mathrm{int}\in\{s,w1,w2,vac\}.
    \end{aligned}
    \end{equation}

    The linear program for $Y^{2,+-}_{2-}$ is
    \begin{equation}
    \begin{aligned}
        \mr{min} &\quad Y^{2,+-}_{2-} \\
        \mr{s.t.} &\quad \sum_{m=0}^{N_c} \Pr(m|\mathrm{int}) Y^{m,+-}_{2-} = E^{\mathrm{int},+-}_{2-}, \\
        &\quad \mathrm{int}\in\{s,w1,w2,vac\}.
    \end{aligned}
    \end{equation}

    The linear program for $Y^{2,-+}_{2+,11}$ is
    \begin{equation}
    \begin{aligned}
        \mr{max} &\quad Y^{2,-+}_{2+,11} \\
        \mr{s.t.} &\quad \sum_{m=0}^{N_c} \Pr(m|\mathrm{int}) Y^{m,-+}_{2+,11} = E^{\mathrm{int},-+}_{2+,11}, \\
        &\quad \mathrm{int}\in\{s,w1,w2,vac\}.
    \end{aligned}
    \end{equation}

    The linear program for $Y^{2,--}_{2+,11}$ is
    \begin{equation}
    \begin{aligned}
        \mr{min} &\quad Y^{2,--}_{2+,11} \\
        \mr{s.t.} &\quad \sum_{m=0}^{N_c} \Pr(m|\mathrm{int}) Y^{m,--}_{2+,11} = E^{\mathrm{int},--}_{2+,11}, \\
        &\quad \mathrm{int}\in\{s,w1,w2,vac\}.
    \end{aligned}
    \end{equation}

Finally, for the two-photon phase-error rate, we have
\begin{equation}
\begin{split}
    (e^X Y)^2_2 &= \frac{1}{2} Y_{2-}^{2,+} + \frac{1}{2}  Y_{2+}^{2,-} \\
    &= 
    \frac{1}{2}( 2 Y^{2,++}_{2-} - Y^{2,+-}_{2-} ) + \frac{1}{2}( 2 Y^{2,-+}_{2+,11} - Y^{2,--}_{2+,11} ).
\end{split}
\end{equation}

\subsection{Applying the Chernoff-Hoeffding bound}\label{app:Chernoff}
From the estimation of the underlying expected values of $Q^*_{0}, Y^1_{1}, Y^2_{2}, (e^X Y)^1_{1}$ and $(e^X Y )^2_{2}$, in this section, we bound the corresponding frequency values in all the accepted $Z\to Z$ rounds with the signal intensity: $M^{Z,s}_{*,0}, M^{Z,s}_{1,1}, M^{Z,s}_{2,2}, M^{eX,s}_{1,1}$ and $M^{eX,s}_{2,2}$. 
Here, the superscript $eX$ indicates the event of an $X$-basis measurement error. As a reminder, the expected values are given by
\begin{equation}\label{eq:ExptoFreq}
\begin{split}
    \mathbb{E}[M^{Z,s}_{*,0}] &= \Pr(n=0,\checkmark|Z,s,Z^B)\Pr(Z^B)\cdot N^{Z,s} \\ 
    &= Q^*_{0}\Pr(Z^B) N^{Z,s}, \\
    \mathbb{E}[M^{Z,s}_{n,n}] &= \Pr(n,\checkmark|Z,s,Z^B,n)\Pr(n|Z,s)\Pr(Z^B)\cdot N^{Z,s} \\
    &= Y^n_{n} \Pr(n|s) \Pr(Z^B) N^{Z,s}, \quad n=1,2, \\
    \mathbb{E}[M^{eX,s}_{n,n}] &= \Pr(n,\checkmark,eX|Z,s,Z^B,n)\Pr(n|Z,s)\Pr(Z^B)\cdot N^{Z,s} \\
    &= (e^X Y)^n_{n} \Pr(n|s) \Pr(Z^B) N^{Z,s}, \quad n=1,2.
\end{split}
\end{equation}
% Here, $m$ and $n$ indicate the emitted photon number and the received photon number, respectively. $\checkmark$ indicates the event which pass the squash filter, and $eX$ indicates the event which introduce a $X$-basis measurement error.
Again, the expected values should be understood as conditioned on proper probability filtrations. 
% We can estimate their frequency values in the virtual protocol by applying Kato's inequality from (conditional) expected values to frequencies. Nevertheless, we can adopt tighter bound based on Chernoff-Hoeffding inequality of Bernoulli's variables.

To calculate the final key length, we need to obtain the frequency values of $M_{n,n}^{Z,s}$ and $M_{n,n}^{eX,s}$. That is, in all the $Z\rightarrow Z$ rounds where Alice emits $n$ photons and Bob accepts $n$ photons, how many of them is chosen to be with signal intensity $s$, and how many of them will be detected to have phase errors if they were measured on the $X$-basis. To this end, we apply the techniques in Ref.~\cite{zhang2017improved}, which applies the Chernoff-Hoeffding bound to Bernoulli random variables to convert the expectation value of $\mathbb{E}[M^{Z,s}_{n,n}]$ and $\mathbb{E}[M^{eX,s}_{n,n}]$ to their frequency values. We first review the setting of Chernoff-Hoeffding bound. Consider $N$ independent Bernoulli random variable $\chi_i\in\{0,1\}$ with probability $p_i$ to be $1$ and $(1-p_i)$ to be $0$. In general, these variables might not be identical. We denote their summation as $\chi = \sum_{i=1}^n \chi_i$. It is linked with its expectation value $\chi_E := \mbb{E}[\chi]$ via the Chernoff-Hoeffding bounds:
\begin{equation} \label{eq:ChernoffBernoulli}
\begin{split}
    \Pr[\chi>(1+\delta^L)\chi_E] &< \left[ \frac{ e^{\delta^L} }{ (1+\delta^L)^{1+\delta^L} } \right]^{\chi_E} = g(\delta^L, \chi_E), \\
    \Pr[\chi<(1-\delta^U)\chi_E] &< \left[ \frac{ e^{-\delta^U} }{ (1-\delta^U)^{1-\delta^U} } \right]^{\chi_E} = g(-\delta^U, \chi_E), \\
\end{split}
\end{equation}
where $\delta^L >0, 0<\delta^U<1$, and $g(\delta,\chi_E):=\left[ \frac{e^\delta}{(1+\delta)^{1+\delta}} \right]$. 
Consider the case when we have obtained bounds on the expected value $\chi_E$. We can thus apply Eq.~\eqref{eq:ChernoffBernoulli} to estimate the frequency values $\chi$. To be more specific, we first set the lower and upper bounds on the estimation of $\chi$ as
\begin{equation}
\begin{aligned}
    \chi^L &:= (1- \delta^L) \chi_E, \\
    \chi^U &:= (1+ \delta^U) \chi_E.
\end{aligned}
\end{equation}
The failure probability, i.e., the probability of $\chi\notin[\chi^L, \chi^U]$, is given by:
\begin{equation} \label{eq:failureProb}
    \epsilon = \exp\left(- \frac{(\delta^L)^2 \chi_E}{2+\delta^L}  \right) +\exp\left(- \frac{(\delta^U)^2 \chi_E}{2+\delta^U}  \right).
\end{equation}
In practice, we can preset the lower and upper bound $[\chi^L, \chi^U]$ by assuming a Gaussian distribution on $\chi$ first,
\begin{equation}
\begin{split}
    \chi^L &= \chi - n_\alpha \sqrt{\chi_E}, \\ \chi^U &= \chi + n_\alpha \sqrt{\chi_E},
\end{split}
\end{equation}
where $n_\alpha$ is a preset parameter to determine the estimation precision. After that, we calculate the failure probabilities by Eq.~\eqref{eq:failureProb}.
Note that, if $\chi_E$ is too small such that $\chi_L$ becomes $0$, the failure probability is also $0$ since $\Pr(\chi<0)=0$.

Next, we show how to apply Eq.~\eqref{eq:ChernoffBernoulli} for the decoy-state problem. We show the estimation of $M_{1,1}^{Z,s}$ and $M_{2,2}^{Z,s}$, and the estimation for other quantities follows the same procedure. First, notice that when the emitted photon number $m$ is given, the intensity variable $\mr{int}$ is independent and identically distributed by the distribution $\Pr(\mr{int}|m)$. In the following, we follow the approach in Ref.~\cite{zhang2017improved} to estimate the frequency values from expectation values.

To be specific, we assume that during parameter estimation, Alice and Bob first determine the frequency of $Z\rightarrow Z$ rounds $M_{m,n}^Z$ where $m$ photons are emitted and $n$ photons are accepted. Conditioned on the emitted photon number $m$, the intensity label $\mr{int}\in\{s,w1,w2,vac\}$ is i.i.d. over the rounds. Among all the rounds $M_{m,n}^Z$, denote the number of rounds with intensity $\mr{int}$ as $M_{m,n}^{Z,\mr{int}}$. We can write $M_{m,n}^{Z,\mr{int}}$ as a summation of Bernoulli random variables
\begin{equation}
    M_{m,n}^{Z,\mr{int}} = \sum_{j=1}^{M_{m,n}^Z} \hat{\chi}_{m,n;j}^{\mr{int}}, \quad \mr{int}\in\{s,w1,w2,vac\}
\end{equation}
where
\begin{equation}
    \hat{\chi}_{m,n;j}^{\mr{int}} = 
\begin{cases}
    1, &\quad \text{with probability } \Pr(\mr{int}|m), \\
    0, &\quad \text{with probability } 1- \Pr(\mr{int}|m), \\
\end{cases}.
\end{equation}
The conditional probability $\Pr(\mr{int}|m)$ is related to the source properties,
\begin{equation}
    \Pr(\mr{int}|m) = \frac{ \Pr(m|\mr{int}) \Pr(\mr{int}) }{ \sum_{\mr{int}'}\Pr(m|\mr{int}') \Pr(\mr{int}') },
\end{equation}
where $\Pr(m|\mr{int})$ is the Poisson distribution given by Eq.~\eqref{eq:poi}.

% We then have,
% \begin{equation}
% \begin{aligned}
%     & \mbb{E}\left[ M_{m,n}^{\mr{int}} \right] = \Pr(\mr{int}|m) M_{m,n}, \\
%     \Rightarrow & \mbb{E}\left[ M_{*,n}^{\mr{int}} \right] = \sum_m \Pr(\mr{int}|m) M_{m,n}
% \end{aligned}
% \end{equation}

% Now, if we divide both side by the number of $ZZ$-basis rounds $N^{\mr{int}}$ when Alice emits the pulse with intensity $\mr{int}$, from the definition of $Y^m_n$ in Eq.~\eqref{eq:Ymn} and $E_n^{\mr{int},Z}$ in Eq.~\eqref{eq:poisson}, we can then retrieve the decoy-state formula in Eq.~\eqref{eq:decoy_Ynn},
% \begin{equation}
% \begin{aligned}
%     & \mbb{E}\left[ \frac{ M_{*,n}^{\mr{int}} }{ N^{\mr{int}} } \right] = \sum_m \Pr(\mr{int}|m) \frac{M_{m,n}}{N^{\mr{int}}}, \\
%     &\Rightarrow E_n^{\mr{int},Z} = \sum_{m=0}^{\infty} \Pr(m|\mr{int}) Y^m_n.
% \end{aligned}
% \end{equation}
% We omit the derivation since it is similar to the one in Eq.~(13) in Ref.~\cite{zhang2017improved}. 

From the decoy-state estimation of yield values $Y^1_1$ and $Y^2_2$ in the previous step, we can learn the expectation values of $\mbb{E}[M_{1,1}^Z]$ and $\mbb{E}[M_{2,2}^Z]$ by 
\begin{equation}
    \mbb{E}[M_{n,n}^Z] = \Pr(n,\checkmark|Z,Z^B,n) \Pr(Z^B) N^{Z} = Y^n_n \Pr(Z^B) N^{Z}.  
\end{equation}
The remaining problem is to determine the number of rounds $M_{1,1}^{Z,s}$ and $M_{2,2}^{Z,s}$ with signal-state intensity $s$. This can be done by applying the Chernoff-Hoeffding bound in Eq.~\eqref{eq:ChernoffBernoulli}, where $\chi_E$ is set as the expectation value $\Pr(\mr{int}|m) \mbb{E}[M_{m,n}^Z]$, which can be obtained from Eq.~\eqref{eq:ExptoFreq}, and $\chi$ is set as the frequency value $M_{m,n}^{Z,\mr{int}}$. 
In a similar fashion, one can also estimate $M^{Z,s}_{*,0}$, $M_{1,1}^{eX,s}$ and $M_{2,2}^{eX,s}$ from the expectation yield values $Q^*_0$, $(e^X Y)^1_1$ and $(e^X Y)^2_2$, respectively.

\subsection{Putting everything together}

To end this section, we give a formal theorem of the finite-size key generation result.

\begin{theorem}
Consider the time-bin CV QKD protocol in Table 3 of the main text with reverse reconciliation. Suppose the total number of rounds is $N$, the number of $ZZ$-basis key generation rounds with intensity $s$ is $M^{Z,s}$, and the quantum bit error rate is $e^Z$. Given the failure probability in parameter estimation $\varepsilon_{\mathrm{pe}}\in(0,1)$, and the failure probability in privacy amplification $\varepsilon_{\mathrm{pa}}\in(0,1)$, conditioned on the success of information reconciliation with efficiency $f$,  except a total failure probability $\varepsilon=\varepsilon_{\mathrm{pe}}+\varepsilon_{\mathrm{pa}}$, the secure key length is lower-bounded by
\begin{equation}
    n\geq n_{\mathrm{key}}=\underline{M_{*,0}^{Z,s}} + \underline{M_{1,1}^{Z,s}} \left[1-h\left(\frac{\overline{M_{1,1}^{eX,s}}}{\underline{M_{1,1}^{Z,s}}}\right)\right] + \underline{M_{2,2}^{Z,s}}\left[1-h\left(\frac{\overline{M_{2,2}^{eX,s}}}{\underline{M_{2,2}^{Z,s}}}\right)\right] - f M^{Z,s} h(e^{Z,s})-\log\frac{1}{\varepsilon_{\mathrm{pa}}},
\end{equation}
where $\underline{M_{*,0}^Z}$ is the lower bound on the number of key generation rounds accepting a vacuum state, $\underline{M_{1,1}^{Z,s}}$ is the lower bound on the number of key generation rounds sending and accepting a single-photon state while Alice emit intensity $s$ pulses, ${\underline{M_{2,2}^{Z,s}}}$ is the lower bound on the number of key generation rounds sending and accepting a two-photon state while Alice emit intensity $s$ pulses, $\overline{M_{1,1}^{eX,s}}$ is the upper bound on the single-photon phase errors, and $\overline{M_{2,2}^{eX,s}}$ is the upper bound on the two-photon phase errors. 
Following the parameter estimation procedure through Appendix~\ref{app:EstProcedure} to~\ref{app:Chernoff}, the estimation failure probability $\varepsilon_{\mathrm{pe}}$ is the summation of those for each term, and each term is composed of the failure probability in applying Kato's inequality [see Eq.~\eqref{eq:martingalefail}] and the one in applying the Chernoff-Hoeffding bound [see Eq.~\eqref{eq:failureProb}]. 
The secure key rate per round is then lower-bounded by $r\geq n_{\mathrm{key}}/N$.
\end{theorem}

\end{document}